\documentclass[10pt,aps,prx,amsmath,amsfonts,amssymb,floatfix,longbibliography,superscriptaddress,notitlepage]{revtex4-1}
\usepackage{mathrsfs} \usepackage{graphicx,xcolor} \usepackage{bbm} \usepackage{multirow}
\usepackage[normalem]{ulem}
\usepackage{lineno}
\usepackage{tabularx}
\usepackage{here}

\usepackage[colorlinks=true]{hyperref}
\hypersetup{
  colorlinks=true,
  linkcolor=blue,
  filecolor=magenta,
  citecolor=blue,
  urlcolor=blue,
}

\newcommand{\red}[1]{\textcolor{black}{#1}}

\newcommand{\blue}[1]{\textcolor{black}{#1}}

\def\to{\rightarrow}

\newcommand{\eq}[1]{Eq.~(\ref{#1})}
\newcommand{\beq}{\begin{equation}} \newcommand{\eeq}{\end{equation}}
\newcommand{\beqn}{\begin{eqnarray}} \newcommand{\eeqn}{\end{eqnarray}}

\renewcommand{\phi}{\varphi}
\newcommand{\bs}{\blacksquare}
\newcommand{\ws}{\square}

\newcommand{\bc}{\begin{center}}
\newcommand{\ec}{\end{center}}

\newenvironment{redtext}{\par\color{black}}{\par}

\begin{document}

%\title{Teacher-student scenario in a deep neural network}
\title{Spatially heterogeneous learning by a deep student machine}
%\title{Statistical mechanics of a deep student machine}
%\title{How a deep student machine learns}
%\title{An anatomy of a deep learning student}

\author{Hajime Yoshino}
\affiliation{Cybermedia Center, Osaka University, Toyonaka, Osaka 560-0043, Japan}
\affiliation{Graduate School of Science, Osaka University, Toyonaka, Osaka 560-0043, Japan}

\begin{abstract}
Despite the spectacular successes, deep neural networks (DNN) with a huge number of adjustable parameters remain largely black boxes.
To shed light on the hidden layers of DNN,
we study supervised learning by a DNN of width $N$ and depth $L$
consisting of $NL$ perceptrons with $c$ inputs by a statistical mechanics approach called the teacher-student setting.
We consider an ensemble of student machines that exactly reproduce
$M$ sets of $N$ dimensional input/output relations provided by a teacher machine.
\red{We show that the statistical mechanics problem becomes exactly solvable in a high dimensional limit
which we call as 'dense limit': $N \gg c \gg 1$ and $M \gg 1$ with fixed $\alpha=M/c$
using the replica method developed in 
(Hajime Yoshino, SciPost Phys. Core 2, 005 (2020)) \cite{yoshino2020complex}.}
 \red{In parallel to the theoretical study,}
 we also study the model numerically performing simple greedy Monte Carlo simulations.
 \red{Simulations reveal} that learning by the DNN is quite heterogeneous in the network space: configurations of \red{the teacher and the student} machines are more correlated within the layers closer to the input/output boundaries while the central region remains much less correlated due to the over-parametrization
\red{in qualitative agreement with the theoretical prediction.}
%Deep enough systems relax faster thanks to the less correlated central region.
\red{
  %Using the exact solution of the theory in the dense limit
  %and fully equilibrated samples of machines obtained by simulations, 
We evaluate the generalization-error of the DNN with various depth $L$ both theoretically and numerically.  
}
Remarkably both the theory and simulation suggest generalization-ability of the student machines,
which are only weakly correlated with the teacher in the center,
does not vanish even in the deep limit $L \gg 1$ where the system becomes heavily over-parametrized.
We also consider the impact of effective dimension $D(\leq N)$ of data by
incorporating the hidden manifold model
(Sebastian Goldt, Marc M\'{e}zard, Florent Krzakala, and Lenka Zdevorov\'{a}, Physical Review X 10, 041044 (2020).
 \cite{goldt2020modeling}) into our model.
The replica theory implies that the loop corrections to the dense limit, which reflect correlations between different nodes in the network, become enhanced by either decreasing the width $N$ or decreasing the effective dimension $D$ of the data. Simulation suggests both leads to significant improvements in generalization-ability.
\end{abstract}
\maketitle

%\tableofcontents

\section{Introduction}
\label{sec-introduction}

The mechanism of machine learning by deep neural networks (DNN)
\cite{lecun2015deep} remains largely unknown. One of the most puzzling points is \red{the issue of over-parametrization:
supervised learning by DNN can work even in the regime where the number of adjustable parameters is larger than the data size by orders of magnitudes.}
This goes sharply against the traditional wisdom of data modeling: for example, one should avoid fitting $10$ data points by a fitting function with $100$ adjustable parameters, which is just nonsense. However empirically it has been found repeatedly that such over-parametrized DNNs can somehow avoid over-fitting and generalize well, i.~e. they can successfully describe new data not used during training. Uncovering the reason for this \red{peculiar} phenomenon is a very interesting and challenging scientific problem \cite{carleo2019machine,geiger2020scaling}. An important point to be noted is that the effective dimension $D$ of the data which can be much smaller than the apparent dimension $N$ of data. It has been shown in studies of shallow networks that the generalization ability improves by increasing $N/D$ due to a kind of self-averaging mechanism \cite{mei2022generalization,loureiro2022fluctuations,d2021triple}.
However, the generalization ability of the deeper system remains unexplained.

\begin{redtext}
Statistical mechanics on neural networks has a long history that dates back to the 1980’s \cite{amit1985spin,gardner1988space,gardner1989three}.
Studies on the single perceptrons \cite{gardner1988space,gardner1989three} and shallow networks \cite{mei2022generalization,gerace2020generalisation}
have provided many useful insights and some progresses have been made also on deeper networks \cite{gabrie2018entropy,aubin2019spiked,schroder2023deterministic,cui2023optimal}. However what is going on in the hidden layers remain largely unknown.
The first attempt to uncover the black box was made in \cite{yoshino2020complex} by the present author 
based on the replica method and predicted unexpected phenomenology of DNN: spatially heterogeneous learning.
Unfortunately the theory suffered a serious problem due to an uncontrolled approximation and the validity of the prediction remained elusive.
%(Amit et al. 1985; Gardner \& Derrida 1987)..
\end{redtext}

      \begin{figure}[h]
    \bc
      \includegraphics[width=0.5\textwidth]{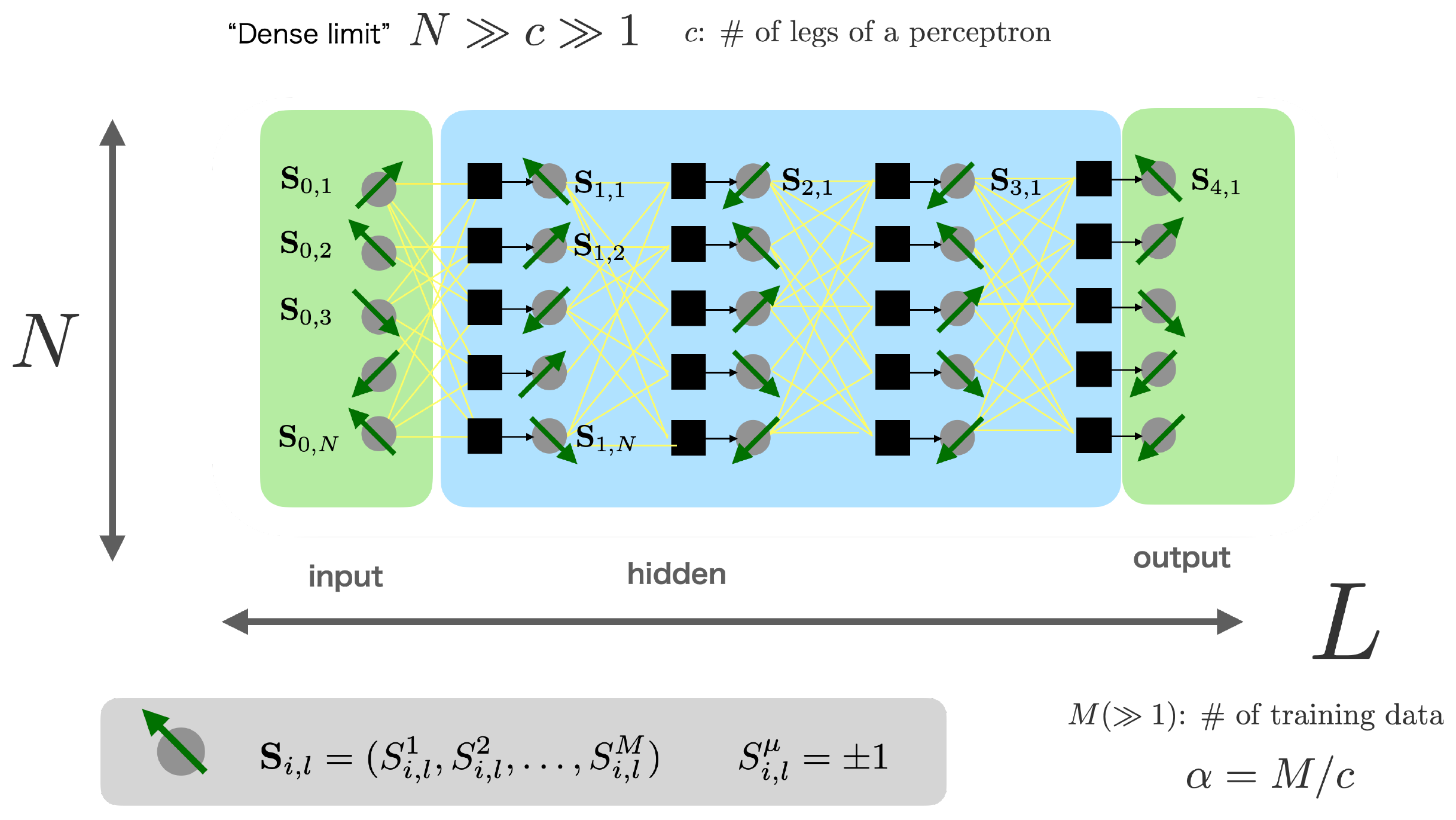}
  \ec
  \caption{
    Schematic picture of the multi-layer perceptron network
    of depth $L$ and width $N$. In this example,
    the depth is $L=4$. 
    Each arrow represents a $M$-component vector spin ${\bf S}_{i}=(S_{i}^{1},S_{i}^{2},\ldots,S_{i}^{M})$
    with its component $S_{i}^{\mu}=\pm 1$ representing the state of a 'neuron' in the $\mu$-th pattern.
  }
   \label{fig_schematic_model}
  \end{figure}

\begin{figure}[t]
    \bc
      \includegraphics[width=0.5\textwidth]{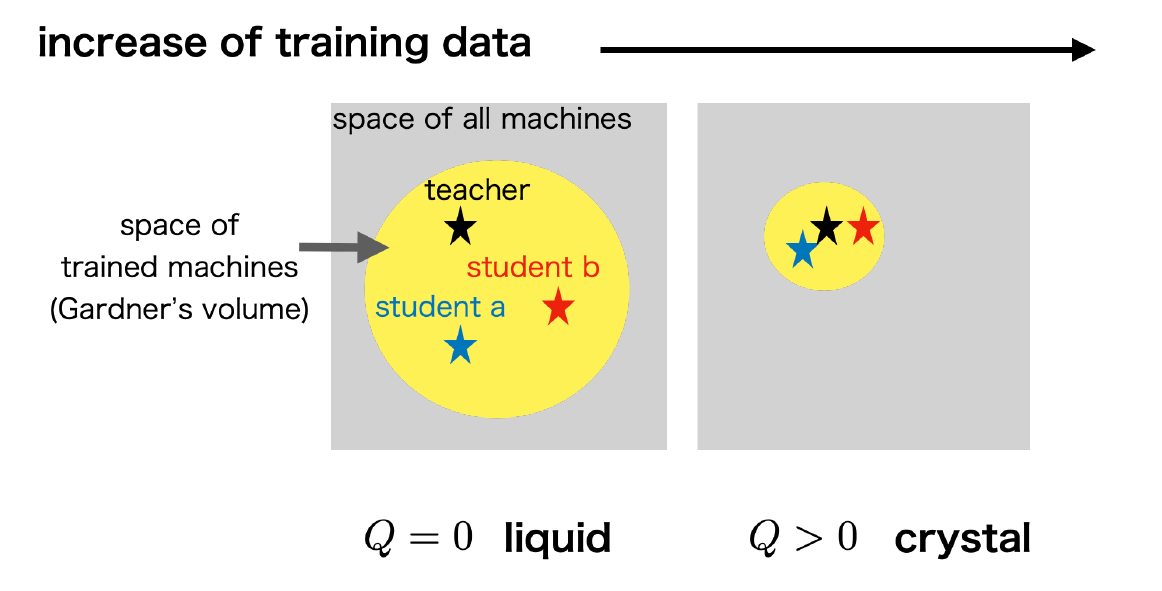}
  \ec
  \caption{
    Schematic picture of the phase space of machines: the gray box represents
    the set of all machines which can be generated varying
    the parameters (e.g. synaptic weights) given a network structure.
    The yellow region resents the subspace in which
    machines agree teacher's machine for a given $M$ set of training data.
    Liquid phase: if the number of the training data $M$ is small,
    the subspace is so large that 
    the machines are typically widely separated and their mutual overlap
    $Q$ is typically zero . 
    Crystalline phase: $M$ is large enough,
    that machines have finite overlap $Q$ with respect to each other.
  }
   \label{fig_gardner_volume}
  \end{figure}

To understand the mechanism for the  generalization ability of deep networks,
we study supervised learning by DNN considering the so-called teacher-student setting which is a canonical setting to study statistical inference problems \cite{engel2001statistical,zdeborova2016statistical} by methods of statistical mechanics.
We consider a prototypical DNN of rectangular shape with width $N$ and depth $L$ consisting of $NL$ perceptrons with $c$ inputs,
which defines a mapping between a $N$ dimensional input vector to a $N$ dimensional output vector \red{(See Fig.~\ref{fig_schematic_model})}.
For the data, we consider $M$ pairs of input/output vectors provided by a teacher machine and
we consider an ensemble of student machines that exactly satisfy the same input/output relations as the teacher.
The phase space volume of such an ensemble is called as Gardner's volume \cite{gardner1988space,gardner1989three}
which should be very large for over-parametrized DNNs.
\red{In fact, it is known that gradient descent dynamics
find such a machine without going over barriers in the loss landscape} \cite{jacot2018neural,mei2018mean,chizat2018global}.
In Fig.~\ref{fig_gardner_volume}, we show a schematic picture of the phase space of the machines. If $M$ is small, typically students will not
find the teacher. This situation would be regarded as {\it liquid phase}.
If $M$ is increased, {\it crystalline phase} may emerge
in which students find the (hidden) crystal, i.~e. teacher.
We also wish consider the impact of effective dimension $D$ of the data by incorporating the hidden manifold model \cite{goldt2020modeling,gerace2020generalisation,goldt2022gaussian}) in our model.
Using methods of statistical mechanics we wish to
investigate how different machines which satisfy the
same set of input/output boundary conditions become
correlated with each other in the hidden layers
and evaluate their generalization ability : the ability of the students to reproduce the teacher's output against new input data not used in training.

\red{  
The first attempt to tackle the statistical mechanics problem of DNN was made recently in \cite{yoshino2020complex}
based on the replica method in a high dimensional limit with $c=N=D \gg 1$ and $M \gg 1$ with fixed $\alpha=M/c$.
Unfortunately it suffered from an uncontrolled 'tree' approximation which is invalid for the global coupling case $c=N$.
In the present paper we show that the problem can be overcome in the limit $N \gg c \gg 1$, which we call as {\it dense limit}.      
As the result we establish an exactly solvable statistical mechanics model of DNN which has been waited for a long time.}

\red{Using the exact solution of the model which can be obtained by the replica method developed in \cite{yoshino2020complex}, we analyze the key question: generalization-ability of DNN including the over-parametrized regime.}
%using the scheme proposed by \cite{levin1990statistical,opper1996statistical} in our replica theory.
We also show that the effect of the finiteness of the width (apparent dimension of the data) $N$
and the effective dimension $D$ similarly enhances loop corrections which induce correlations between distant layers.
\red{In parallel to the theoretical study, we also perform extensive numerical simulations to examine the theoretical predictions.} We use the Monte Carlo method which allows more efficient exploration of the solution space compared with the usual gradient descent algorithms.

\begin{redtext}
The following sections are organized as follows.
In sec~\ref{sec-summary-and-results} we summarize the main results of this paper.
In sec~\ref{sec-model} we introduce our model.
%the dense coupling $N \gg c \gg 1$ which plays key roles in this work 
%and the teacher-student setting. Then we define the partition function of the
%machines called Gardner's volume.
%We point out a gauge symmetry in the problem.
%Finally, we introduce the hidden manifold model.
We discuss the replica approach in sec \ref{sec-replica-theory}
and numerical simulations in sec \ref{sec-simulation}.
%We first show the formalism of our replica theory. We introduce the overlaps which are the order parameters and the scheme to evaluate the generalization error. Then we show the result of an analysis on DNN with various depths. Finally, we discuss the effects of finite width $N$, hidden dimension $D$, and finite connectivity $c$ as corrections to the dense limit.
%In sec \ref{sec-simulation} we discuss numerical simulations.
%We first introduce the two setups: The Bayes-optimal scenario and hidden manifold scenario,
%our greedy Monte Carlo method, and different initializations. Then we introduce our observables: overlaps and generalization errors. For the overlaps, we introduce squared overlaps which are invariant under gauge transformations
%and permutations. Then we discuss the results of the simulations.
In sec~\ref{sec-conclusions} we conclude this paper with perspectives.
In appendix \ref{sec-transfermatrix} we discuss a connection between some layered spinglass models and DNNs and in appendix \ref{sec-replica-appendix} we present some details of the replica theory.
\end{redtext}

\begin{redtext}
\section{Summary of results}
\label{sec-summary-and-results}

Let us summarize below the main results of this work.
On the theoretical side we find the following.
\begin{itemize}
\item We establish an exactly solvable statistical mechanics
  model of DNN in the dense Limit $N \gg c \gg 1$.
  The exact solution of the model is obtained using replica the approach.
  It is shown that the correction to the dense limit due to finiteness of the width $N$ can be expressed by loop-corrections.
Fortunately, it turns out that
the theoretical results presented in \cite{yoshino2020complex}
are essentially valid in the dense limit $N \gg c \gg 1$
although they are unjustified for the global coupling $c=N$ assumed there.

\item We show that smallness of the effective dimension $D(< N)$ of the hidden manifold model  \cite{goldt2020modeling} enhances the loop corrections.
Thus finite dimension $D$ effect is predicted to be similar to finite width $N$ effect.
    
\item The learning curve $\epsilon=\epsilon_{L}(\alpha)$
of the DNN with various depth $L$ is analyzed
evaluating the generalization error $\epsilon_{L}(\alpha)$ in the case of Bayes optimal teacher-student setting where the replica symmetry holds.
It becomes independent of the depth $L$, i.~e.
$\epsilon_{L}(\alpha)=\epsilon_{\infty}(\alpha)$
as long as the network is deep enough such that liquid phase, where
students are de-correlated from the teacher,
remain in the center reflecting strong over-parametrization.
%which decreases increasing $\alpha$,
%The generalization ability improves even in such
%a situation increasing $\alpha$, by which the crystallie
%phase,  where the students are correlated with the teacher,
%grow from the input/output boundaries.  
\end{itemize}

On the numerical side we find the following.
\begin{itemize}
\item We simulated the model with finite connectivity $c$
and width $N$
in the Bayes-optimal teacher-student setting and found that a simple greedy Monte Carlo algorithm allows the student machines to equilibrated after sufficiently long times.
Thus typical equilibrium states are accessible starting from 
typical random initial configurations without going over barriers in the loss landscape.
\item Observation of the overlap between the machines reveal
  spatially in-homogeneous learning  in qualitative agreement with the theory.
  While the theory in the dense limit $N \gg c \gg 1$ predicts, in the case of strong over-parametrization,
  crystalline regions with finite overlap close to the input/output
  layers separated by a liquid region with zero overlap in the center, 
  the distinction between the crystalline and the liquid phases become
  blurred in systems with finite width $N$ and finite connectivity $c$.
   Nonetheless, the presence of the liquid like region in the center becomes clearer
   by making the width $N$ large and the connectivity $c$ large or the depth $L$ large. We consider that the remnant overlap left in the center
   by the finite width $N$ and finite connectivity $c$ effects
   play the role of symmetry breaking field which connect the two crystalline regions attached to the boundaries.
\item Observation of the learning curve
 $\epsilon=\epsilon_{L,c,N,D}(\alpha)$
  reveal that it becomes independent of the depth $L$ in deep enough systems in agreement with the theoretical prediction.
\item The observation reveal that finite effective dimension $D$ effect and finite width $N$ effects are indeed very similar as suggested by the consideration of the loop effects in the theory.
  The generalization error
$\epsilon_{L,c,N,D}(\alpha)$  decreases significantly
  decreasing either the width $N$ or the effective dimension $D$.
\end{itemize}

\end{redtext}

\section{Model}
\label{sec-model}

      \subsection{Multi-layer perceptron network}
      \label{subsec-model-dnn}

We consider a simple multi-layer neural network of a rectangular shape
with  width $N$ and depth $L$ (see Fig.~\ref{fig_schematic_model}).
The input and output layers are located at the boundaries $l=0$ and $L$
respectively while $l=1,2,\ldots,L-1$ are hidden layers.
On each layer $l=0,1,2,\ldots,L$ there are $N$ neurons labeled as
$(l,i)$ with $i=1,2,\ldots,N$.
The state of the neuron $(l,i)$
is represented by an Ising spin $S_{l,i}$: it is active if $S_{l,i}=1$ and inactive if $S_{l,i}=-1$.

\red{The network is constructed as follows.}
There are $N_{\bs}=NL$ perceptrons.
Consider a perceptron $\bs=(l,i)$ which is the $i$-th neuron in the $l$-th layer.
%i.~e. ${\bf S}_{\bs}={\bf S}_{l,i}$,
It receives $c$ inputs from the outputs of the perceptrons $\bs(k)$ $(k=1,2,\ldots,c)$ in the previous $l-1$-th layer, weighted by ${\bf J}_{\bs}=(J_{\bs}^{1},J_{\bs}^{2},\ldots,J_{\bs}^{c})$. 
(For the special case $l=1$, $\bs(k)$ should be understood as one of the spins in the input layer.)
The $c$ perceptrons are selected randomly out of  $N$ possible perceptrons in the $l-1$ th layer.

The output of the perceptron $\bs$, which we denote as $S_{\bs}$, is given by,
 \beq
S_{\bs}={\rm sgn} \left(\frac{1}{\sqrt{c}}\sum_{k=1}^{c} J_{\bs}^{k}S_{\bs(k)}\right)
 \label{eq-perceptron}
 \eeq
where ${\rm sgn}(y)=y/|y|$ is our choice for the activation function. We assume that the synaptic weights $J^{k}_{\bs}$ take real numbers normalized such that,
\beq
\sum_{k=1}^{c}(J^{k}_{\bs})^{2}=c.
\label{eq-J-normalization}
\eeq 

For convenience, we call the state of the neurons $S_{l,i}$'s  as 'spins', and the synaptic weights $J_{\bs}^{k}$s as 'bonds' in the present paper.
We denote the set of perceptrons in the $l$-th layer as $\bs \in l$
and denote the set of perceptrons whose outputs become input
for $\bs$ as $\partial \bs$, i.~e. $\partial \bs=\{ \bs(1), \bs(2),\ldots, \bs(c) \}$. For convenience we introduce also $\bs \in 0$
so that we can write the set of spins in the input layer as $S_{\bs \in 0}$.

\subsection{\protect\red{Dense coupling}}
\label{sec-dense-coupling}

As stated above, $c$ legs of a perceptron $\bs$ at $l$-th layer is connected to $c$ neurons $S_{\bs(k)}$ $(k=1,2,\ldots,c)$
in the previous $l-1$-th layer. The $c$ neurons out of $N$ possible neurons are selected randomly.
Thus our graph becomes  a sort of sparse (layered) random graph when $c$ is finite.
\red{We will find in sec.~\ref{sec-replica-theory} that this construction enables us to obtain an exactly solvable statistical mechanics model of
DNN because of the following reasons.}

\begin{itemize}
\item The graph becomes locally tree-like as in the case of Bethe-lattices so that contributions of 'loops' can be neglected in the wide limit $N \to \infty$ with fixed $c$.
  This can be seen as follows. For instance, consider a loop $0 \to 1 \to 2 \to 3 \to 0$ shown in Fig.~\ref{fig_loop}.
  Starting from $0$, choose any $1$ connected to $0$.
  Then choose any $2$  connected to $1$.
  Then choose any $3$ (different from $1$) connected to $0$.
In the case of global coupling $c=N$, $2$ is certainly connected to $3$ completing a loop. However, in the case of dense coupling, in a given realization of the random graph, 
$2$ is connected to $3$ only with a probability $\sim c/N$.
Thus in the limit $N \to \infty$ with fixed $c$, the probability to complete the loop vanishes.
This argument can be generalized for 2-loops, 3-loops,....which happen with probability $O(c/N)^{2})$, $O(c/N)^{3})$, ...
\red{Note that the loops cannot be neglected.
In the case of global coupling $c=N$ (assumed in \cite{yoshino2020complex}).}

      \begin{figure}[h]
    \bc
      \includegraphics[width=0.5\textwidth]{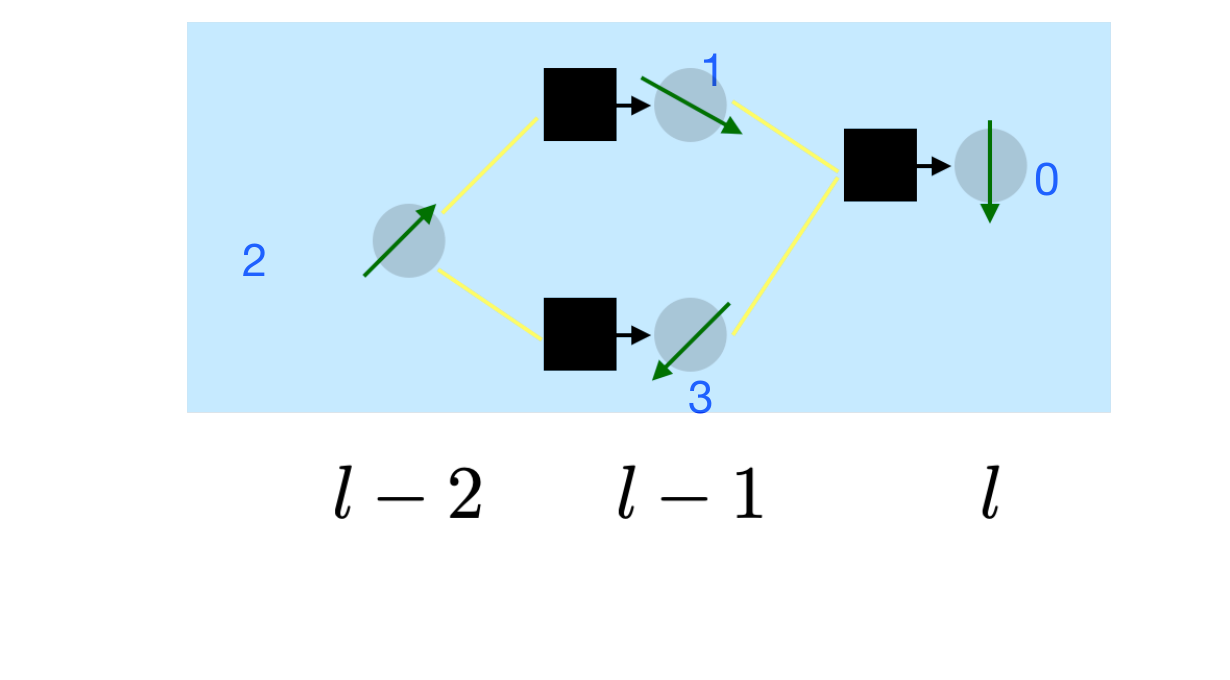}
  \ec
  \caption{
    A loop of interactions in a DNN
      extended over 3 layers, through 3 perceptrons
      and 4 bonds.
  }
   \label{fig_loop}
    \end{figure}

    \item In the case of the global coupling $c=N$, the system is symmetric under permutations of the perceptrons within each layer so that one has to consider whether this symmetry becomes broken
      spontaneously \cite{barkai1992broken}.
      In the case of sparse coupling $c < N$, we can eliminate this symmetry by choosing the connections in stochastic ways, i.~e. random graph.
      
    \item In the setup of our theory, we finally consider  $c \to \infty$ (and $M=\alpha c \to \infty$ (see \eq{eq-alpha}) ({\it after} $N \to \infty$).
      This greatly simplifies the theory as it allows us to use the saddle point method in theoretical analysis.  We call such
      intermediately dense coupling with
  \beq
  N \gg c \gg 1
  \label{eq-dense-coupling}
  \eeq
  as {\it dense coupling}.
\end{itemize}

\subsection{Connection to spinglasses}
\label{sec-connection-to-SG}

The feed-forward network made of perceptrons is equivalent to the zero-temperature limit
of the transfer-matrix of a spin-glass with Hamiltonian,
\beq
H=-\frac{1}{\sqrt{c}}\sum_{\bs} \sum_{k=1}^{c} J_{\bs}^{k}S_{\bs}S_{\bs(k)}
\label{eq-hamiltonian-layered-SG}
\eeq
\red{as shown in appendix \ref{sec-transfermatrix}.}
This is a spin-glass model put in a layered structure. Specifying the spin configuration
on the boundary $l=0$, spin configurations at layers $l=1,2,\ldots,L$ become specified
deterministically in the $T \to 0$ limit of the transfer matrix.
The perceptrons \eq{eq-perceptron} just do this operation.

\red{An important point is that there are no direct interactions within
each layer much as the restricted Boltzmann machines (RBMs)\cite{ackley1985learning} which make the operations of the $T=0$ transfer matrices equivalent to
the simple feed-forward non-linear mappings \eq{eq-perceptron}.
In appendix \ref{sec-transfermatrix}  we also show that
such representations are possible for generic activation functions
including the function ${\rm sgn}(y)$ which we employ in this paper
just as a special case.}

For a given set of interactions $J_{\bs}^{k}$, the ground state of the system
is unique if the boundaries are allowed to relax. But here we are considering ground
states with different realizations of frozen boundaries. Specifying the boundary condition
on one side, the configurations on the other side becomes fixed deterministically.

From this view point, the exponential expressibility of DNN \cite{poole2016exponential}
can be traced back to the chaotic sensitivity of spinglass ground state \cite{bray1987chaotic,newman1992multiple}.
Even if a change of the configuration on the boundary
$S_{\bs \in 0}^{\mu} \to  S_{\bs \in 0}^{\nu(\neq \mu)}$ is small, the resultant changes of the
spin configurations become larger going deeper into the system $l=1,2,\ldots$.
This can be viewed as an avalanche process. In deeper layers $l=1,2,\ldots$ 
larger number of nodes $i=1,2,\ldots$ will be involved in a single avalanche event.
In sec.~\ref{sec-squared-overlaps-simulation-2}
we discuss a quantity which reflects the avalanche sizes, i.~e. the number of nodes
involved in a same avalanche caused by $S_{\bs \in 0}^{\mu} \to  S_{\bs \in 0}^{\nu(\neq \mu)}$.

\subsection{Teacher-Student setting}
\label{sec-teacher-student-setting}

\begin{figure}[h]
    \bc
      \includegraphics[width=0.5\textwidth]{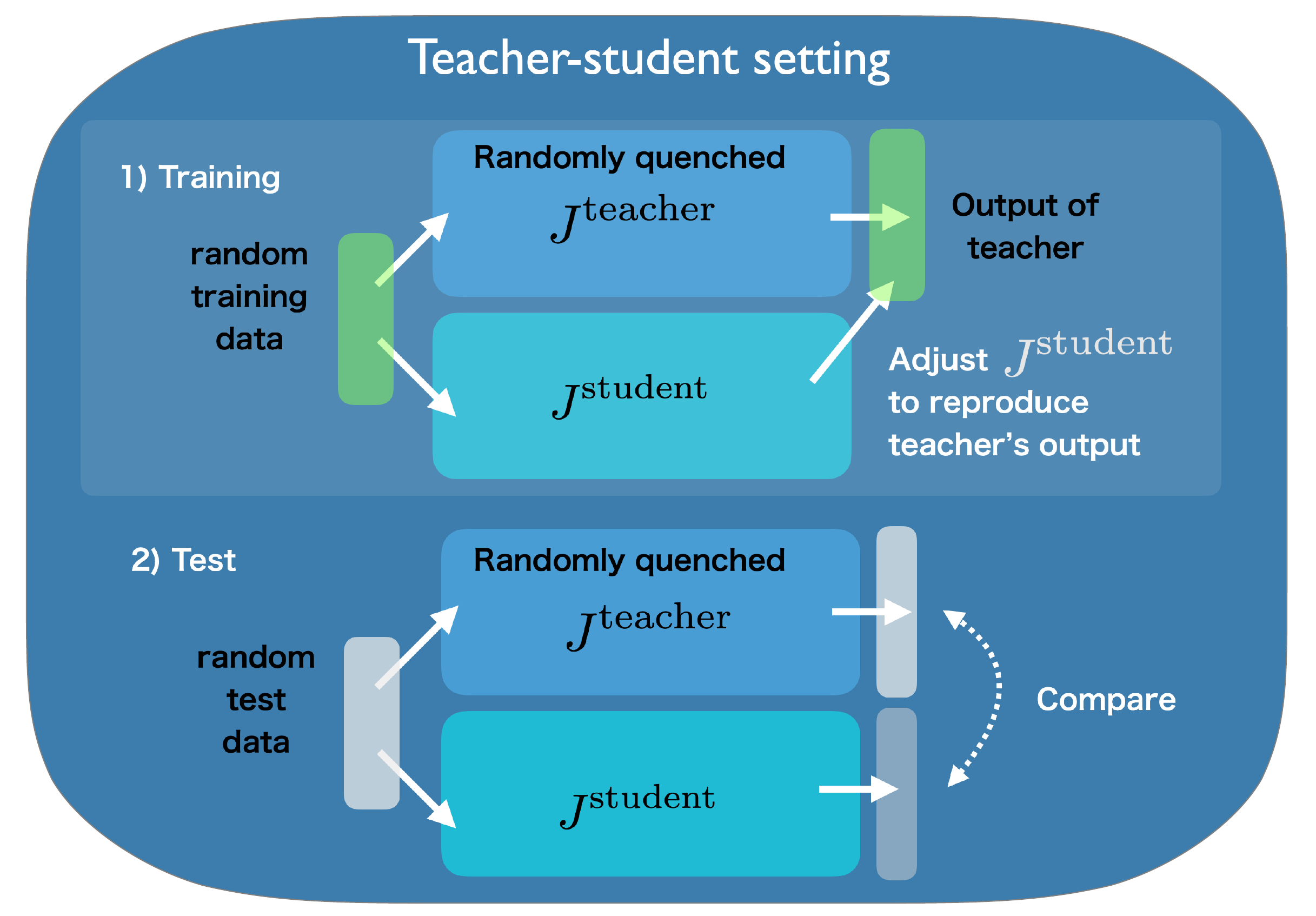}
  \ec
  \caption{Schematic pictures of the teacher-student setting. 
  }
   \label{fig_teacher_student}
  \end{figure}

As shown in Fig.~\ref{fig_teacher_student} we consider a learning scenario
by a teacher machine and a student machine.
For simplicity we assume that the teacher is a 'quenched-random teacher':
its synaptic weights $\{(J_{\bs}^{k}\}_{\rm teacher}\}$ are iid random variables
which take continuous values subjected to the normalization condition \eq{eq-J-normalization}. 

{\bf Training:} we generate $M$ sets of training data labeled as
$\mu=1,2,\ldots,M$ as follows. The values of the
spins in the input layer $S_{\bs \in 0}^{\mu}=\{S^{\mu}_{0,1}\}_{\rm teacher}$
are set as iid random Ising numbers $\pm 1$
($i=1,2,\ldots,N$, $\mu=1,2,\ldots,M$) and the corresponding output of the teacher $\{S^{\mu}_{L,i}\}_{\rm teacher}$
are obtained. The student does training by adjusting its own synaptic weights $\{(J_{\bs}^{k})_{\rm student}\}$ such that
it reproduces perfectly the $M$ sets of the input-output relations of the teacher.
More precisely we consider an idealized setting that 1) the student has exactly the same architecture
as the teacher \red{including the specific realization of the random network between adjacent layers}
2) student knows exactly the $M$ sets of the input/output relations of the teacher.
In short, the student knows everything about the teacher except for
its actual values of $\{(J_{\bs}^{k})_{\rm teacher}\}$.
Within the framework of Bayesian inference,
this is a so-called Bayes optimal setting \cite{iba1999nishimori,zdeborova2016statistical}.

  The configurations of the spins associated with the $M$-patterns of the training data
may be represented by $M$-component vectors ${\bf S}_{l,i}=(S_{l,i}^{1},S_{l,i}^{2},\ldots,S_{l,i}^{M})$ (see Fig.~\ref{fig_schematic_model}).
In the theory, we will consider $M \to \infty$ limit with,
 \beq
 \alpha\equiv
 \frac{M}{c}.
  \label{eq-alpha}
  \eeq
fixed. Note that our network is parametrized by $NcL$ variational bonds
and the $NM$ constrained spin components on the input and output boundaries.
The ratio of the two scales as,
\beq
r\equiv\frac{NcL}{NM}=\frac{L}{\alpha}
\label{eq-beta}
\eeq

{\bf Test (validation)}: the generalization ability of the student can be examined empirically using a set of test data.
Preparing $M'$ sets test data as new iid random data $(S^{\mu}_{0,i})_{\rm teacher}$ ($i=1,2,\ldots,N$, $\mu=1,2,\ldots,M'$)
(not used for the training) we compare the output of the teacher and student machines.
The probability that the student makes an error can be measured as,
\beq
\epsilon=\frac{1}{2}
\left(1-\frac{1}{NM'}\sum_{\mu=1}^{M'}\sum_{i=1}^{N} (S^{\mu}_{L,i})_{\rm teacher}(S^{\mu}_{L,i})_{\rm student}
\right).
\label{eq-generaization-error-def}
\eeq
If the student is just making random guesses $\epsilon=1/2$ while $\epsilon=0$ if it perfectly reproduces
the teacher's output.

\subsection{Gardner's volume}

Following the pioneering work by Gardner \cite{gardner1988space,gardner1989three} we investigate the ensemble of all possible machines (choices of the synaptic weights $J_{\bs}^{k}$s) of the student which are perfectly compatible with the $M$ set of the input
${\bf S}_{0}$ and output data ${\bf S}_{L}$ provided by the teacher machine
(See Fig.~\ref{fig_gardner_volume}).
As we noted in sec.~\ref{sec-connection-to-SG}, each machine with
the feed-forward propagation of signals can be viewed as a zero-temperature limit
of the transfer-matrix of a spin-glass with a set of $J_{\bs}^{k}$s.
So the ensemble of machines is an ensemble of such transfer matrices, which are typically chaotic.

The phase space volume, which is called Gardner's volume,
can be expressed for the present DNN as\cite{yoshino2020complex},
\beqn
&& V_{M}\left({\bf S}_{0},{\bf S}_{L}\right)= e^{N M {\cal S}\left({\bf S}_{0},{\bf S}_{l}\right)} \nonumber \\
& =&\left(\prod_{\bs} {\rm Tr}_{{\bf J}_{\bs}}\right)
\left( \prod_{\bs\backslash {\rm output}}{\rm Tr}_{{\bf S}_{\bs}}  \right)
\prod_{\mu=1}^{M}
\prod_{\bs}e^{-\beta v(r^{\mu}_{\bs})}
\label{eq-gardner-volume-DNN}
\qquad
\eeqn
where $v(r)$ is a hardcore potential,
\beq
e^{-\beta v(r)}=\theta(r)
\label{eq-hardcore}
\eeq
\red{with $\theta(r)$ being the Heviside step function} 
and we introduced the 'gap' variable,
\beq
r^{\mu}_{\bs} \equiv
S^{\mu}_{\bs}
\sum_{k=1}^{c}\frac{J_{\bs}^{k}}{\sqrt{c}}S^{\mu}_{\bs(k)}
\label{eq-gap}
\eeq
The trace over the spin and bond configurations can be written explicitly as,
\beq
    {\rm Tr}_{\bf S} =\prod_{\mu=1}^{M}\sum_{S^{\mu}=\pm 1}
    \qquad
        {\rm Tr}_{\bf J} =\int_{-\infty}^{\infty}
\prod_{j=1}^{c}
dJ^{j}
\delta\left(\sum_{k=1}^{c}(J^{k})^{2}-c \right)
    \label{eq-spin-and-bond-trace}
    \eeq

The key idea behind the expression \eq{eq-gardner-volume-DNN} is 
the {\it internal representation} \cite{monasson1995weight}: we are considering the spins (neurons)
in hidden layers ($l=1,2,\ldots,L-1$) as dynamical variables in addition to the synaptic weights.
This is allowed because the input-output relation of the perceptrons \eq{eq-perceptron} is forced to be satisfied
by requiring the gap to be positive $r_{\mu}^{\bs} > 0$
for all perceptrons $\bs=1,2,\ldots,N$ in the network for all training data $\mu=1,2,\ldots,M$
in \eq{eq-gardner-volume-DNN}. \red{As shown in appendix \ref{sec-transfermatrix},
the expression \eq{eq-gardner-volume-DNN} can be also obtained considering the transfer matrix representation.}

\red{The main quantity of our interest in the present paper
is the generalization error $\epsilon$
 \eq{eq-generaization-error-def}. }
The Gardner's volume $V_{M}$ provides a way to estimate the generalization ability of the network
for the test data \cite{levin1990statistical,opper1996statistical}.
The probability that
the network which perfectly satisfies the constraint put by $M$ sets of training data happens to be 
compatible with one more unseen data is given by the ratio $V_{M+1}/V_{M}$.
Then the generalization error, namely the error probability $\epsilon$, the probability that the configuration of one spin in the output layer $l=L$
of the student machine is wrong (different from the teacher) for a test data can be expressed as,
\beq
\epsilon=1-\left(\frac{V_{M+1}}{V_{M}}\right)^{1/N}
\label{eq-epsilon}
\eeq

\subsection{Symmetries}
\label{sec-symmetry}

Let us note here that there are some symmetries (besides the replica symmetry which we discuss later)
in the present problem. The following becomes important, especially in numerical simulations.

\subsubsection{Gauge symmetry}
\label{sec-gauge-symmetry}

        \begin{figure}[h]
    \bc
      \includegraphics[width=0.5\textwidth]{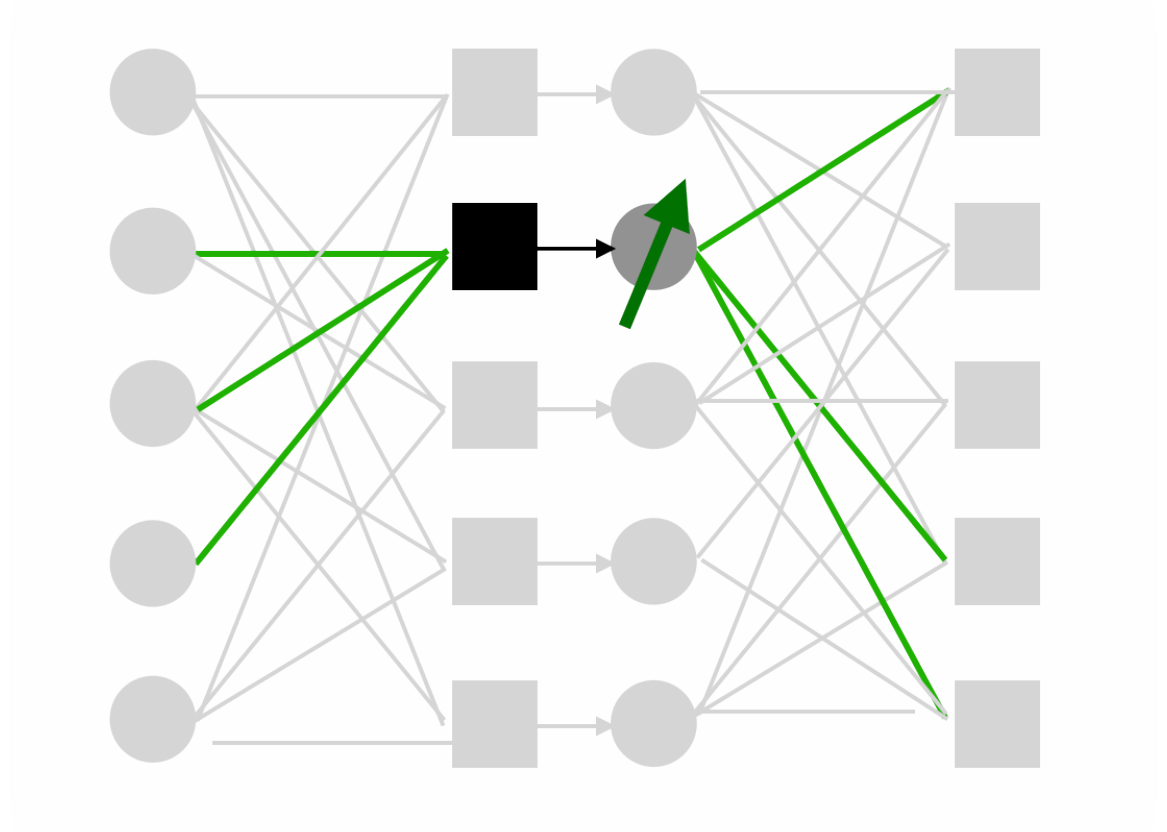}
  \ec
  \caption{Variables associated with a perceptron $\bs$ which changes sign
   by the flip of the gauge variable $\sigma_{\bs} \to -\sigma_{\bs}$
  }
   \label{fig_gauge_transform}
  \end{figure}

For any $\bs$, the system is invariant under gauge transformation 
\beqn
 S_{\bs}^{\mu} &\to  & \sigma_{\bs} S_{\bs}^{\mu} \qquad \mu=1,2,\ldots,M \\
J_{\bs}^{k}& \to& \sigma_{\bs}J_{\bs}^{k}\sigma_{\bs(k)} \qquad k=1,2,\ldots,c
\eeqn
specified by gauge variables
 \beq
  \sigma_{\bs}=\pm 1 \qquad \bs=1,2,\ldots,N(L-1)
  \eeq
  Note that we do not have a gauge transformation in the output layer $l=L$
  \red{since the output layer is constrained.}
It can be easily seen that the gap variables $r_{\bs}^{\mu}$ (see  \eq{eq-gap}) are invariant under
the gauge transformation.

  Thus for a given realization of a  machine with a set of synaptic weights, there are $2^{(L-1)N}$
completely equivalent machines specified by $2^{(L-1)N}$ possible realizations of the gauge variables:
all of them operate exactly in the same way yielding the same output for any input.
If the synaptic weights only take Ising values $J_{\bs}^{k}=\pm 1$,
the number of possible configurations of the machines modulo the gauge symmetry
is $2^{NLc^{2}-N(L-1)}$.

The presence  of the gauge invariance
is natural given the connection to the spinglass as mentioned in sec.~\ref{sec-connection-to-SG}.
While the gauge variables are frozen in spin-glass problems with quenched bonds \cite{nishimori2001statistical}, here the bonds are dynamical variables so that the gauge variables also evolve in time during learning.

Importantly this is a local symmetry in the sense that change of any $\sigma_{\bs}$
induce changes only in the neighborhood of $\bs$ (see Fig.~\ref{fig_gauge_transform}):
$S_{\bs}^{\mu} \to -S_{\bs}^{\mu}$ for $\forall \mu$, 
$J_{\bs}^{k} \to -J_{\bs}^{k}$ for $\forall k$, 
$J_{\ws}^{l} \to -J_{\ws}^{l}$  for $\forall( \ws,l)$ such that $\ws(l)=\bs$.
This means that in sparse systems with finite connectivity $c$ and $M(=\alpha c)$, the evolution of the machine from one to another connected by a local gauge transformation takes only a finite time in dynamics.
Only in the limit $c\to \infty$, do such gauge transformations become frozen in time.

\subsubsection{Permutation symmetry in globally coupled systems}

\red{As we already noted in sec.~\ref{sec-dense-coupling}}, in globally coupled systems with $c=N$, the system is invariant under permutations of
  perceptrons $\bs \in l$ within each layer $l=1,2,\ldots,L$. This symmetry can be removed
  if the coupling is not global $c < N$ since we can construct random networks (see sec.~\ref{sec-dense-coupling}).

\subsection{Hidden manifold}
\label{sec-hidden-manifold}

\begin{figure}[h]
    \bc
      \includegraphics[width=0.5\textwidth]{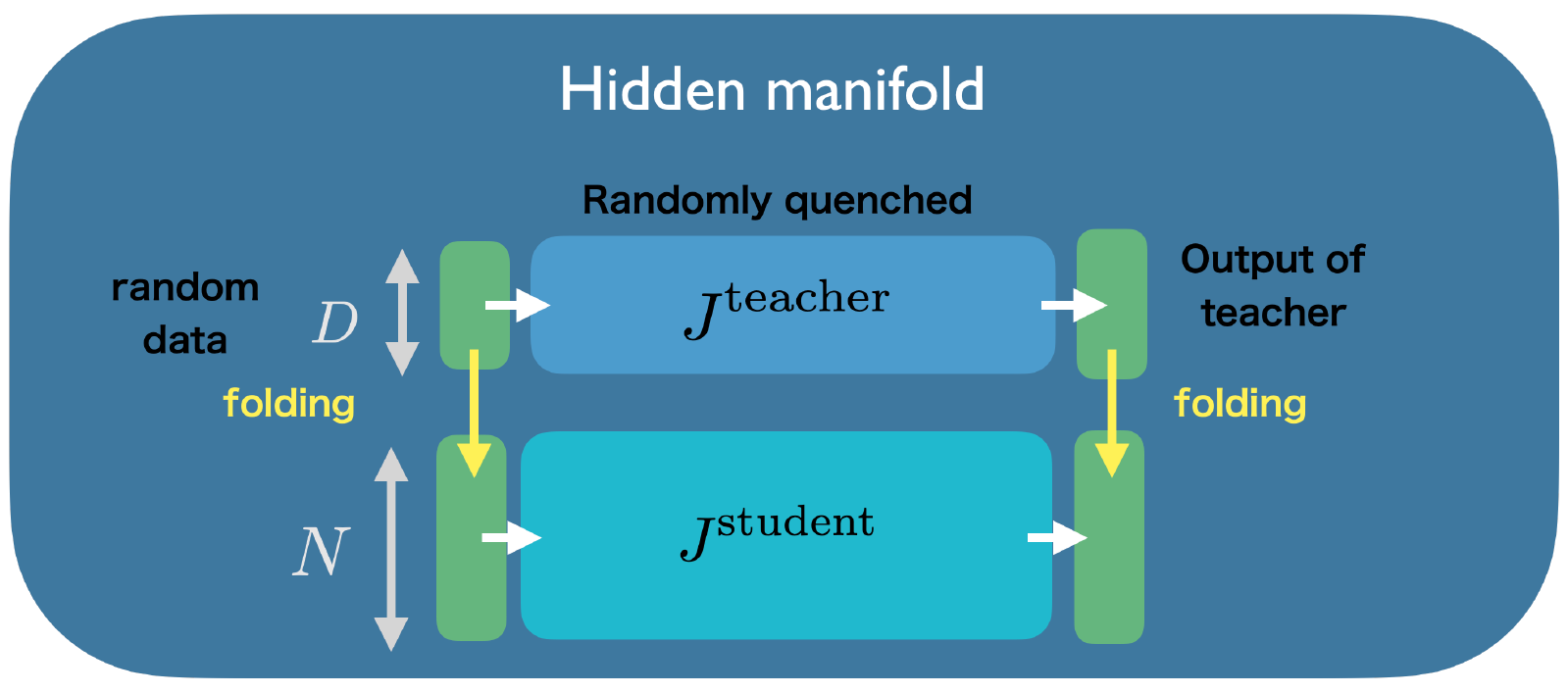}
  \ec
  \caption{  Schematic picture of the hidden manifold model.
  }
   \label{fig_schematic_hidden_manifold}
  \end{figure}

We incorporate the hidden manifold model for the data (S. Goldt et al (2020) \cite{goldt2020modeling} in our model as the following. We replace the original teacher machine of width $N$ with a narrower teacher machine of with $D ( \leq N)$ (see Fig.~\ref{fig_schematic_hidden_manifold}). The teacher is working entirely in $D$ dimensional space
being subjected to $D$ dimensional input data and produces $D$ dimensional output.
Student machines are provided $N$ dimensional input/output data
which are obtained from the $D$ dimensional input/output of the teacher via
folding matrices $F_{i,k}$ of size $N \times D$,
 \beqn
 (S_{\rm student})^{\mu}_{0,i}={\rm sgn} \left(\sum_{k=1}^{D} F_{i,k}(S_{\rm teacher})^{\mu}_{0,k}\right) \nonumber \\
 (S_{\rm student})^{\mu}_{L,i}={\rm sgn} \left(\sum_{k=1}^{D} F_{i,k}(S_{\rm teacher})^{\mu}_{L,k}\right) 
 \label{eq-hidden-manifold}
 \eeqn
for $i=1,2,\ldots,N$ and $\mu=1,2,\ldots,M$.
 For the folding matrix, we consider a simple model, 
 \beq
 F _{i,k}= \left \{ \begin{array}{ll}
   1 & k={\rm mod}(i-1,D)+1 \\
   0 & \mbox{otherwise} 
 \end{array}
 \right.
 \label{eq-simple-folding-matrix}
\eeq
In this model $N$ elements of the data for students are created
simply by making $N/D$ copies of the $D$ elements of the teacher's data.

            \section{Replica theory}
            \label{sec-replica-theory}

            \red{Now we develop and analyze a replica theory for the statistical mechanics problem of DNN in the dense limit $N \gg c \gg 1$ introduced in sec.~\ref{sec-dense-coupling}. We first show that it can be solved exactly overcoming the issue of uncontrolled approximation made in \cite{yoshino2020complex}. This is the first main result of this paper. For clarity we repeat the steps made in \cite{yoshino2020complex} and indicate how the problem is resolved.  Then we revisit the replica symmetric solution in the teacher-student setting presented in \cite{yoshino2020complex} and analyze it in more details.  Using the exact solution we analyze the generalization-ability of DNN evaluating the generalization error $\epsilon$ via \eq{eq-epsilon},  which is second main result of this paper.
The technical details of the theory is presented in appendix ~\ref{sec-replica-appendix}.            }

            \subsection{Formalism}

%We first revisit the formulation of the replica theory \cite{yoshino2020complex}
%considering dense coupling $N \gg c \gg 1$ (see \eq{eq-dense-coupling})
%introduced in sec. \ref{sec-dense-coupling} replacing the global coupling $N=c \gg 1$
%originally considered in \cite{yoshino2020complex}. Then we set up the
%computation of the error probability for the test data (see \eq{eq-epsilon}).
%In the replica theory, we consider $1+s$ replicas $a=0,1,2,\ldots,s$ where $a=0$ is the teacher machine. All of the student's machines $a=1,2,\ldots,s$ are subjected to the same boundary condition
%on the input ${\bf S}_{0}$ and output ${\bf S}_{L}$ provided by the teacher machine.
%
%{\bf comment on connection to cloned liquid, disorder-free system}
%
\subsubsection{Order parameters}
\label{sec-order-parameters}

We are considering the dense coupling \eq{eq-dense-coupling} 
in which 1) perceptrons have large connectivity $c \gg 1$ and 2) the permutation symmetry
of the perceptrons which exist in globally coupled systems is removed.
We are also considering a large number of
training patterns $M=\alpha c \gg 1$  (see \eq{eq-alpha}).
Then we can naturally introduce 'local' order parameters associated with each perceptron $\bs$,
\beq
Q_{ab,\bs}=\frac{1}{c}\sum_{k=1}^{c}(J_{\bs}^{k})^{a}(J_{\bs}^{k})^{b}
\qquad q_{ab,\bs}=\frac{1}{M}\sum_{\mu=1}^{M}(S_{\bs}^{\mu})^{a}(S_{\bs}^{\mu})^{b}
\label{eq-glass-order-parameters}
\eeq
The overlaps between the teacher and student machines are represented by $Q_{0b,\bs}=Q_{b0,\bs}$
and $q_{0b,\bs}=q_{b0,\bs}$ ($b=1,2,\ldots,s$) while those between the student machines are
are represented by $Q_{ab,\bs}=Q_{ba,\bs}$ and $q_{ab,\bs}=q_{ba,\bs}$ ($a,b=1,2,\ldots,s$).

It is important to note that the order parameters $Q_{ab,\bs}$ and $q_{ab,\bs}$ defined above changes sign under the change of the gauge variable $\sigma_{\bs}^{a}$ which can be defined independently for each replica $(a=1,2,\ldots,n)$ (see Fig.~\ref{fig_gauge_transform}).
Thus they trivially vanish in thermal equilibrium in sparse systems with finite connectivity $c$.
Only in the dense limit $c \to \infty$ the gauge variables $\sigma_{\bs}^{a}$ can be considered as slow variables.

It is natural to expect that order parameters are homogeneous within each layer \red{since we will
  take the average over realization of random connections
between adjacent layers (see \eq{eq-F-teacher-student})}.
Thus we assume they only depend on the index $l$ of layers,
\beqn
Q_{ab,\bs}=Q_{ab}(l)  && \qquad l=1,2,\ldots,L-1 \nonumber \\
q_{ab,\bs}=q_{ab}(l)  && \qquad l=0,1,2,\ldots,L-1,L
\label{eq-order-parameters-at-each-layers}
\eeqn
Here we have included, for our convenience,  the spin overlaps at the boundaries $l=0,L$
where spins of all student replicas $a=1,2,\ldots,s$ are forced take
the same values as the teacher $a=0$,
\beq
q_{ab}(0)=1 \qquad q_{ab}(L)=1.
\eeq
Note also that the normalization condition for the bonds \eq{eq-J-normalization} and the spins (which take Ising values $\pm 1$)
implies $Q_{aa}(l)=q_{aa}(l)=1$ for $\forall a$ and $\forall l$.

The order parameters also vanish in thermal equilibrium in globally coupled system with $c=N$
due to the permutation symmetry - the 2nd symmetry mentioned in sec \ref{sec-symmetry}.
%The latter symmetry may become spontaneously broken only in the limit $N \to \infty$ \cite{barkai1992broken}.
\red{This issue is removed by using the dense coupling by selecting connections between adjacent layers randomly.}

\subsubsection{Replicated Gardner volume, Free-energy}
Let us introduce the replicated Gardner's volume, where the teacher machine is included as the $0$-th replica,
\beqn
&&V^{1+s}\left({\bf S}_{0},{\bf S}_{L}\right)
=e^{N M {\cal S}_{1+s}\left({\bf S}_{0},{\bf S}_{l}\right)} \nonumber\\
&& =\prod_{a=0}^{1+s}
\left(\prod_{\bs} {\rm Tr}_{{\bf J}^{a}_{\bs}}\right)
\left( \prod_{\bs\backslash {\rm output}}{\rm Tr}_{{\bf S}^{a}_{\bs}}  \right)
\left\{ \prod_{\mu,\bs,a}
e^{-\beta v(r_{\bs,a}^{\mu})}
\right \} \nonumber\\
 \label{eq-replicated-gardner-volume}
\eeqn
with
\beq
r^{\mu}_{\bs,a} \equiv
(S^{\mu}_{\bs})^{a}
\sum_{k=1}^{c}\frac{(J_{\bs}^{k})^{a}}{\sqrt{c}}(S^{\mu}_{\bs(k)})^{a}
\label{eq-gap-replica}
\eeq
Here the output ${\bf S}_{L}$ is the output of the teacher so that ${\bf S}_{L}={\bf S}_{L}({\bf S}_{0}, \{(J^{k}_{\bs})_{\rm teacher}\})$.
The main object we are interested in is the free-energy \red{functional} (Franz-Parisi's potential \cite{FP95}),
 \beqn
 \frac{-\beta F[\{{\hat Q}(l),{\hat q}(l)\}]}{NM}
 && =\frac{\left. \partial_{s} \overline{V^{1+s}({\bf S}_{0},{\bf S}_{L}({\bf S}_{0},{\cal J}_{\rm teacher})))}\right |_{s=0}}{NM}\nonumber \\
&&  =
  \left.  \partial_{s}s_{1+s}[\{{\hat Q}(l),{\hat q}(l)\}]\right |_{s=0}.
  \label{eq-F-teacher-student}
\eeqn
where the over-line denotes the average over 1) the random inputs ${\bf S}_{0}$ imposed commonly on all machines
and 2) realization of the random synaptic weights $\{(J^{k}_{\bs})_{\rm teacher}\}$ of the teacher and \red{ 3) realization of random connections between adjacent layers.}

\red{
  It turns out that the dense coupling
\eq{eq-dense-coupling} $N \gg c \gg 1$ allows us to obtain  
the  exact expression for the replicated Gardner volume for $n=1+s$
replicas in terms of the order parameters \eq{eq-order-parameters-at-each-layers},
}
\beqn
 s_{n}[\{{\hat Q}(l),{\hat q}(l)\}]
&=&\frac{1}{\alpha} \sum_{l=1}^{L} s_{\rm ent,bond}[\hat{Q}(l)]
+\sum_{l=1}^{L-1} s_{\rm ent,spin}[\hat{q}(l)] \nonumber \\
&& -\sum_{l=1}^{L} {\cal F}_{\rm int}[\hat{\lambda}(l)]
\qquad 
\label{eq-S-total}
\eeqn
with 
\beq
\lambda_{ab}(l)=q_{ab}(l-1)Q_{ab}(l)q_{ab}(l)
\label{eq-def-lambda-l}
\eeq
Here $s_{\rm ent,bond}[\hat{Q}(l)]$ and $s_{\rm ent,spin}[\hat{q}(l)]$ and the entropic part of the free-energy
associated with bonds and spins respectively and $-{\cal F}_{\rm int}[\hat{q}(l-1),\hat{Q}(l),\hat{q}(l)]$ is the interaction part of the
free-energy.
\red{Fortunately 
the free-energy functional obtained in \cite{yoshino2020complex}
turns out to be valid in the dense limit $N \gg c \gg 1$
although it is unjustified for the global coupling $c=N$ assumed there.
The details of the expressions and the derivation 
are presented in the appendix ~\ref{sec-replica-summary}.}

\red{The main reasons for the success,
which allows us to overcome the problems in \cite{yoshino2020complex},
are the three points discussed in sec.~\ref{sec-dense-coupling}.
First, the sparseness of the network allows us
to safely neglect contribution of loops as we explain in detail in sec.~\ref{sec-cumulant}. Second, the random connections between adjacent layers
eliminate the permutation symmetry which exist in globally coupled
system $c=N$. Third the limit $c \to \infty$ allows us to use the
standard saddle point method to evaluate thermodynamic quantities exactly.
}

%was obtained in \cite{yoshino2020complex} for the case of global coupling $c=N$ 

%free-energy functional for the dense coupling case,

%under the assumption of 'tree-approximation' which amounts to neglecting contributions from closed loops in the network as the one
%shown in Fig.~\ref{fig_loop}. Now with our dense coupling \eq{eq-dense-coupling} $N \gg c \gg 1$ the contributions of such loops can be

%Consequently we obtain the exact free-energy functional for the dense coupling case,
%which have precisely the same form as the free-energy obtained in  \cite{yoshino2020complex}, with $\alpha=M/c$ (see \eq{eq-alpha}),

\subsubsection{Replica symmetric ansatz}

Since our current problem is a Bayes optimal inference problem,
 we can safely assume a replica symmetric (RS) solution,
\beqn
(a,b=1,\ldots,s)  \qquad
&& Q_{ab}(l)=(1-Q(l))\delta_{ab}+Q(l) \nonumber \\
&& q_{ab}(l)=(1-q(l))\delta_{ab}+q(l)\nonumber \\
(a=1,\ldots,s) \qquad && Q_{0a}(l)=Q_{a0}(l)=R(l) \nonumber \\
&& q_{0a}(l)=q_{0a}(l)=r(l) 
\eeqn
for $l=1,2,\ldots, L$ \red{and the Nishimori condition  (see sec.~\ref{sec-nishimori-condition})},
\beq
Q(l)=R(l) \qquad q(l)=r(l)
\label{eq-identity-bayes-optimal-replica}
\eeq
which must hold in Bayes optimal cases.
The saddle point equations which extremize the replicated free energy are obtained in \cite{yoshino2020complex}.
It can be checked that the saddle point equations can verify the relation \eq{eq-identity-bayes-optimal-replica}.

\subsubsection{Generalization error}

Based on the above results we can analyze the
error probability  \eq{eq-epsilon} \red{which is 
the main object of our interest in the present paper.}
  Using the free-energy \eq{eq-F-teacher-student}
and \eq{eq-S-total} we readily find it as,
\beq
\epsilon=1-\exp \left(
\sum_{l=1}^{L-1} \partial_{s} \left. s_{\rm ent,spin}[\hat{q}(l)] \right|_{s=0}
-\sum_{l=1}^{L} \partial_{s} \left. {\cal F}_{\rm int}[\hat{\lambda}(l)]\right|_{s=0}
\right)
\label{eq-generalization-error-replica}
\eeq
Explicit expressions of the free-energy needed to evaluate the above quantity are given in the appendix ~\ref{sec-FP-potential-RS}.

\begin{figure}[t]
  \bc
   \includegraphics[width=0.5\textwidth]{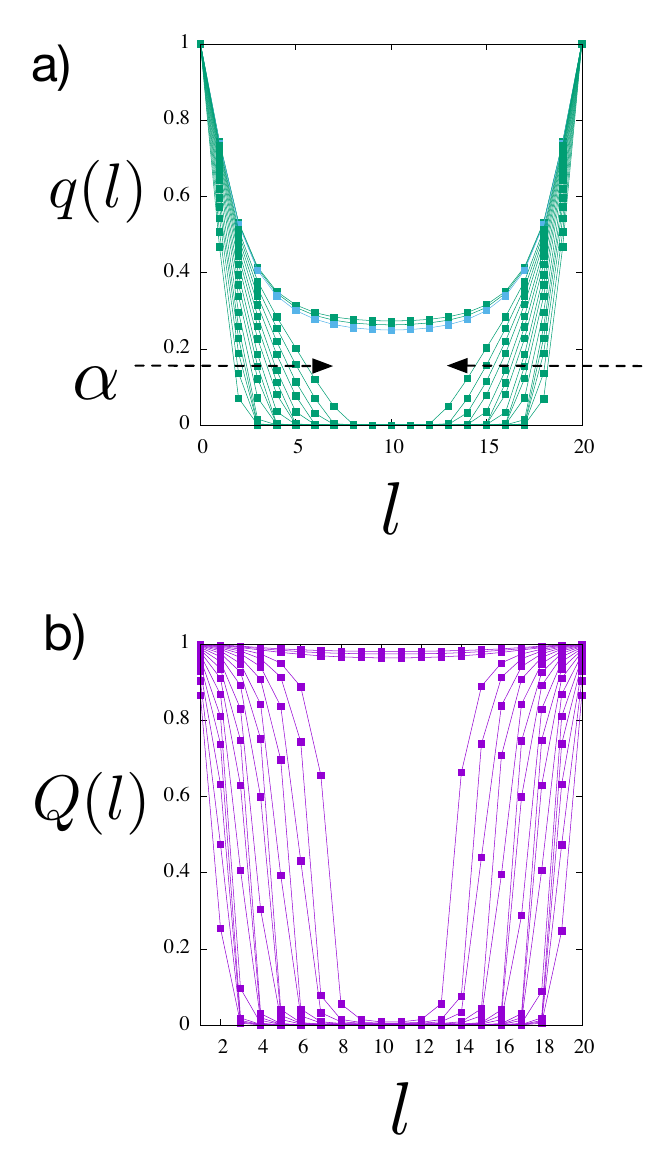}
      \ec
      \caption{Spatial profile of the order parameter obtained by solving the replica symmetric saddle point equations. (top) the overlap of spins (neurons) (bottom) and the overlap of bonds (synaptic weights).
        Here $L=20$. Different lines corresponds to  $\alpha=16-10^{3}$ with equal spacing in $\ln \alpha=0.23..$.
      }
\label{fig-order-parameter-profile-replica}    
\end{figure}

  \begin{figure*}[t]
    \bc
      \includegraphics[width=0.95\textwidth]{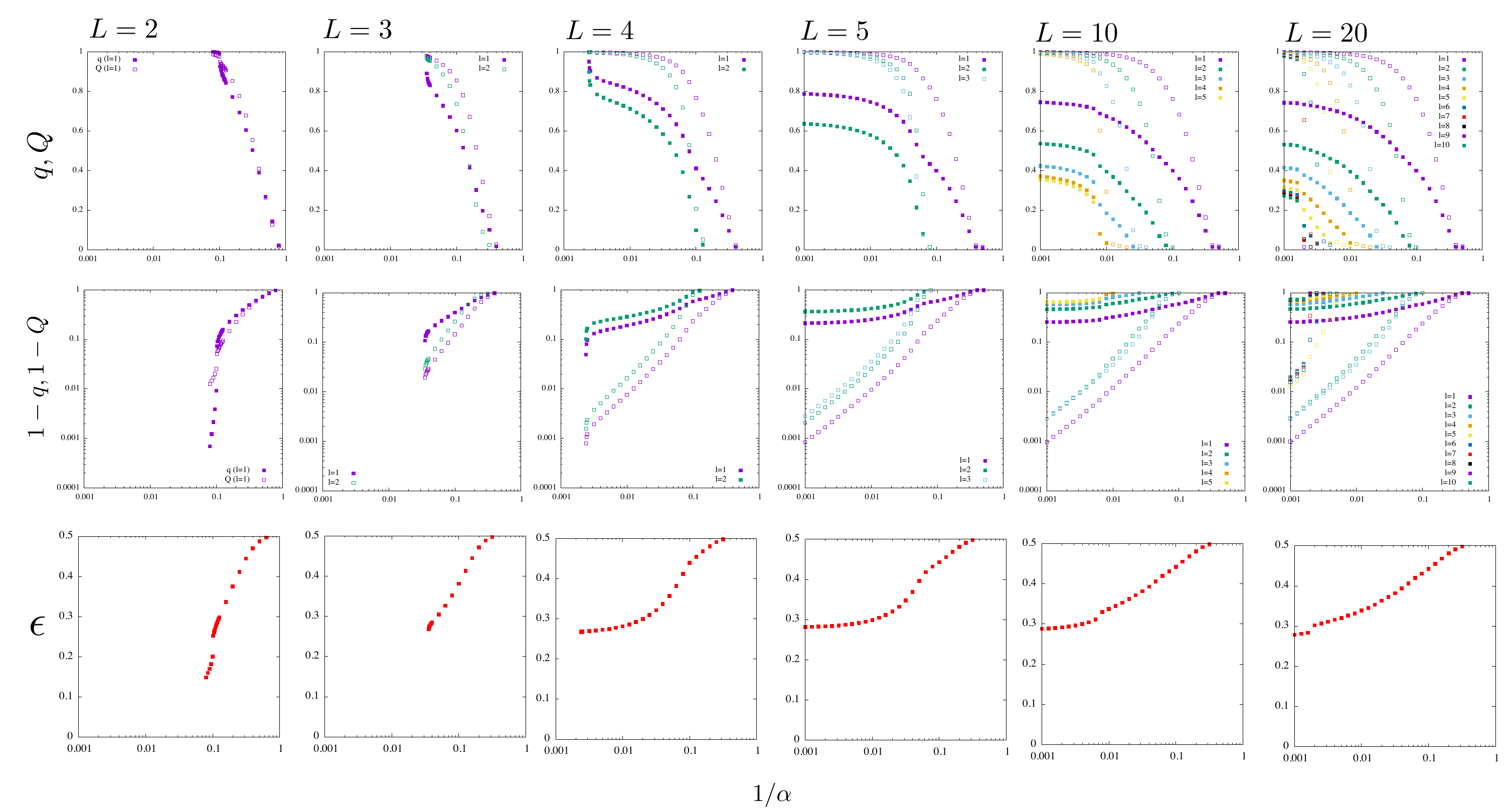}
  \ec
  \caption{Order parameters and generalization error
\red{obtained by solving the replica symmetric saddle point equations.}
    In the panels on the 1st and 2nd lows,
    overlaps of spins $q(l)$ (filled symbols) and $Q(l)$ bonds (open symbols) are shown.
    In the bottom low the generalization error $\epsilon$ is shown.
  }
   \label{fig_replica_plots}
  \end{figure*}

\subsection{Analysis}

\subsubsection{Order parameters}
\label{sec-analysis-order-parameter}

We numerically solved the saddle point equations
\eq{eq-SP-general} to obtain the order parameters \red{repeating
the analysis in \cite{yoshino2020complex} but in a wider parameter space.}
In Fig.~\ref{fig-order-parameter-profile-replica} we show the spatial profile of the order parameters.
\red{As already shown in \cite{yoshino2020complex}, the theory predicts
spatially heterogeneous learning.}
It can be seen that the 'crystalline' phase with finite order
parameter (inference of the teacher's configuration is successful)
grows increasing $\alpha$ starting from the input/output boundaries.
This is reminiscent of wetting transitions
\cite{de1985wetting,krzakala2011melting1,krzakala2011melting2}.
Details of the behavior of the order parameters are displayed in Fig.~\ref{fig_replica_plots}.

The central region remains in the liquid phase with zero order parameter
(where inference of the teacher's configuration is impossible) until the two crystalline phases meet in the center \red{at a critical point $\alpha_{c1}(L)$. Naturally  $\alpha_{c1}(L)$ increases with  $L$.
For $\alpha > \alpha_{c1}(L)$ the central liquid phase is absent.
 As far as $\alpha < \alpha_{c1}(L)$ we find the the crystalline parts attached to the two opposite boundaries grow with $\alpha$ but the profile of the order parameters remain independent of $L$.
}

\red{ Now for $\alpha > \alpha_{c1}(L)$, where the liquid phase is absent, the order parameters depend explicitly on the depth $L$.   One may regard this as a "finite depth $L$ effect". At some larger $\alpha$ it becomes difficult to follow the saddle point solution numerically. Presumably this implies spinodal instability associated with a first order transition to another solution $q=Q=1$ (which is a saddle point solution) at some $\alpha_{c2}(L) ( > \alpha_{c1}(L))$. Such a discontinuous 'perfect recovery' behavior has been found in the case of single perceptron with binary couplings \cite{gyorgyi1990first}. We skip detailed analysis of the 1st order transition and leave it for future works.
Up to the discontinuous change, the evolution of the order parameters with increasing $\alpha$ is continuous.}

\subsubsection{Generalization errors}
\label{sec-generalization-error-theory}

\begin{figure}[h]
    \bc
       \includegraphics[width=0.5\textwidth]{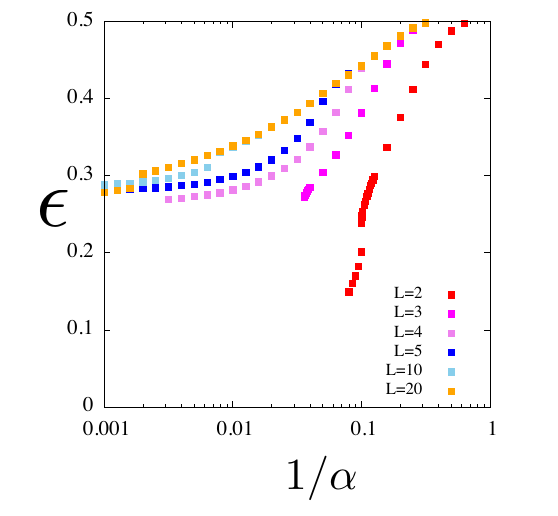}
  \ec
  \caption{Learning curves of DNN with various depth $L$
\red{obtained by solving the replica symmetric saddle point equations.}    
  }
   \label{fig_replica_all_epsilon}
    \end{figure}

\red{Now we turn to the generalization error $\epsilon$ which is of
our main interest in the present paper.}
It is obtained as shown in the bottom panels of
Fig.~\ref{fig_replica_plots} and Fig.~\ref{fig_replica_all_epsilon}.
The relation $\epsilon$ vs $\alpha$ is called often as {\it learning curves}. Without learning $\alpha=0$, $\epsilon=1/2$ because the student just makes random guesses. 
%Increasing $\alpha$ the generalization error $\epsilon\red{_{L}(\alpha)}$ decreases. It can be seen that for a fixed $\alpha$, the generalization error $\epsilon\red{_{L}(\alpha)}$ first increases increasing the depth $L$ but then saturates to a value $\epsilon\red{_{\infty}(\alpha)} < 1/2$.
The learning curves $\red{\epsilon=\epsilon_{L}(\alpha)}$ consist of two parts as follows.
\begin{itemize}
\item Let us recall that for sufficiently small $\alpha$,
the two crystalline phases at the boundaries remain disconnected from each other
separated by the liquid phase in the center and that the profile of the order parameters are independent of $L$ since
the two crystalline regions do not meet  \red{ as we discussed in sec.~\ref{sec-analysis-order-parameter}.
In this regime the learning curve does not depend on the depth $L$, i.~e. \red{$\epsilon=\epsilon_{\infty}(\alpha)$}.
The reason is that the contribution from the liquid region where $q(l)=Q(l)=0$
to $\epsilon$ \eq{eq-generalization-error-replica} is just zero: it contributes neither positively nor negatively to $\epsilon$. On the other hand, the crystalline region where $q(l),Q(l) >0$ contribute negatively
to $\epsilon$ \eq{eq-generalization-error-replica} and it is is independent of $L$ as long as 
the two crystalline regions do not meet.
It is remarkable that $\epsilon_{\infty}(\alpha) < 1/2$ and decreases with increasing $\alpha$:
the system generalizes even though the central part is in the liquid phase due to over-parametrization.}
\item  \red{ 
  Increasing $\alpha$, the crystalline phases meet at some critical point $\alpha_{c1}(L)$ and the central liquid phase disappear. Note again that $\alpha_{c1}(L)$ is larger for larger $L$.}
For sufficiently large $\alpha \red{> \alpha_{c1}(L)}$ where the central liquid gap is filled up by the crystalline phase, the learning curve depend on the depth $L$ as the order parameters now depend on $L$.
\red{For even larger  $\alpha > \alpha_{c2}(L)$,
  we speculate that $\epsilon$ jumps to $0$ due to the 1st order transition
mentioned in sec.~\ref{sec-analysis-order-parameter}}
\end{itemize}

\red{The $L$ independent behavior of the learning curve can be seen in Fig.~\ref{fig_replica_all_epsilon} as follows. For instance one can see that $\epsilon_{L}(\alpha)$ of $L=10,20$ are indistinguishable for $1/\alpha > 0.01$ 
and $L=10,20,5$ are indistinguishable for $1/\alpha > 0.1$.}

        \subsection{Finite width $N$ / dimension $D$ effects and finite connectivity $c$ effects}
        \label{sec-finite-N-finite-D-effect}

\red{In reality, DNNs have some finite width $N$ and finite connectivity $c$ while in the theory we assumed an idealized situation:  the dense limit $N \gg c \gg 1$ and $M \gg 1$ with fixed $\alpha=M/c$. It is very important to consider the effects of finite with $N$ and finite connectivity $c$ (and $M$).        }

        \subsubsection{Finite width $N$ effect}
        \label{sec-finite-N-effect}

      \begin{figure}[h]
    \bc
      \includegraphics[width=0.5\textwidth]{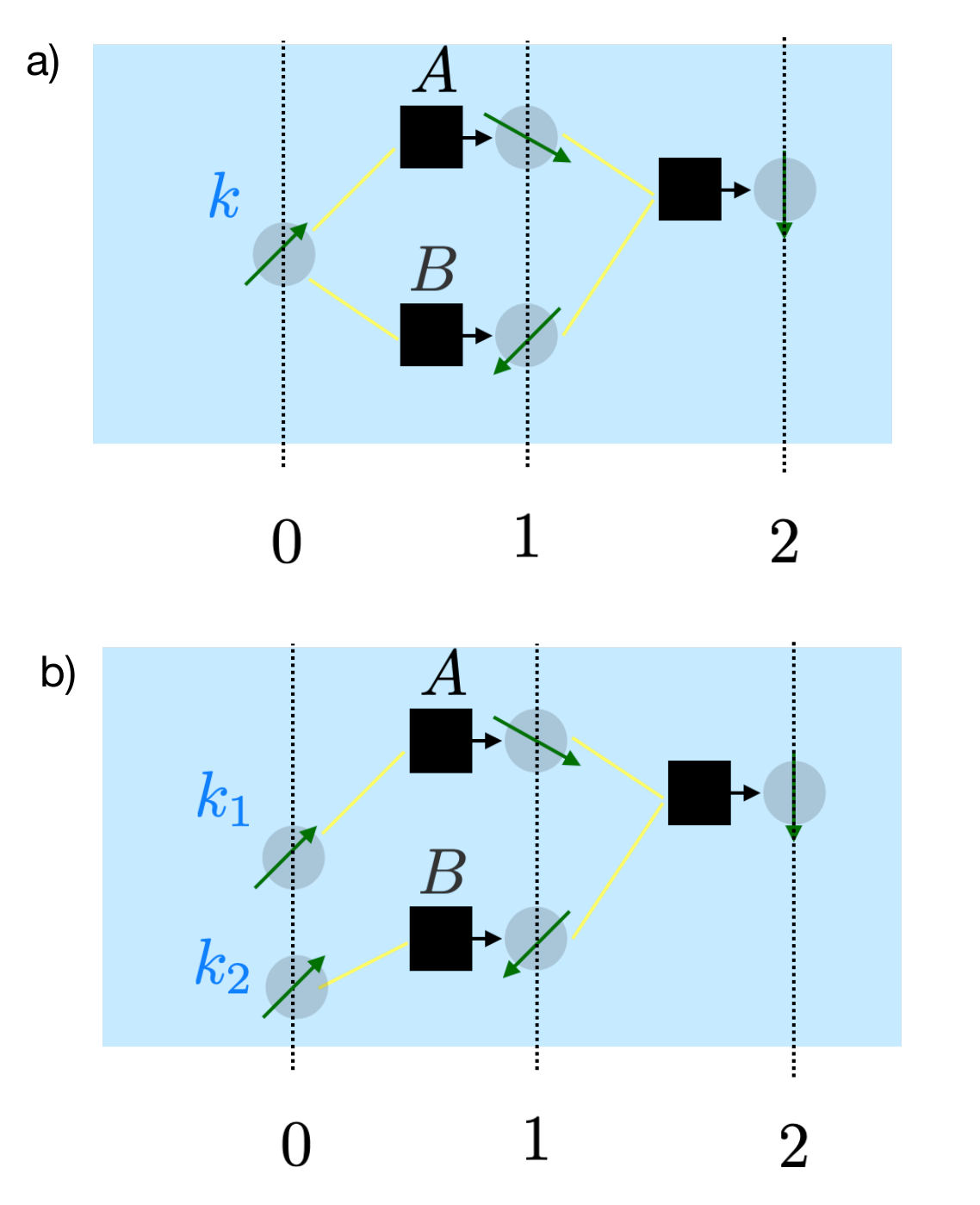}
  \ec
  \caption{Schematic picture of the closed and unclosed loop at the boundary
  }
   \label{fig_loop_closing}
      \end{figure}

 The effects of finite width $N$ can be attributed to the corrections due
 to geometrically closed loops in the network which becomes non-negligible
 when the width $N$ is finite as we discussed in sec.~\ref{sec-dense-coupling}.
 The simplest is the one shown in Fig.~\ref{fig_loop} which connects adjacent three layers.
 More extended ones exist as shown in Fig.~\ref{fig_loop_appendix_extended} which connect many
 layers. As we showed in sec.~\ref{sec-dense-coupling}, the probability to have such a geometrically closed
 single loop is proportional to $c/N$ \red{no matter how extended it is}.
 The probability vanishes in $N \to \infty$ but exists as long as $N$ is finite.

 Most importantly the loops connect different layers and different nodes within the same layer inducing correlations inside the network. Indeed as discussed in detail sec. \ref{sec-cumulant} in the appendix, the loops yield finite width $N$ corrections to the interaction part of the
free-energy. 
%Let us note that the loop corrections exist not only in the crystalline phase
%but also in the liquid phase where $q(l)=Q(l)=0$.
We also note that the symmetry concerning the exchange of input/output sides
present in the saddle point solutions (See Fig.~\ref{fig-order-parameter-profile-replica})
becomes lost in the presence of such loop correction terms.

\subsubsection{Finite hidden dimension $D$ effect}
        \label{sec-finite-D-effect}

It is interesting to discuss here the hidden manifold model \cite{goldt2020modeling}
introduced in sec.~\ref{sec-hidden-manifold}.
Let us recall that our original model contains no correlations within the boundaries. We can consider the effect of the correlations put in the
input/output boundaries by the hidden manifold model in a perturbative manner \red{around the replica symmetric saddle point solution as the following.} 

Within the simplest model \eq{eq-simple-folding-matrix} for the folding matrix $F$, the same values are repeated in the input (output) data on different nodes $i(=1,2,\ldots,N)$. This amount to induce additional closed loops. For example the unclosed loop in panel b) of Fig.~\ref{fig_loop_closing} becomes closed if the input data at $k_{1}$ and $k_{2}$ are forced to take the same value by the simplest hidden manifold model. This means that finite width $N$ effects become enhanced as the effective dimension $D$ becomes smaller. This consideration implies finite width $N$ effect and finite hidden dimension $D$ effect will be similar.
Both will lead to increase of correlations inside the network.

\red{Let us note that the teacher and students have different architectures in the hidden manifold model
so that the inference by the students is not Bayes optimal. In such a circumstance the replica symmetry is not guaranteed.
We leave the analysis beyond the perturbative analysis for future works.}

\subsubsection{Finite connectivity $c$ effect}
        \label{sec-finite-c-effect}

Finally, in $N \to \infty$ limit, we will be still left with finite connectivity $c$ effects.
In our theoretical analysis
%, which predicted the wetting-like phase transitions,
we assumed $c \to \infty$ which allowed us to perform the saddle point computations.
One can naturally consider $1/c$ corrections taking into account contributions from fluctuations
around the saddle point as sketched in sec.~\ref{sec-hessian} in the appendix.

Naturally the fluctuating field around the saddle point induce correlations inside network.  
Let us also note that the cubic term in the expansion breaks the symmetry concerning the exchange of input/output sides (see sec.~\ref{sec-cubic-expansion} in the appendix). 

\begin{redtext}
\subsubsection{Discussions}
\label{sec-finite-N-c-D-effect-discussion}

The corrections due to the loops ( sec.~\ref{sec-finite-N-effect})
and those due to the fluctuations around the saddle points
(sec.~\ref{sec-finite-D-effect})
can be easily separated considering the dense limit $N \gg c \gg 1$, which is difficult for the case of global coupling $c=N$.
Nonetheless we found that the two corrections bring qualitatively
similar effects: 1) correlations inside the network and 2) asymmetry with respect to the exchange of input/output sides.

We consider that the two effects, which disappear in the dense limit, are important in practice in the following respects.
\begin{itemize}
\item{ Remnant symmetry breaking field}

One would wonder: how a student machine can recognize the existence of the two crystalline regions (teacher's configuration) if the two are separated by the liquid region  as in Fig.~\ref{fig-order-parameter-profile-replica}?". For an algorithm to work in this situation, some remnant symmetry breaking field should help the student. We consider the corrections due to the loops and the fluctuations around the saddle point play this role.

\item{ Input-output asymmetry}
  
One would also wonder: how a DNN with the feed-forward propagation of information can have such spatial profile
which is completely symmetric concerning the exchange of input/output sides as in Fig.~\ref{fig-order-parameter-profile-replica} ?
We consider the corrections due to the loops and the fluctuations around the saddle point are responsible for the breaking
of this symmetry.

\end{itemize}

\end{redtext}

%  \clearpage
\section{Simulation}
\label{sec-simulation}

Now let us turn to discuss Monte Carlo simulations on the same model we analyzed theoretically.
We first explain the simulation method in sec.~\ref{sec-method-simulation}, introduce the observables in
sec.~\ref{sec-observables-simulation} and then present the results in sec.~\ref{sec-results-simulation}.

\subsection{Method}
\label{sec-method-simulation}

%introduced in sec \ref{sec-model}:
%the multi-layer perceptron network model in the teacher-student setting.

\subsubsection{Learning scenarios}

We simulate the teacher-student scenario (see sec.~\ref{sec-teacher-student-setting})
in the Bayes-optimal setting and the setting with the hidden manifold model (see  sec.~\ref{sec-hidden-manifold}).
%In the theoretical analysis, we mainly considered the Bayes-optimal scenario and
%briefly discussed the hidden manifold scenario. Here we explicitly conduct
%simulations on both scenarios.

\begin{itemize}
\item {\bf Bayes-optimal scenario}
\begin{itemize}
\item {\bf Network}: teacher and student machines
have the same rectangular network of width $N$ and depth $L$
(see Fig.~\ref{fig_schematic_model}).
The rectangular network is created as a random graph as the following.
Every $\bs \in l$ is given $c$ arms. Each of the arms is  connected to a $\bs \in l-1$
chosen randomly out of $N$ possible ones.
%The paring of the arms in the adjacent layers is done stochastically. 

\item {\bf Synaptic weights of teacher machine}: the teacher's synaptic weights
$\{(J_{\bs}^{k})_{\rm teacher}\}$ for $\bs \in 1,2,\ldots,L$ and $k=1,2,\ldots,c$
are prepared as iid random numbers
drawn from the Gaussian distribution with zero mean and unit variance.

\item {\bf Data}: $M$ set of training data is prepared as follows. First
the input data for the teacher are prepared as iid random numbers 
$(S_{\rm teacher})^{\mu}_{0,i}=\pm 1$ for $i=1,2,\ldots,N$ and $\mu=1,2,\ldots,M$.
Then the output $(S_{\rm teacher})^{\mu}_{L,i}$ for $i=1,2,\ldots,N$ and $\mu=1,2,\ldots,M$
are obtained by the feed-forward propagation of the signal using  \eq{eq-perceptron}.
These outputs are used as the target outputs $(S_{*})^{\mu}_{L,i}$ to train the student machines (see below),
i.~e. $(S_{*})^{\mu}_{L,i}=(S_{\rm teacher})^{\mu}_{L,i}$. Another $M'$ set of data for the test (validation)
are created in the same way.

\end{itemize}

\item {\bf Hidden manifold scenario} \cite{goldt2020modeling}
  \label{sec-hidden-minifold-simulation}

\begin{itemize}
\item {\bf Network}: the networks of the teacher and student machines are the rectangular,
random regular network as in the Bayes optimal scenario
but the teacher machine is narrower than the student machine, i.~e. $D < N$ (see Fig.~\ref{fig_schematic_hidden_manifold}). 

\item {\bf Synaptic weights of teacher machine}: the teacher's synaptic weights are prepared in the same manner
  as in the case of the Bayes optimal scenario.

\item {\bf Data}: $M$ sets of data for training and another $M'$ sets of data
for the test (validation) are created in the same way as the following.
Pairs of input/output data of the teacher's machine
is created just as in the case of Bayes optimal scenario but with $D$ replacing $N$.
Then the $N$ dimensional inputs $(S_{\rm student})_{0,i}$ for $i=1,2,\ldots,N$
to be given to the student machines
are created using the simple folding matrix $F$ \red{given by \eq{eq-simple-folding-matrix}}.
Similarly the $N$ dimensional target output  $(S_{*})_{L,i}$ for $i=1,2,\ldots,N$
for the student machines are created using the same folding matrix $F$.
%For the folding matrix, we consider
%both the 'random' and 'simple' models.

\end{itemize}
  \end{itemize}

\subsubsection{Learning algorithm: greedy Monte Carlo method}

For a set of temporal synaptic weights of a student machine
$\{(J_{\bs}^{k})_{\rm student}\}$ for $\bs \in 1,2,\ldots,L$ and $k=1,2,\ldots,c$,
we obtain the output data $(S_{\rm student})_{L,i}$ ($i=1,2,\ldots,N$) for a given input data
$(S_{\rm student})_{0,i}$($i=1,2,\ldots,N$) using the feed-forward propagation based on  \eq{eq-perceptron}.

To train the student machines we use a simple zero-temperature or greedy Monte Carlo algorithm.
We introduce the loss function defined as,
\beq
E=\sum_{i=1}^{N}\sum_{\mu=1}^{M} |(S_{\rm student})^{\mu}_{L,i}-(S_{*})^{\mu}_{L,i}|
\label{eq-loss-function}
\eeq
where $(S_{*})^{\mu}_{L,i}$ is the target output data defined above.
Note that the loss function takes discrete values. In particular, we are interested with
the ensemble of student machines in the $E=0$ space whose phase space volume is nothing but the Gardner's volume.

Starting from
a set of initial synaptic weights, the student machines are updated as the following.
\begin{enumerate}
\item Select a perceptron $\bs$ randomly out of the $N_{\bs}$ possible ones
  and select a link $k$ randomly out of the $c$ possible ones $k=1,2,\ldots,c$.
  Then propose a new synaptic weight,
  \beq
  (J_{\bs}^{k})^{\rm new}_{\rm student}=\frac{(J_{\bs}^{k})_{\rm student}+\delta x}{\sqrt{1+\delta^{2}}}
  \label{eq-J_new}
  \eeq
  where $\delta$ is a parameter and $x$ is an iid random number drawn from the Gaussian distribution
  with zero mean and unit variance. Note that $(J_{\bs}^{k})^{\rm new}_{\rm student}$
  is normalized such that its variance remains to be $1$.
\item  Accept the proposed one if the resultant loss function {\it does not increase}.
  Otherwise reject it and go back to 1.
  Importantly we accept updates by which the loss function remains unchanged. This is crucial to allow exploration of the
  $E=0$ (SAT) space.
\end{enumerate}
Within one Monte Carlo step (MCS) we repeat the above procedure for $N_{\bs}c$ times.

We simulate learning by two student machines '1' and '2'
which are subjected to the same training data but evolve independently from each other
using statistically independent random numbers for the step 1. and 2. explained above.

\subsubsection{Learning and unlearning}
\label{sec-learning-and-unlearning}

For the training, we consider the following two protocols
\begin{itemize}
\item {\bf Learning}: the initial synaptic weights of the student machines
$\{ (J_{\bs}^{k})_{\rm student}\}$ are prepared just as iid Gaussian random numbers
  totally uncorrelated with the teacher's weights $\{ (J_{\bs}^{k})_{\rm teacher}\}$.

To facilitate the training, we perform a sort of 'annealing'.
At a given time $t$ (MCS), perform the greedy Monte Carlo update
using a subset of the training data of size $M_{\rm batch}(t) (< M)$.
Starting from $M_{\rm batch}(0)=1$, increase $M_{\rm batch}(t)$ logarithmically
in time $t$ progressively adding more data to the training data-set
such that $M_{\rm batch}(t_{\rm max})=M$ in the end of the simulation at $t_{\rm max}$ (MCS).

\item {\bf Unlearning (or planting)}:
    the initial synaptic weights of the student machines
  $\{ (J_{\bs}^{k})_{\rm student}\}$ are set to be exactly the same as
    the teacher's weights $\{ (J_{\bs}^{k}\}_{\rm teacher}\}$.
    The student machine explores the $E=0$ (SAT) space.
\end{itemize}

If the greedy Monte Carlo method equilibrated the system, the two protocols should
yield the same results for macroscopic observables, which we explain below,
after averaging over time and/or initial configurations in the
stationary state.

   \subsection{\protect\red{Observables}}
   \label{sec-observables-simulation}

\subsubsection{Simple overlaps}
\label{sec-overlaps-simulation}

We are interested in the similarity between different machines in the hidden layers $l=1,2,\ldots,L-1$.
To quantify this we first introduce,  between the two student machines '1', '2' and the teacher machine '0', \red{the following 'simple' overlaps},
\beqn 
  q(l)=\frac{1}{NM}\sum_{i=1}^{N}\sum_{\mu=1}^{M}(S_{1})^{\mu}_{l,i}(S_{2})^{\mu}_{l,j} \\
  r(l)=\frac{1}{2NM}\sum_{i=1}^{N}\sum_{\mu=1}^{M}(S_{0})^{\mu}_{l,j}((S_{1})^{\mu}_{l,i}+(S_{2})^{\mu}_{l,i}).
 \label{eq-def-overlap-simple}
\eeqn
Here $\mu=1,2,\ldots, M$ for the training data and $\mu=1,2,\ldots,M'$ for the test data (and replace the factor $1/M$ by $1/M'$ in the latter case).
 
These are the same as the order parameters for the spins used in the replica theory  ( see \eq{eq-glass-order-parameters}).
 However, as mentioned in sec \ref{sec-order-parameters} the expectation value of the
 simplest overlaps defined above vanish in  thermal equilibrium because of the local gauge symmetry (and
 the permutation symmetry in the case $c=N$) as discussed in sec~\ref{sec-symmetry}.

\subsubsection{Squared Overlaps}
\label{sec-squared-overlaps-simulation}

To overcome the above problem we define the following order parameters
which we call as {\it squared overlaps} which are invariant under the
symmetry operations. Let us first introduce,
\beqn 
  q_{ab,ij}(l)&=&\frac{1}{M}\sum_{\mu=1}^{M}(S_{a})^{\mu}_{l,i}(S_{b})^{\mu}_{l,j} \\
%  r_{ij}(l)&=&\frac{1}{M}\sum_{\mu=1}^{M}(S_{\rm student})^{\mu}_{l,i}(S_{\rm teacher})^{\mu}_{l,j}.
  \label{eq-def-overlap-qij-rij}
  \eeqn
  Here $a$ and $b$ are indices for machines: $0$
  for the teacher machine, $1$ and $2$ for the student machines.
Then we introduce the squared overlaps as,
  \beq
  q_{2,ab}(l) = \frac{1}{N}\sum_{i,j=1}^{N}(q_{ab,ij}(l))^{2}-\frac{N}{M}
% =\frac{1}{NM^{2}}\sum_{i,j=1}^{N}
%  \sum_{\mu,\nu=1}^{M}(S_{\rm student-a})^{\mu}_{l,i}(S_{\rm student-a})^{\nu}_{l,i}
%  (S_{\rm student-b})^{\mu}_{l,j}  (S_{\rm student-b})^{\nu}_{l,j}
%  \qquad
%  \tilde{r}_{2}(l) = \frac{1}{N}\sum_{i,j=1}^{N}r_{ij}^{2}(l)-\frac{N}{M}
   \label{eq-def-overlap-gauge-invariant-0}
   \eeq
We note that this is analogous to the order parameter used in
numerical simulations of vectorial spinglass models which have
the rotational symmetry in spin space \cite{PhysRevE.61.R1008}.

\red{Finally we introduce the normalized version of the squared overlap,
   \beqn
   q_{2}(l)&=&\frac{q_{2,12}(l)}{\sqrt{q_{2,11}(l)}\sqrt{q_{2,22}(l)}} \nonumber \\
   r_{2}(l)&=&\frac{q_{2,01}(l)+q_{2,02}(l)}{\sqrt{q_{2,00}(l)}(\sqrt{q_{2,11}(l)}+\sqrt{q_{2,22}(l)}}.
      \label{eq-def-overlap-gauge-invariant}
   \eeqn
Interestingly these are very similar to the measure 
proposed in \cite{kornblith2019similarity} call as 'centered kernel alignment'.}
   
% Note that $q_{2}$ and $r_{2}$ become $0$ if the machines are totally uncorrelated. 
% It can be verified that these are invariant under the two symmetry operations
% mentioned in sec~\ref{sec-symmetry}. 

%Let us also note that these squared overlaps can be regarded also
 %as a kind of 4-point susceptibility which measure correlations of fluctuations within each layer.

   \subsubsection{\protect\red{Nishimori condition}}
 \label{sec-nishimori-condition}

 Since our teacher-student scenario is a Bayes optimal inference,
we have,
\beq
q(l)=r(l) \qquad  q_{2}(l)=r_{2}(l) \qquad  l=1,2,\ldots,L
\label{eq-identity-bayes-optimal}
\eeq
This is a Nishimori condition which must hold in Bayes optimal inferences \cite{nishimori2001statistical,iba1999nishimori,zdeborova2016statistical}.
These relations are useful to check the equilibration of the system.

\subsubsection{Physical meaning of the squared overlaps}
\label{sec-squared-overlaps-simulation-2}

Let us discuss more closely the significance of the squared overlap
defined in  \eq{eq-def-overlap-gauge-invariant-0} \red{and
           its normalized version  \eq{eq-def-overlap-gauge-invariant}}. In the following we denote
the average over different realization of the inputs as $\overline{\cdots}^{\rm input}$.
From \eq{eq-def-overlap-qij-rij} we can write,
\beqn
&& \overline{(q_{ab,ij}(l))^{2}}^{\rm input}=\frac{1}{M}+\frac{1}{M}\sum_{\mu=1}^{M}\frac{1}{M}\sum_{\nu ( \neq \mu)} \overline{r_{a,i}^{\mu \to \nu}r_{b,j}^{\mu \to \nu}}^{\rm input} \nonumber \\
&& \simeq \frac{1}{M}+\overline{r_{a,i}^{\mu \to \nu(\neq \mu)}r_{b,j}^{\mu \to \nu(\neq \mu)}}^{\rm input}
\eeqn
with 
\beqn
r_{a,i}^{\mu \to \nu}=(S_{a})^{\mu}_{li}(S_{a})^{\nu}_{li}
%\qquad
%r_{b,j}^{\mu \to \nu}=(S_{b})^{\mu}_{lj}(S_{b})^{\nu}_{lj} 
\eeqn
Here $r_{a,i}^{\mu \to \nu}$ can be regarded as change of the sign
of the spin (neuron) at the node $(l,i)$ of the student-a
when the input pattern is changes from $\mu$ to $\nu$.  Using the above expression
we find \eq{eq-def-overlap-gauge-invariant-0} with the subtraction term $-N/M$ becomes,
 \beq
  q_{2,ab}(l) \simeq \frac{1}{N}\sum_{j=1}^{N}
  \overline{r_{a,i}^{\mu \to \nu(\neq \mu)}r_{b,j}^{\mu \to \nu(\neq \mu)}}^{\rm input}
  \eeq
  This can be viewed as a kind of {\it correlation volume} within layer $l$ in the following sense.

%  Note that the $\mu$-th and the $\nu$-th
%input patterns are typically uncorrelated random patterns.

\begin{itemize}
\item{Totally uncorrelated random machines}
  
Suppose that the student-a and student-b are totally uncorrelated (far beyond the trivial
difference by the gauge transformation and the permutation) randomly generated machines.
Then we naturally expect
$\overline{r_{a,i}^{\mu \to \nu(\neq \mu)}r_{b,j}^{\mu \to \nu(\neq \mu)}}^{\rm input}=0$.
%so that $\overline{(q_{ab,ij}(l))^{2}}^{\rm input}=1/M$.
This means that the minimum value of the squared overlap $q_{2,ab}(l)$ is $0$.

\item{Same random machine modulo gauge transformation and permutation}

On the other hand, if the two machines are the same machine modulo the gauge transformation
and permutation we can write
  \beq
  \overline{r_{a,i}^{\mu \to \nu(\neq \mu)}r_{b,j}^{\mu \to \nu(\neq \mu)}}^{\rm input}=\delta_{ij}+(1-\delta_{ij})
    \overline{r_{a,i}^{\mu \to \nu(\neq \mu)}r_{b,j}^{\mu \to \nu(\neq \mu)}}^{\rm input}
    \eeq
Thus in this case the squared overlap $q_{2,ab}(l)$ is at least $1$ and can be larger.

In the case of the perceptrons with random synaptic weights and
the highly non-linear activation function (see \eq{eq-perceptron}),
we expect $\overline{r_{a,i}^{\mu \to \nu(\neq \mu)}r_{b,j}^{\mu \to \nu(\neq \mu)}}^{\rm input}$
becomes significant also between different nodes $i \neq j$.
This is because of the chaos effect which we discussed in sec.~\ref{sec-connection-to-SG}:
it is known that in such a non-linear random feed-forward network
a slight change of the
input induces chaotic changes in the state of spins (neuron) as the signal propagates deeper
into the network \cite{poole2016exponential}.
This is an avalanche-like process so that the correlation
$\overline{r_{a,i}^{\mu \to \nu(\neq \mu)}r_{b,j}^{\mu \to \nu(\neq \mu)}}^{\rm input}$ for
$i \neq j$ becomes more significant increasing $l$.
In this case the squared overlap
$q_{2,ab}(l)$ can be viewed as a measure of avalanche size within layer $l$.

\item General case

  Based on the above consideration, we naturally expect that 
  in general $q_{2,ab}(l)$ quantifies the avalanche size and
  similarity of the avalanche patterns taking place
  in machines a and b through changes of inputs $\mu \to \nu(\neq \mu)$.
  \red{Then it becomes clear that  normalized version \eq{eq-def-overlap-gauge-invariant} quantifies the  quantifies the
  similarity of the avalanche patterns in machines a and b.}
  
  \end{itemize}

\subsubsection{Generalization error}

 To measure the generalization ability of the student machines
 (see sec.~\ref{sec-teacher-student-setting}) we measure,
 \beq
 r_{\rm out}= \frac{1}{NM}\sum_{i=1}^{N} \sum_{\mu=1}^{M'}(S_{\rm student})^{\mu}_{l,i}(S_{\rm teacher})^{\mu}_{l,i}
\eeq
Here we use the $M'$ sets of the outputs of the teacher and student machines
for the test data (not used for training). The generalization error (see \eq{eq-epsilon}) can be evaluated as,
\beq
\epsilon=\frac{1}{2}(1-r_{\rm out}).
\label{eq-epsilon-simulation}
\eeq
In the above expression, we used simple overlap defined on the output layer $l=L$.
Note that there are no gauge transformations or permutations on the output layer.

\begin{figure*}[t]
    \includegraphics[width=0.95\textwidth]{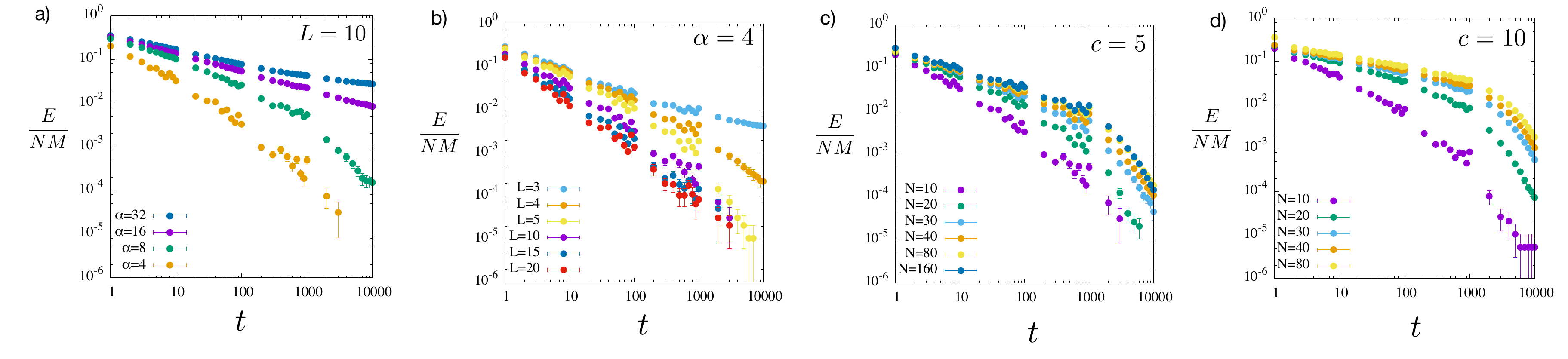}
    \caption{Relaxation of the loss function in learning \red{(annealing with $t_{\rm max}=10^{4}$ (MCS))}
      \red{observed by the MC simulation.}
In all cases $N=10$. 
(a) various $\alpha=M/c$ with $L=10$ and $c=5$. (b) various $L$ with $\alpha=M/c=4$
and $c=5$. (c) various $N$ with $L=10$, $\alpha=4$ and $c=5$.
(d) the same as (c) but with $c=10$.
  The unit of time $t$ is $1$ (MCS). }
\label{fig_relaxation_loss}
\end{figure*}

\begin{figure*}[t]
  \begin{center}
        \includegraphics[width=0.95\textwidth]{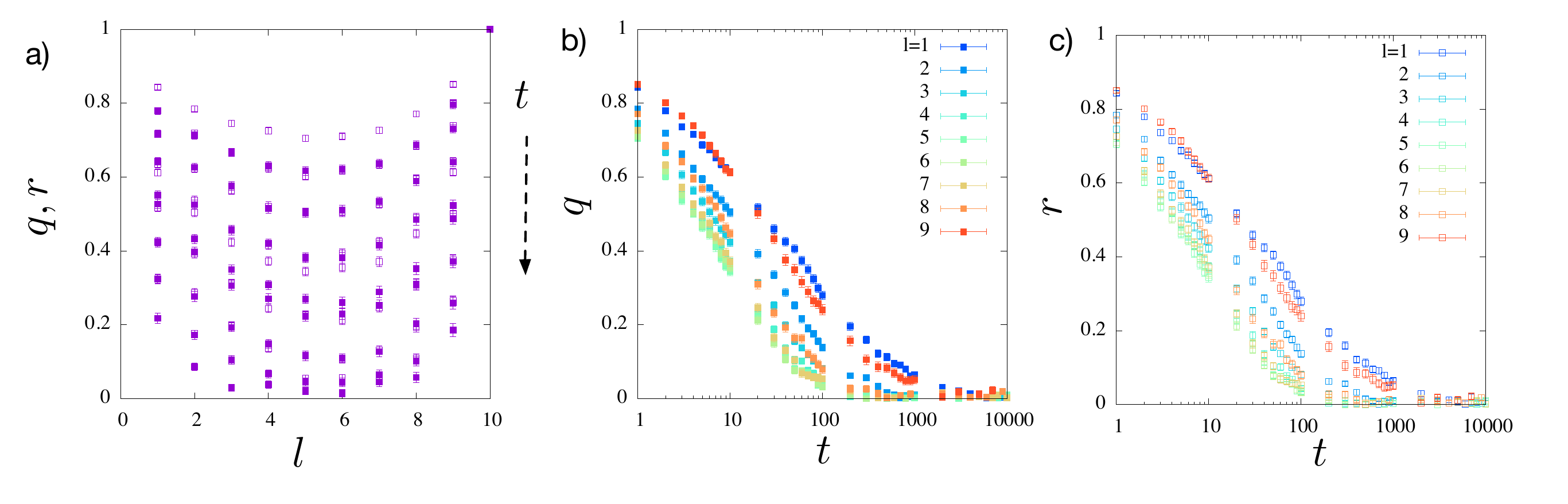}
    \end{center}
  \caption{\red{Time evolution of} simple student-student overlap $q(l)$ (filled symbols) and  teacher-student overlap $r(l)$ (open symbols)
           observed in \red{MC simulations of} the unlearning protocol. Here $N=10$, $\alpha=4$, $c=5$.
           Panel a) shows data at $t=1,2,4,8,10,20,40,80$. Panels b) and c) show the simple student-student overlap $q(l)$
         and simple teacher-student $r(l)$ overlap respectively.}
\label{fig-vanishing-simple-overlap}
\end{figure*}

\subsection{Results}
\label{sec-results-simulation}

Now let us discuss the results of the simulations.
First, we discuss the equilibration process through the learning and unlearning protocols (see sec.~\ref{sec-learning-and-unlearning}). Next, we discuss the equilibrium properties of macroscopic observables.

\red{In the simulations we used $\delta=0.1$ to generate new weights by \eq{eq-J_new}.
For the test we used $M'=M$ data uncorrelated with the training data.}
In the following observables are averaged
are took over statistically independent $240$ samples (different realizations of the teacher machine, initial configurations of student machines for learning, realizations of random numbers used in Monte Carlo updates).

%\subsubsection{Learning and Unlearning}

\subsubsection{Learning}

In Fig.~\ref{fig_relaxation_loss} we present the relaxation of the loss function
\eq{eq-loss-function} in the learning protocol (see sec.~\ref{sec-learning-and-unlearning}).
It can be seen in panel a) that relaxation of the loss function slows down by increasing the
number of the training data $M=c\alpha$.
On the other hand, it can be observed in panel b) that relaxation becomes faster increasing the
depth $L$ of the network.

As shown in panel c), the relaxation depends also on the width
$N$ but converges in large enough $N$ with fixed $c$ and $\alpha$
suggests that
relaxation time is finite in systems with finite connectivity $c$ even in $N \to \infty$ limit.
For larger $c$, the relaxation curves converge to a slower curve as shown in panel d)
suggesting that the relaxation time becomes larger for larger connectivity $c$.

\subsubsection{Unlearning}

 In Fig.~\ref{fig-vanishing-simple-overlap} we show the 
 simple overlaps defined in \eq{eq-def-overlap-simple} observed in the unlearning protocol
 which explores the $E=0$ landscape (SAT phase)
(see sec.~\ref{sec-learning-and-unlearning}). Note that $q(l)=r(l)=1$ at the beginning.
The student machines become de-correlated from the teacher machine
and also de-correlated from each other as time $t$ elapses. 
It is interesting to note that relaxation is in-homogeneous in space:
relaxation is faster in the central part of the network and slower closer to the input/output boundaries.

It is important to note that the complete vanishing of the simple overlaps does not necessarily mean that the solution space is completely in a liquid state as the overlaps are not gauge invariant.
Because of the gauge symmetry
(see sec~\ref{sec-gauge-symmetry}), even machines that are completely the same as 
the teacher-machine modulo the gauge transformation can have vanishing simple overlap
with the teacher-machine. Indeed we will find below that normalized squared overlaps \eq{eq-def-overlap-gauge-invariant} (which are gauge invariant) instead indicate correlations between different machines.

The in-homogeneity of the relaxation observed here suggests that the system is more constrained closer to the boundary while the center is freer. We have also observed that the deeper system
relax faster as shown in Fig.~\ref{fig_relaxation_loss} b).
These may be interpreted as an echo of
the  'crystal-liquid-crystal' sandwich structure predicted by the theory (Fig.~\ref{fig-order-parameter-profile-replica}).

%:  the sandwich
%structure 'solid-liquid-solid' predicted by the theory is replaced by a spatially inhomogeneous
%liquid which is more viscous closer to the boundaries.

\begin{figure*}[t]
    \bc
             \includegraphics[width=0.95\textwidth]{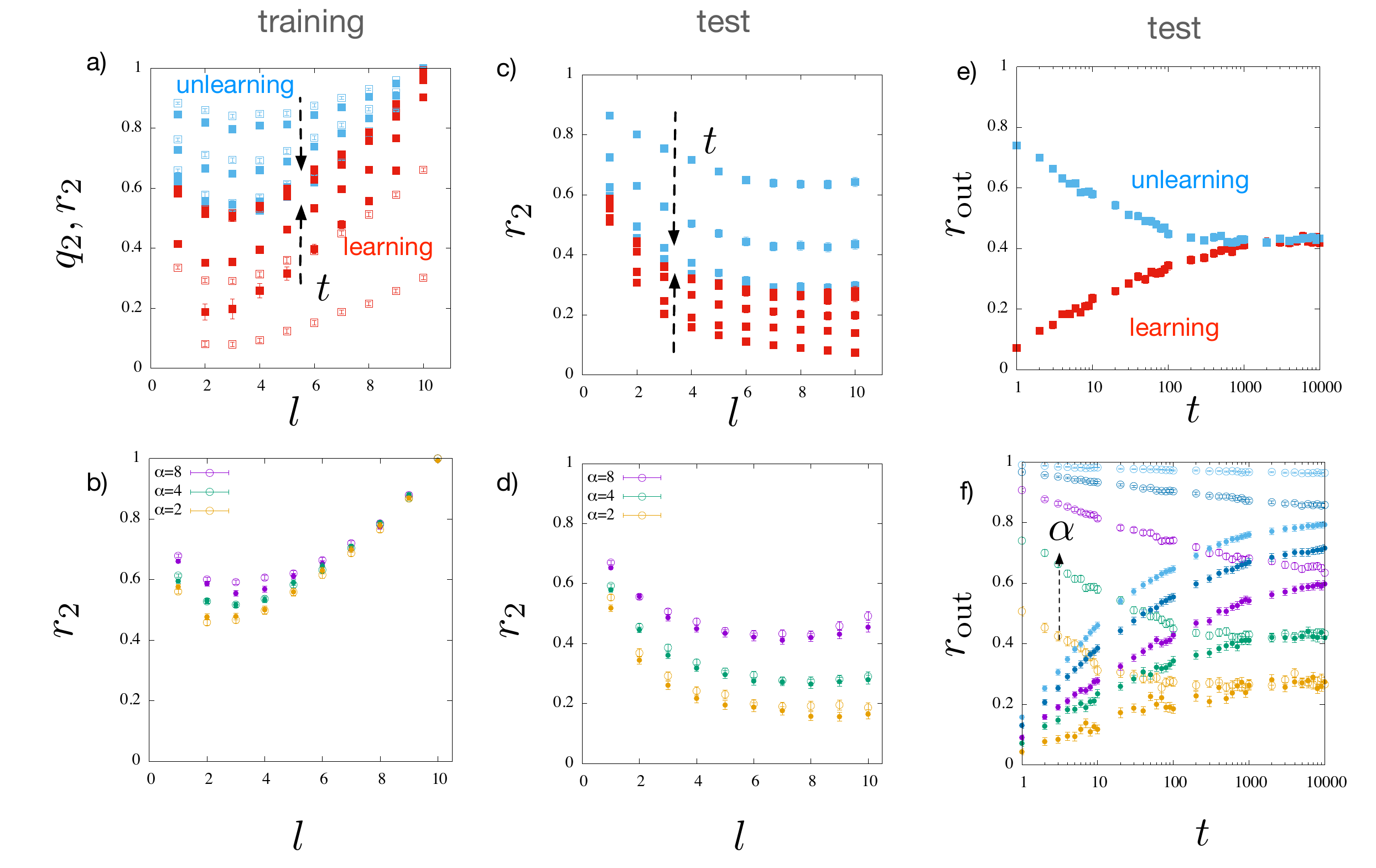}
  \ec
  \caption{Spatial profile of the normalized squared teacher-student overlaps $r_{2}(l)$
    and student-student overlaps $q_{2}(l)$ (see \eq{eq-def-overlap-gauge-invariant})
    for training a),b) and test c),d) and time evolution of the simple teacher-student overlap $r(L)$
    in the output layer ($l=L$) for test e), f). \red{All data are obtained by
    the MC simulations.}
    In panels a), c), and e), data of $q_{2}(l)$ (open symbols) and $r_{2}(l)$ (filled symbols)
    of learning/unlearning are represented by red/blue points ($\alpha=4$) .
    Panels a) and c) show the normalized squared overlaps at various times
    $t=1,10,100,1000,10000$ (increasing along the arrows) at each layer.
    Panel e) shows the time evolution of the simple teacher-student overlap $r(L)$
    at the output layer ($l=L$) for the test data.
    Panel b) and d) show the normalized squared teacher-student overlap for unlearning (open symbols)/learning (filled symbols)
    at $\alpha=2,4,8$ at $t=10^{4}$ (MCS).
    Panel f) show the time evolution of the simple teacher-student overlap $r(L)$
     at the output layer ($l=L$)  for the test data,  obtained by unlearning (open symbols) /learning
    (filled symbols) protocols with $\alpha=2,4,8,16,32$.
Here $N=10$, $L=10$, and $c=5$ for all data.
}
   \label{fig_N=10_M=40_L=10}
\end{figure*}

\subsubsection{Equilibration}

In equilibrium learning and unlearning protocols should give the same results for macroscopic observables after sufficiently long times. This is indeed verified as shown in the top panels a) c) e) of Fig.~\ref{fig_N=10_M=40_L=10}. In panels a) and c) we show the normalized squared overlaps defined in  \eq{eq-def-overlap-gauge-invariant}
which are invariant under the gauge transformations.
The normalized squared overlaps of unlearning and learning protocols agree
suggesting the establishment of equilibrium.
Furthermore,  it can be seen that the Nishimori condition $q_{2}(l)=r_{2}(l)$ (see \eq{eq-identity-bayes-optimal}) expected for the Bayes optimal inferences become satisfied after sufficiently long times. This is another evidence of thermal equilibration. Equilibration can also be seen in panel e) where we show the simple overlap between the teacher and student machines in the output layer $l=L$ for the test data.

As can be seen in Fig.~\ref{fig_N=10_M=40_L=10}, the spatial profile of the normalized squared overlaps
$q_{2}(l)$ and $r_{2}(l)$ are strongly in-homogeneous in space.
As discussed in sec.~\ref{sec-squared-overlaps-simulation-2}, we consider the normalized squared overlaps quantifies
\red{the similarity of the avalanche patterns} taking place
  in different machines through changes of inputs $\mu \to \nu(\neq \mu)$.
  At the beginning of unlearning, which starts from the teacher's configuration, the normalized squared overlaps
  take high values.
  %increase nearly linearly with $l$. This essentially reflect the avalanches
%of the teacher machine.
  On the other hand,
  they are small at the beginning of learning which is not surprising 
because teacher and student machines are totally uncorrelated at the beginning.
In equilibrium, they converge to a non-trivial, spatially non-monotonic function.
%$q_{2}(l)(=r_{2}(l))$ still grows with $l$. 
This implies that the equilibrium phase is not just a liquid as we might have thought based on the observation of the vanishing
simple overlap (Fig.~\ref{fig-vanishing-simple-overlap}).
On the contrarily, the gauge invariant quantity show that the student machines are strongly correlated with each other and with the teacher machine in equilibrium.
The spatial non-monotonicity means that they become less correlated
with each other
in the center (beyond the trivial difference by the gauge transformations)
while they are similar to each other (modulo the gauge transformation) closer to the input and output boundaries. This observation can be regarded as another echo of the spatial in-homogeneity predicted by the theory (Fig.~\ref{fig-order-parameter-profile-replica}). From the theoretical point of view, the strong asymmetry concerning the exchange of input/output sides, which is absent in the saddle point solution, may be attributed to the finiteness of the width $N$ and the connectivity $c$ \red{as we discussed in sec.~\ref{sec-finite-N-c-D-effect-discussion}}.

As shown in the bottom panels b) d) f) in Fig.~\ref{fig_N=10_M=40_L=10},
the overlaps increase as $\alpha$ increases as expected. In panel f) it can be seen that dynamics
of both learning and unlearning slow down as $\alpha$ increases.

\subsubsection{Typical student machines}

 \begin{figure*}[h]
  \includegraphics[width=0.6\textwidth]{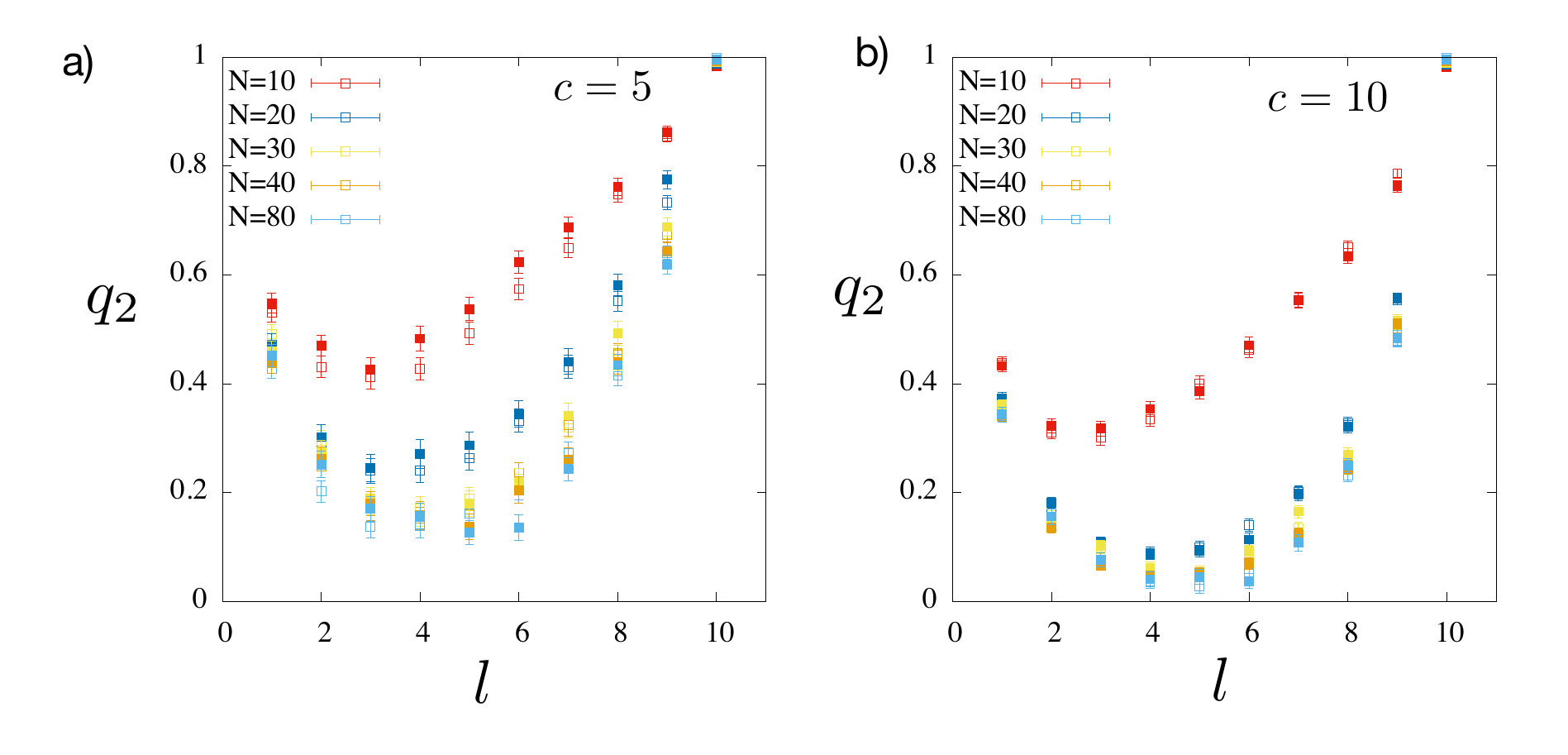}
  \caption{Finite $N$ effect and finite $c$ effect: the asymmetry becomes smaller as $N$ increases for unlearning (open symbols) /learning  (filled symbols). $N=10,20,30,40,80$ with  $\alpha=1.0$ and (left) $c=5$, (right) $c=10$ .  All data are obtained by \red{MC simulations of}  $t=10^{4}$ (MCS).}
\label{fig_finite_N_effect}
 \end{figure*}

 \begin{figure*}[h]
  \bc
  \includegraphics[width=0.95\textwidth]{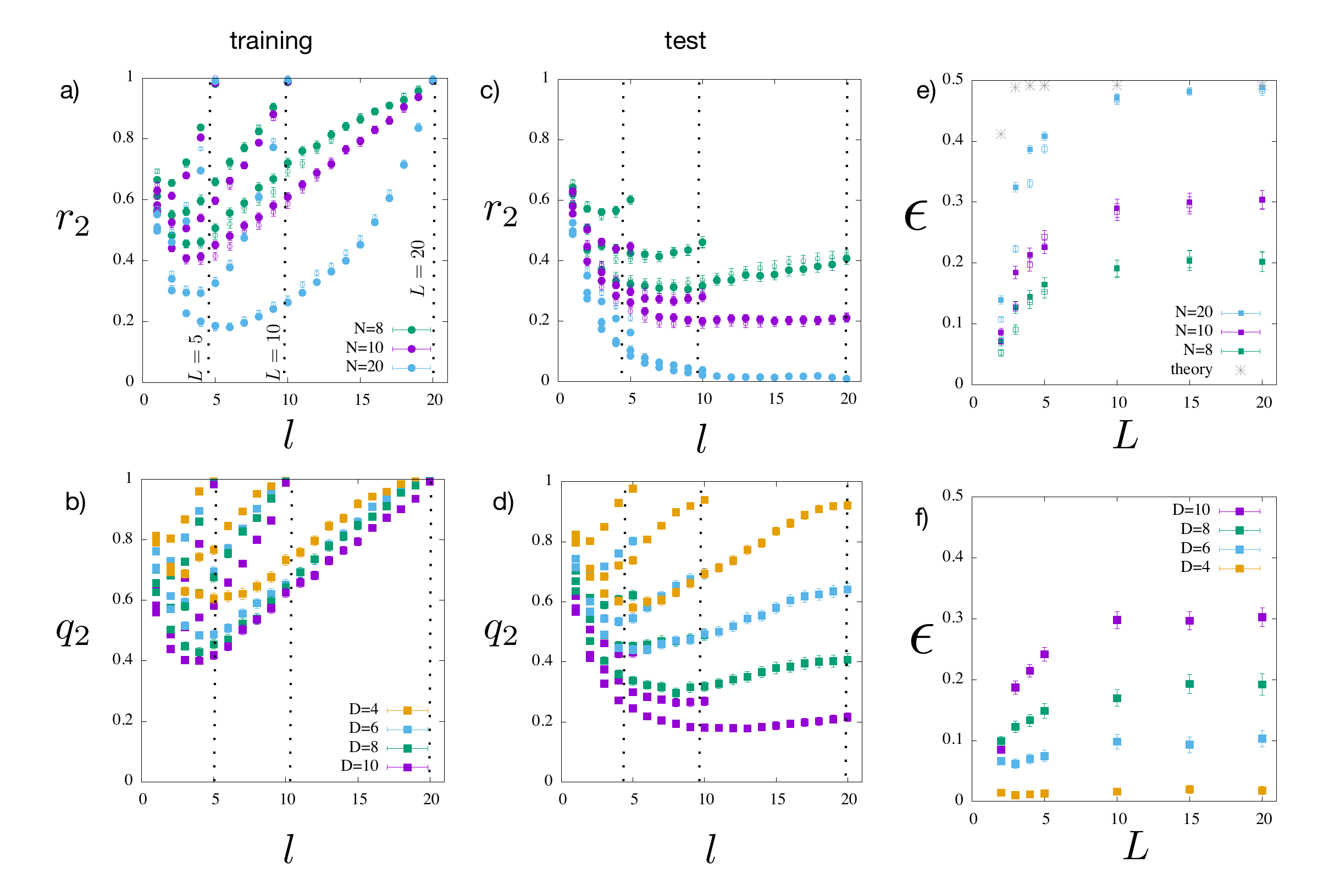}

  \ec
  \caption{Spatial profile of the normalized squared overlaps and the generalization error
in systems with various width $N$ (top panels) and hidden dimension $D (\leq N)$ (bottom panels) \red{obtained by MC simulations}.
In the top panels, data obtained by both learning (filled symbols) and unlearning (open symbols) are shown.
Panels a) and b) show the normalized squared overlaps for training, c) and d) show those for the test.
Panels e) and f) show the generalization error $\epsilon$ (see \eq{eq-epsilon-simulation}).
In panel e) we also show the generalization error $\epsilon$ obtained by the theory
in the dense limit: $N \to \infty$ \red{followed by $N \to \infty$} (see \eq{eq-epsilon} and Fig.~\ref{fig_replica_all_epsilon}).
In panels a), b), c), and d) data with $L=5,10,20$ are shown.
In panels a), c), and e) data with $N=8,10,20$ are shown.
In panels b), d) and f) data with $D=4,6,8,10$ are shown.
In all cases $\alpha=4$, $c=5$ and $t=10^{4}$.
  }
  \label{fig_variousN_variousL}
\end{figure*}

 Now let us examine further the equilibrium properties, i.~e. properties of typical student machines
 sampled in the solution space.
 We show in Fig. ~\ref{fig_finite_N_effect}
some data of the normalized squared overlaps $q_{2}(l)$. It can be seen again
that the data obtained by both learning (filled symbols)
and unlearning (open symbols) agree confirming that the system is equilibrated.
Quite remarkably the equilibrium normalized squared overlap $q_{2}(l)$ evolves non-monotonically in space for large enough $N$ and $c$. It first decreases with $l$ but finally increases with $l$.
This means avalanches taking place in different machines become de-correlated in the middle of the network but strongly correlated closer to the input and output boundaries.
It appears that the situation has become closer to the 'crystal-liquid-crystal' sandwich structure predicted by the theory (Fig.~\ref{fig-order-parameter-profile-replica}) \red{increasing $N$ and $c$}.

\red{
In sec.~\ref{sec-finite-N-c-D-effect-discussion} we discussed that the corrections due to the loops and fluctuations around
the saddle point can be separated considering the dense limit $N \gg c \gg 1$ but bring similar effects 1) remnant symmetry breaking
field 2) input-output asymmetry.}
Indeed it can be seen in Fig. ~\ref{fig_finite_N_effect} that for fixed connectivity $c$, the normalized squared overlaps decrease around the center of the
network and the asymmetry with respect to the exchange of input/output becomes weaker as $N$ increases.
Furthermore, the data suggests convergence in  $N \to \infty$ limit with fixed $c$.
%From the theoretical point of view (see sec.~\ref{sec-finite-N-effect}),
%a natural interpretation is that the finite width $N$ effects
%are due to the loop corrections which induce spatial correlations 
%inside the network and break the symmetry concerning the exchange of the input and output sides.
Comparing panels a) and b) it can be seen that the remnant asymmetry becomes weaker as the connectivity $c$ increases.
%As discussed in sec.~\ref{sec-finite-c-effect}, we expect finite $c$ corrections also induce such asymmetry due to higher order terms in the expansion around the saddle point.
Remarkably de-correlation in the center also becomes clearer increasing $N$ and $c$, 
suggesting the emergence of liquid like region in the center due to over-parametrization as suggested theoretically
in the dense limit $N \gg c \gg 1$.

In Fig.~\ref{fig_variousN_variousL} we show the normalized squared overlaps $q_{2}(l)$, $r_{2}(l)$
and the generalization error $\epsilon$ in systems with $\alpha=4$, $c=5$ observed after $t=10^{4}$ (MCS) in systems with different depth $L=5,10,20$. As shown in the top panels a) c) and e)
data obtained by both learning (filled symbols) and unlearning (open symbols) agree proving again that the system is
equilibrated. As shown in panels a) and c) we find again that the normalized squared overlaps are strongly in-homogeneous in space. The machines de-correlate more concerning each other in the central region in deeper systems but correlations recover approaching the output layer.
We also find again that normalized squared overlaps increase significantly and that the asymmetry concerning the exchange of the input and output sides becomes stronger decreasing the width $N$.

Now let us turn to the effect of finite dimension $D$ introduced by the hidden manifold model (see  sec.~\ref{sec-hidden-minifold-simulation}).
The results of simulations on the hidden manifold model are
displayed in panels b) and d) of Fig.~\ref{fig_variousN_variousL}. Here we used the simplest
folding matrix $F$ of the form \eq{eq-simple-folding-matrix} but we obtained qualitatively the same results (not shown) also in the case of random matrices.
Comparing the panels b) to a) and d) to c) we immediately notice that the effect of hidden dimension $D$
is quite similar to the effect of finite width $N$: decreasing $D$ with fixed $N$ is much like decreasing $N(=D)$. We conjecture that this is due to the enhancement of the loop corrections induced by
the closing of the loops by the correlated inputs as discussed in sec.~\ref{sec-finite-D-effect}.

Finally, let us discuss the generalization error $\epsilon$ shown in panels e) and f) of Fig.~\ref{fig_variousN_variousL}. In panel e) we also show the generalization error $\epsilon$ obtained by the theory in the dense limit $c \to \infty$ (see \eq{eq-epsilon} and Fig.~\ref{fig_replica_all_epsilon}). Remarkably the effect of finite width $N$ and hidden dimension $D$ is very similar again. The generalization error improves significantly either by decreasing $N(=D)$ or $D$ with fixed $N$. Presumably this is due to the increase of correlations inside the network
induced by the loop corrections.
Moreover, the generalization error becomes independent of the depth $L$ at sufficiently
deep systems much like the theoretical prediction. The result implies the
generalization ability first decreases making the system deeper but does not vanish even in $L \to \infty$ limit.
\red{This is consistence with the $L$ independent learning curve $\epsilon=\epsilon_{\infty}(\alpha)$ predicted theoretically
  (see sec.~\ref{sec-generalization-error-theory}).}

% \clearpage

\section{Conclusions}
\label{sec-conclusions}

\begin{redtext}
In the present paper we obtained an exactly solvable statistical mechanics model
of machine learning by a deep neural network (DNN) in the dense limit $N \gg c \gg 1$.
Exact solutions are obtained using the replica method developed in \cite{yoshino2020complex}.
We used the replica theory to analyze the generalization-ability of the DNN in the Bayes-optimal
teacher-student setting. The learning curve $\epsilon=\epsilon_{L}(\alpha)$ becomes
independent of of the depth $L$ as long as the two crystalline phases attached to input/output
boundaries are separated by the liquid phase in the center. Thus the system is predicted
to generalize even in the limit $L \to \infty$ where the system becomes extremely over-parametrized.
We discussed the loop corrections to the dense limit and argued that
finite width $N$ and finite hidden dimension $D$ effects appear similarly.
Both should lead to increase of correlations \red{inside the network}.
\end{redtext}

In simulations, the simple greedy Monte Carlo method turned turned out to work efficiently to
enable sampling of typical machines in equilibrium suggesting the simplicity of the
loss landscape. The main obstacle in simulations is the gauge invariance of the system
by which order parameters in the original simple form vanish.
To overcome the difficulty we measured normalized squared overlap 
which quantifies the correlation of avalanches concerning changes in input data between
different machines. It is a gauge (and permutation) invariant quantity 
that reflects the similarity between machines modulo the gauge (and permutation) symmetries. 
\red{The result is qualitatively consistent with the theoretical prediction that
over-parametrization in the DNN lead to spatially in-homogeneous learning: students become close to the
teacher around the input/output boundaries while remain only weakly correlated in the center.}
We note that liquid like central region was also noticed in \cite{zou2021data}.
Furthermore, \red{somewhat counter-intuitively but} in agreement with the theoretical prediction,
the generalization error first increases increasing the depth $L$ but then
becomes independent of the depth $L$ suggesting that the generalization ability 
survives in $L \to \infty$ limit.
Simulations confirm that finite width $N$
and finite hidden dimension $D$ effects are quite similar and lead similarly to
significant improvements in the generalization ability. Presumably this reflects
increase of correlations inside the network due to the loop corrections.
\red{As we noted in sec.~\ref{sec-finite-N-c-D-effect-discussion} we consider that the
  corrections due to the loops and fluctuations around the saddle point play the role of
symmetry breaking field which allows the student to recognize again the teacher in spite of the liquid like center.}

After all what is the advantage of making the system deeper? One important advantage is that
the learning dynamics become faster increasing the depth as we found numerically.
This should be due to the presence of the central region where the system is less constrained.
We believe that this point will become more important as we move away from the idealized, Bayes optimal teacher-student setting
we considered in the present work. From the theoretical point of view, there is no guarantee that replica symmetry continue
to hold as we move away from the Bayes optimal situation toward the situations in the real world.
For example, one can consider a noisy teacher-student scenario
by adding noise to the training data provided by the teacher.
Then the situation becomes closer to the random scenario considered in 
\cite{yoshino2020complex} where complex replica symmetry breaking (RSB) was found in the DNN. In the latter case,
RSB evolves in space such that the hierarchy of RSB becomes simplified layer-by-layer
approaching the center so that the central region can remain in replica symmetric liquid phase
if the network is made deep enough. This implies deeper system will relax faster even in the presence of
the RSB around the boundaries.

\red{There are numerous directions to generalize and extend the present work.
  Let us mention a few of them below.
  It is straightforward to study the model exactly with the parameter $\alpha$ depending on space
  $\alpha=\alpha(l)$. This amount to make the width $N$ of the network to vary in space $N=N(l)$.
  It  will be interesting to study how one can control the
  spatial heterogeneity of learning by changing $\alpha(l)$.
  The dense limit $N \gg c \gg 1$ will be useful not only for the replica
theory but other theoretical approaches. For instance, it should be possible to develop cavity
approaches in the dense limit. It will also be very interesting to generalize our theory
considering more general activation functions as we noted in sec.~\ref{sec-connection-to-SG}.
}

\section*{Acknowledgments}

We thank Giulio Biroli, Angelo Giorgio Cavaliere, Yoshiyuki Kabashima and Ryo Karakida for useful discussions. 
Numerical simulations presented in this work has been done using the supercomputer systems SQUID at the Cybermedia Center, Osaka University. This work was supported by KAKENHI (No. 21K18146) (No. 22H05117) from MEXT, Japan.

%\clearpage
\begin{appendix}

\begin{redtext}
    
\section{Transfer matrix representation of feed-forward networks}
\label{sec-transfermatrix}

Here we show that transfer-matrix representations of a family of spin-glass models put in the layered geometry
(like in Fig.~\ref{fig_schematic_model}) become feed-forward DNNs in the zero temperature limit.

In the following we denote the configuration of the set of spins
in the $l$-th layer as ${\bf S}_{l}^{\mu}= \{S_{\bs \in l}^{\mu}\}$
where $\mu$ is the label to specify a data set.
Let us write the conditional probability to realize a
spin configuration ${\bf S}^{\mu}_{L}$ on the output boundary 
given a spin configuration ${\bf S}^{\mu}_{0}$ on the input boundary 
as,
%$P({\bf S}^{\mu}_{0} \to {\bf S}^{\mu}_{L})$
%\beq
%P({\bf S}^{\mu}_{0} \to {\bf S}^{\mu}_{L})=
%%\frac{\langle {\bf S}_{0}^{\mu} |\prod_{l=1}^{L}{\bf T}_{l}|{\bf S}_{L}^{\mu}\rangle}{{\rm Tr}_{{\bf S}^{\mu}_{L}}\langle {\bf S}_{0}^{\mu} |\prod_{l=1}^{L}{\bf T}_{l}|{\bf S}_{L}^{\mu}\rangle  }
%\label{eq-conditional-prob-from-input-to-output}
%\eeq
%where ${\bf T}_{l}$ is the transfer matrix from the $l-1$ to $l$ th layer. We can naturally decompose the conditional probability \eq{eq-conditional-prob-from-input-to-output} using $1={\rm Tr}_{S_{l}^{\mu}}| {\bf S}_{l}^{\mu}\rangle\langle {\bf S}_{l}^{\mu}|$ as,
\beq
P({\bf S}^{\mu}_{0} \to {\bf S}^{\mu}_{L})
= \left(\prod_{l=1}^{L-1}{\rm Tr}_{{\bf S}^{\mu}_{l}}\right)
\prod_{l=1}^{L}P({\bf S}^{\mu}_{l-1} \to {\bf S}^{\mu}_{l})
\label{eq-conditional-prob-from-input-to-output}
\eeq
where $P({\bf S}^{\mu}_{l-1} \to {\bf S}^{\mu}_{l})$
is the conditional probability to realize a
spin configuration ${\bf S}^{\mu}_{l}$ on the $l$-th layer
given a spin configuration ${\bf S}^{\mu}_{l-1}$ on the $l-1$-th layer,
\beqn
P({\bf S}^{\mu}_{l-1} \to {\bf S}^{\mu}_{l})&=&
\frac{\langle {\bf S}_{l-1}^{\mu} |{\bf T}_{l}|{\bf S}_{l}^{\mu}\rangle}{
  {\rm Tr}_{{\bf S}^{\mu}_{l}}\langle {\bf S}_{l-1}^{\mu} |{\bf T}_{l}|{\bf S}_{l}^{\mu}\rangle   } 
\label{eq-transfer-matrix-repsentation-for-conditional-prob}
\eeqn
where ${\bf T}_{l}$ is the transfer matrix from the $l-1$ to $l$ th layer.

\subsection{Layered Ising spin-glass model}
\label{sec-layered-Ising-SG}

Let us first consider the layered Ising spin-glass model
\eq{eq-hamiltonian-layered-SG}
with the restricted Boltzmann machine (RBM) like architecture
\cite{ackley1985learning}, 
\beq
H=-\frac{1}{\sqrt{c}}\sum_{\bs} \sum_{k=1}^{c} J_{\bs}^{k}S_{\bs}S_{\bs(k)}
\eeq
where the spins are Ising variables $S^{\mu}=\pm 1$.
The matrix elements of the transfer matrix are given by,
\beq
\langle {\bf S}^{\mu}_{l-1} | {\bf T}_{l} |  {\bf S}^{\mu}_{l} \rangle
=e^{\sum_{\bs \in l} \sum_{k=1}^{c} \frac{\beta J_{\bs}^{k}}{\sqrt{c}}S_{\bs}^{\mu}S^{\mu}_{\bs(k)}}
\label{eq-transfer-matrix-ising}
\eeq
Here $\beta=1/T$ is the inverse of the temperature $T$.
The size of the matrix is $2^{N} \times 2^{N}$.
Then we find the conditional probability
\eq{eq-transfer-matrix-repsentation-for-conditional-prob}
becomes,
\beq
P({\bf S}^{\mu}_{l-1} \to {\bf S}^{\mu}_{l})
= \prod_{\bs \in l}P({\bf S}^{\mu}_{l-1} \to S^{\mu}_{\bs})
\label{eq-transfer-matrix-repsentation-for-conditional-prob-ising}
\eeq
where
\beq
\prod_{\bs \in l}P({\bf S}^{\mu}_{l-1} \to S^{\mu}_{\bs})
=\prod_{\bs \in l}\frac{e^{\beta r_{\bs}^{\mu}}}{
  {\rm Tr}_{S_{\bs}^{\mu}}  e^{\beta r_{\bs}^{\mu}}}
\eeq
with $r_{\bs}^{\mu}$ being the gap variable \eq{eq-gap},
\beq
r_{\bs}^{\mu}=S_{\bs}^{\mu} 
\sum_{k=1}^{c} \frac{J_{\bs}^{k}}{\sqrt{c}}S^{\mu}_{\bs(k)}
\eeq
It is important to notice the factorization of the conditional probability
in \eq{eq-transfer-matrix-repsentation-for-conditional-prob-ising}.
It is the consequence of the RBM type architecture:
there are no direct interactions within each layer.

Taking the zero temperature limit $T\to 0$ ($\beta \to \infty$) we find,
\beq
P({\bf S}^{\mu}_{l-1} \to {\bf S}^{\mu}_{l})
 \xrightarrow[T \to 0]{}
\prod_{\bs \in l}\theta(r_{\bs}^{\mu})
\eeq
where $\theta(r)$ being the Heviside step function.
Thus in this limit the configuration
$S_{\bs}^{\mu}$ in $\bs \in \l$
is determined deterministically given ${\bf S}^{\mu}_{\ws}$
in $\ws \in l-1$ by the perceptron's rule  \eq{eq-perceptron},
\beq
S_{\bs}^{\mu}={\rm sgn} \left(
\sum_{k=1}^{c} \frac{J_{\bs}^{k}}{\sqrt{c}}S^{\mu}_{\bs(k)} \right)
\eeq
Note that the operation of the transfer matrix of size
$2^{N} \times 2^{N}$ \eq{eq-transfer-matrix-ising} is now replaced
in the $T \to 0$ limit
by simple non-linear mapping of much lower computational
cost of $O(Nc)$ thanks to the 1) RBM like network structure and 2) $T \to 0$ limit.

Similarly we find,
\beq
\lim_{T \to 0} 
P({\bf S}^{\mu}_{0} \to {\bf S}^{\mu}_{L})=
\left( \prod_{\bs\backslash {\rm output}}{\rm Tr}_{{\bf S}^{\mu}_{\bs}}  \right)
\prod_{\bs}
\theta(r_{\bs}^{\mu})
\eeq
which means $S_{\bs \in L}$ in the output layer becomes determined
by the multi-layer perceptron for a given  $S_{\bs \in 0}$ in the output layer. Note that the Gardner's volume \eq{eq-gardner-volume-DNN} can be
written as,
\beq
V_{M}\left({\bf S}_{0},{\bf S}_{L}\right)=\lim_{T \to 0}
\left(\prod_{\bs} {\rm Tr}_{{\bf J}_{\bs}}\right)
\prod_{\mu}  P({\bf S}^{\mu}_{0} \to {\bf S}^{\mu}_{L})
\eeq
In this representation it becomes clear that
the traces over the spins in the hidden
layers used in the Gardner's volume for DNN
\eq{eq-gardner-volume-DNN} (the internal representation
\cite{monasson1995weight}) 
are equivalent to computation of the products of transfer-matrices
in \eq{eq-conditional-prob-from-input-to-output}.

\subsection{Layered spin-glass models with continuous spins}

Now let us consider a class of slightly more
generalized models on the RBM type network with the Hamiltonian,
\beq
H=-\frac{1}{\sqrt{c}}\sum_{\bs} \sum_{k=1}^{c} J_{\bs}^{k}S_{\bs}S_{\bs(k)}
+\sum_{\bs}U(S_{\bs})
\label{eq-hamiltonian-layered-SG-general}
\eeq
Here we consider spins which take continuous values
$-\infty < S  < \infty$. The function $U(x)$ in the 2nd term
of the Hamiltonian
\eq{eq-hamiltonian-layered-SG-general} represents a confining
potential to regularize the spins.
Then the conditional probability 
\eq{eq-transfer-matrix-repsentation-for-conditional-prob}
becomes,
\beqn
P({\bf S}^{\mu}_{l-1} \to {\bf S}^{\mu}_{l})
&=&\prod_{\bs \in l}\frac{e^{\beta (h_{\bs}^{\mu} S_{\bs}^{\mu}  - U(S_{\bs}^{\mu}))}
 }{
  {\rm Tr}_{S_{\bs}^{\mu}}
e^{\beta (h_{\bs}^{\mu} S_{\bs}^{\mu}  - U(S_{\bs}^{\mu}))}}  
%\label{eq-transfer-matrix-repsentation-for-conditional-prob}
\eeqn
where
\beq
h_{\bs}^{\mu}=
\sum_{k=1}^{c} \frac{J_{\bs}^{k}}{\sqrt{c}}S^{\mu}_{\bs(k)}
\eeq

In $T \to 0$ limit we find,
\beqn
P({\bf S}^{\mu}_{l-1} \to {\bf S}^{\mu}_{l})
& \to & \prod_{\bs \in  l}
\delta\left ( {\bf S}^{\mu}_{\bs}- f(h^{\mu}_{\bs})\right)
\eeqn
where the function $f(h)$ is determined such that
\beq
f(h)={\rm argmin}_{S}
\left ( \blue{-hS+U(S)} \right)
\eeq
Assuming that the potential $U(s)$ is differentiable we find,
\beq
f^{-1}(S)=\blue{\frac{d U(S)}{dS} }
\eeq
or
\beq
U(S)=\int_{-\infty}^{S} f^{-1}(S) dS
\eeq
Thus given a layered spin-glass model with a confining local potential $U(S)$
\eq{eq-hamiltonian-layered-SG-general} we find a corresponding feed-forward neural network
with the activation function $f(h)$ in $T \to 0$ limit.

In Fig.~\ref{fig_transfer_matrix_models} we display some examples
of activation functions $f(h)$ and the associated confining potentials $U(S)$.
In sec~\ref{sec-layered-Ising-SG} we showed that the transfer-matrix of layered Ising spin-glass model becomes
equivalent to the feed-forward network with the activation function $f(h)={\rm sgn}(h)$ in $T \to 0$ limit.
The same activation function can be obtained also from the continuous spin model using confining potential
\beq
U(S)=\left \{
\begin{array}{cc}
  0 & (-1 < u < 1)  \\
  \infty &  (u < -1 \qquad\mbox{or} \qquad u > 1)
\end{array}
\right.
\eeq
Similarly, for a piece wise linear function
parametrized by $u > 0$,
\beq
f(h)=\left \{
\begin{array}{cc}
  1 & (h > u)\\
  \frac{h}{u} & (-u < h < u) \\
  -1 & (h < -u)
\end{array}
 \right.
 \eeq
we find
\beq
U(S)=\left \{
\begin{array}{cc}
  \frac{u}{2}\left(S-\frac{h}{u}\right)^{2}+\frac{h^{2}}{u}
& (-1 < u < 1)  \\
  \infty &  (u < -1 \qquad\mbox{or} \qquad u > 1)  
\end{array}
 \right.
 \eeq
Finally in the case of $f(h)=\tanh(h)$ one can easily find $U(S)=S\tanh^{-1}(S)+(1/2)\ln(1-S^{2})$.

      \begin{figure}[t]
    \bc
      \includegraphics[width=0.5\textwidth]{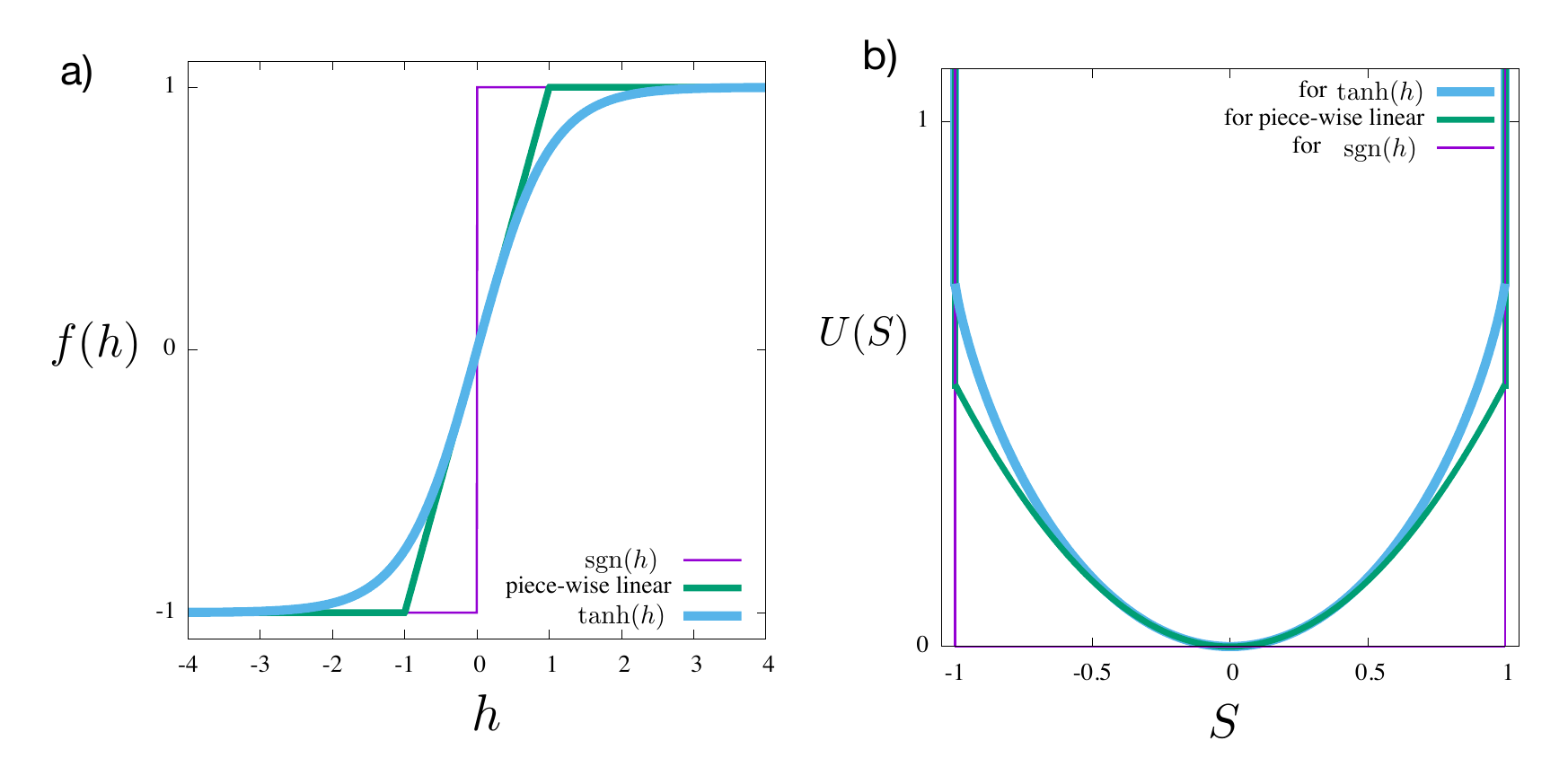}
  \ec
  \caption{Some examples of the activation function $f(h)$
    and the associated confining potential $U(S)$.
Here $u=1$ for the piece-wise linear function.
  }
   \label{fig_transfer_matrix_models}
  \end{figure}

\end{redtext}
\clearpage

\section{Details of the replica theory}
\label{sec-replica-appendix}
%Here we discuss the derivation of the free-energy functional \eq{eq-S-total}.

\subsection{Replicated Gardner volume, Fourier transformation, and Legendre transformation}

By introducing a Fourier representation of the Boltzmann factor, 
\beq
e^{-\beta v(r)}=
\int \frac{d\eta}{\sqrt{2\pi}}
W_{\eta}e^{-i \eta r}.
\eeq
the replicated Gardner volume \eq{eq-replicated-gardner-volume} can be rewritten as,
\beqn
&&V^{n}\left({\bf S}_{0},{\bf S}_{L}\right)
=e^{N M {\cal S}_{n}\left({\bf S}_{0},{\bf S}_{l}\right)} \nonumber \\
& = &\prod_{a}
\left(\prod_{\bs} {\rm Tr}_{{\bf J}^{a}_{\bs}}\right)
\left( \prod_{\bs\backslash {\rm output}}{\rm Tr}_{{\bf S}^{a}_{\bs}}  \right)
\left\{ \prod_{\mu,\bs,a}
%\int \frac{d\eta_{\mu,\bs,a}}{\sqrt{2\pi}}
e^{-\beta v(r_{\bs,a}^{\mu})}
\right \} \nonumber \\
&=& \prod_{\mu,\bs,a}
 \left \{ 
 \int \frac{d\eta_{\mu,\bs,a}}{\sqrt{2\pi}}
 W_{\eta_{\mu,\bs,a}}\right\}
 \tilde{V}^{n}\left({\bf S}_{0},{\bf S}_{L}\right)
\label{eq-replicated-gardner-volume-appendix}
 \eeqn
 where we introduced the Fourier transform of the
replicated Gardner volume
\beqn
 \tilde{V}^{n}\left({\bf S}_{0},{\bf S}_{L}\right) &=&
 \prod_{a}
\left(\prod_{\bs} {\rm Tr}_{{\bf J}^{a}_{\bs}}\right)
\left( \prod_{\bs\backslash {\rm output}}{\rm Tr}_{{\bf S}^{a}_{\bs}}  \right)
\nonumber \\
&& %\left\{
\prod_{\mu,\bs,a}
%\int \frac{d\eta_{\mu,\bs,a}}{\sqrt{2\pi}}
e^{i \eta_{\mu,\bs,a} r_{\bs,a}^{\mu}}
%\right \}
%\label{eq-replicated-gardner-volume}
\label{eq-replicated-gardner-volume-appendix-tilde}
\eeqn
 with the gap variable $r^{\mu}_{\bs,a}$ defined as
\beq
r^{\mu}_{\bs,a} \equiv
(S^{\mu}_{\bs})^{a}
\sum_{k=1}^{c}\frac{(J_{\bs}^{k})^{a}}{\sqrt{c}}(S^{\mu}_{\bs(k)})^{a}
%\label{eq-gap-replica}
\eeq

%basically following sec A of \cite{yoshino2020complex}, the basic strategy is as follows.

Introducing identities
\begin{widetext}
\beqn
1&=&\prod_{a <  b} 
\int_{-\infty}^{\infty}\int_{-i \infty}^{i \infty}
    \left(\frac{c}{2\pi i}\right) dQ_{ab,\bs}d\epsilon_{ab,\bs}
e^{c \sum_{a < b} \epsilon_{ab,\bs}\left(Q_{ab,\bs}-c^{-1}\sum_{k=1}^{c}(J_{\bs}^{k})^{a}(J_{\bs}^{k})^{b}\right)}\nonumber \\
1&=&\prod_{a < b} 
\int_{-\infty}^{\infty}\int_{-i\infty}^{i\infty}
\left(\frac{M}{2\pi i}\right)dq_{ab,\bs}d\varepsilon_{ab,\bs}
e^{ M\sum_{a < b} \varepsilon_{ab,\bs}\left(q_{ab,\bs}-M^{-1}\sum_{\mu=1}^{M}(S_{\bs}^{\mu})^{a}(S_{\bs}^{\mu})^{b}\right)} 
\eeqn
\end{widetext}
we can express the Fourier transformation of the
replicated Gardner volume $\tilde{V}^{n}$ as,
\begin{widetext}
\beqn
 \tilde{V}^{n}\left({\bf S}_{0},{\bf S}_{L}\right)
 &=&
\prod_{a <  b,\bs}
 \left\{
   \int_{-\infty}^{\infty}
   dQ_{ab,\bs}\right \}
\prod_{a <  b,\bs\backslash {\rm output}}
 \left\{
   \int_{-\infty}^{\infty}
   dq_{ab,\bs}\right \} e^{-\beta \tilde{F}_{n}[\hat{Q},\hat{q}]}
%\label{eq-replicated-gardner-volume-appendix}
   \eeqn
where we introduced
\beqn
e^{-\beta \tilde{F}_{n}[\hat{Q},\hat{q}]}&=&
 \prod_{a <  b,\bs}
     \left\{
 \int_{-i \infty}^{i \infty}
 \left(\frac{c}{2\pi i}\right)d\epsilon_{ab,\bs}
 \right \} 
  \prod_{a <  b,\bs\backslash {\rm output}}
     \left\{
 \int_{-i \infty}^{i \infty}
\left(\frac{M}{2\pi i}\right)d\varepsilon_{ab,\bs}
            \right \} \nonumber \\
&&    
                e^{c \sum_{\bs}\sum_{a<b} \epsilon_{ab,\bs}Q_{ab,\bs}+ M \sum_{\bs\backslash {\rm output}}\sum_{a<b} \varepsilon_{ab,\bs}q_{ab,\bs}}
                e^{-\beta \tilde{G}_{n}[\hat{\epsilon},\hat{\varepsilon}]}
                \label{eq-def-F}
            \eeqn
            with
\beqn
-\beta \tilde{G}_{n}[\hat{\epsilon},\hat{\varepsilon}]&=&
-\beta G^{\rm bond}_{n,0}[\hat{\epsilon}]
-\beta G^{\rm spin}_{n,0}[\hat{\varepsilon}]
%\nonumber \\
%&+&
+
\ln \left(
%\prod_{\mu,\bs,a}
% \left \{ 
% \int \frac{d\eta_{\mu,\bs,a}}{\sqrt{2\pi}}
% W_{\eta_{\mu,\bs,a}}\right\}
\left \langle
 \exp \left[ i  \sum_{\mu,\bs,a}\eta_{\mu,\bs,a}
(S^{\mu}_{\bs})^{a}
  \sum_{k=1}^{c}\frac{(J_{\bs}^{k})^{a}}{\sqrt{c}}
  (S^{\mu}_{\bs(k)})^{a}\right] 
\right \rangle_{\epsilon,\varepsilon}
\right)
\label{eq-def-G}
\eeqn
and
\beq
-\beta G^{\rm bond}_{n,0}[\hat{\epsilon}]= \sum_{\bs} \ln
\left(\prod_{a} {\rm Tr}_{{\bf J}^{a}}\right)
e^{-c\sum_{a< b}\epsilon_{ab,\bs}J^{a}J^{b}}
\qquad
-\beta G^{\rm spin}_{n,0}[\hat{\varepsilon}]= 
\sum_{\bs \backslash {\rm output}}
\ln \left( \prod_{a}{\rm Tr}_{{\bf S}^{a}}  \right) e^{-M\sum_{a< b}\varepsilon_{ab,\bs}S^{a}S^{b}}
\label{eq-def-G0}
\eeq
We also introduced,
\beq
\langle \ldots \rangle_{\epsilon,\varepsilon} =
\frac{
 \prod_{\bs\backslash {\rm output}}{\rm Tr}_{{\bf S}_{\bs}^{a}}e^{-\sum_{\mu}\sum_{a< b}\varepsilon_{ab}(S_{\bs}^{\mu})^{a}(S_{\bs}^{\mu})^{b}}
  \prod_{\bs}{\rm Tr}_{{\bf J}{a}}e^{-\sum_{k}\sum_{a< b}\epsilon_{ab}(J^{k}_{\bs})^{a}(J_{\bs}^{k})^{b}}
\ldots
}{
  \prod_{\bs\backslash {\rm output}}{\rm Tr}_{{\bf S}_{\bs}^{a}}e^{-\sum_{\mu}\sum_{a< b}\varepsilon_{ab}(S_{\bs}^{\mu})^{a}(S_{\bs}^{\mu})^{b}}
  \prod_{\bs}{\rm Tr}_{{\bf J}{a}}e^{-\sum_{k}\sum_{a< b}\epsilon_{ab}(J^{k}_{\bs})^{a}(J_{\bs}^{k})^{b}}
}
\label{eq-def-av-free}
\eeq
which represents an averaging using a non-interacting system with polarizing field $\epsilon_{ab}$ and $\varepsilon_{ab}$
conjugated to the order parameters $Q_{ab}$ and $q_{ab}$ \cite{parisi1989mechanism}.

\end{widetext}

Note that \eq{eq-def-F} defines $-\beta \tilde{F}_{n}[\hat{Q},\hat{q}]$ by a Legendre transformation
of $-\beta \tilde{G}_{n}[\hat{\epsilon},\hat{\varepsilon}]$ defined by \eq{eq-def-G}.
The integrations over $\epsilon$ and $\varepsilon$ can be done by the saddle point method for $c \gg 1$ and $M \gg 1$ yielding
\beqn
&& -\beta \tilde{F}_{n}[\hat{Q},\hat{q}]=-\beta \tilde{G}_{n}[\hat{\epsilon}^{*},\hat{\varepsilon}^{*}] \nonumber \\
&+&c\sum_{a < b,\bs}\epsilon_{ab,\bs}^{*}Q_{ab,\bs}+M\sum_{a < b,\bs}\varepsilon_{ab,\bs}^{*}q_{ab,\bs}
\label{eq-G-to-F-replica}
\eeqn
where the saddle points $\epsilon^{*}=\epsilon^{*}[\hat{Q}]$ and $\varepsilon^{*}=\varepsilon^{*}[\hat{q}]$
satisfy
\beqn
Q_{ab,\bs}&=&-\left. \frac{1}{c}\frac{\partial}{\partial \epsilon_{ab,\bs}}(-\beta \tilde{G}_{n}[\hat{\epsilon},\hat{\varepsilon}]) \right |_{\epsilon=\epsilon^{*},\varepsilon=\varepsilon^{*}} \nonumber \\
&=&\frac{1}{c}\sum_{k=1}^{c}\langle (J_{\bs}^{k})^{a}(J_{\bs}^{k})^{b} \rangle_{\epsilon^{*},\varepsilon^{*}}  \nonumber \\
q_{ab,\bs}&=&-\left. \frac{1}{M}\frac{\partial}{\partial \varepsilon_{ab,\bs}}(-\beta \tilde{G}_{n}[\hat{\epsilon},\hat{\varepsilon}]) \right |_{\epsilon=\epsilon^{*},\varepsilon=\varepsilon^{*}} \nonumber \\
&=&\frac{1}{M}\sum_{\mu=1}^{M}\langle (S_{\bs}^{\mu})^{a}(S_{\bs}^{\mu})^{b} \rangle_{\epsilon^{*},\varepsilon^{*}}
\label{eq-q-as-saddlepoint}
\eeqn
The latter implies
\beqn
\langle (J^{k})^{a} (J^{k})^{b} \rangle_{\epsilon} = Q_{ab} \qquad \forall k  \qquad
\langle (S^{\mu})^{a} (S^{\mu})^{b} \rangle_{\varepsilon} = q_{ab} \qquad \forall \mu
\label{eq-epsilon-and-q}
\eeqn
since different components $\mu$'s and $k$'s are equivalent and independent in the averaging \eq{eq-def-av-free}.

Note that $-\beta \tilde{G}_{n}[\hat{\epsilon},\hat{\varepsilon}]$ \eq{eq-def-G} consists of a non-interacting part (entropic term)
$-\beta G^{\rm bond}_{n,0}[\hat{\epsilon}]$ and $-\beta G^{\rm spin}_{n,0}[\hat{\varepsilon}]$ defined in \eq{eq-def-G0} and contribution of interactions which involves
an evaluation using the non-interacting system \eq{eq-def-av-free}. Certainly, the latter is the crucial one.
Our strategy is to analyze the effect of interactions using a combination of the Plefka expansion (sec.~\ref{sec-plefka})
and the cumulant expansion (sec.~\ref{sec-cumulant})

\subsection{Plefka expansion}
\label{sec-plefka}

Suppose that the effect of the interactions can be treated perturbatively which enables the following decompositions \cite{plefka1982convergence},
\beqn
       \tilde{F}_{n}&=&F_{n,0}+\lambda \tilde{F}_{n,1}+\frac{\lambda^{2}}{2} \tilde{F}_{n,2}+\ldots \nonumber \\
\tilde{G}_{n}&=&G^{\rm bond}_{n,0}+G^{\rm spin}_{n,0}+\lambda \tilde{G}_{n,1}+\frac{\lambda^{2}}{2} \tilde{G}_{n,2} +\ldots \nonumber \\
\epsilon_{ab}&=&(\epsilon_{0})_{ab}+\lambda (\epsilon_{1})_{ab}+\frac{\lambda^{2}}{2} (\epsilon_{2})_{ab} \ldots \nonumber \\
 \varepsilon_{ab}&=&(\varepsilon_{0})_{ab}+\lambda (\varepsilon_{1})_{ab}+\frac{\lambda^{2}}{2} (\varepsilon_{2})_{ab} \ldots
\label{eq-plefka}
\eeqn
where we introduced a parameter  $\lambda$ to keep track of the expansion.
Here the quantities with the suffix $0$ represent those that are present in the absence of interactions
and those with suffixes $1,2,\ldots$ represent those due to interactions.

The Legendre transform \eq{eq-G-to-F-replica} becomes, at $O(\lambda^{0})$,
\beqn
&&-\beta F_{n,0}[\hat{Q},\hat{q}]
=-\beta G^{\rm bond}_{n,0}[\hat{\epsilon}_{0}^{*}]-\beta G^{\rm spin}_{n,0}[\hat{\varepsilon}_{0}^{*}] \nonumber \\
&&+c \sum_{a< b,\bs} (\epsilon^{*}_{0})_{ab,\bs}Q_{ab,\bs}
+ M \sum_{a< b,\bs} (\varepsilon^{*}_{0})_{ab,\bs}q_{ab,\bs}
\label{eq-F-0}
\eeqn
where $(\epsilon^{*}_{0})_{ab}$ and $(\varepsilon^{*}_{0})_{ab}$ are defined such that,
\beqn
Q_{ab}&=&-\frac{1}{c}\left. \frac{\partial}{\partial \epsilon_{ab}} (-\beta G^{\rm bond}_{n,0}[\hat{\epsilon}])\right
|_{\hat\epsilon=\hat\epsilon_{0}^{*}[\hat{Q}]} \nonumber \\
q_{ab}&=&-\frac{1}{M}\left. \frac{\partial}{\partial \varepsilon_{ab}} (-\beta G^{\rm spin}_{n,0}[\hat{\varepsilon}])\right
|_{\hat\varepsilon=\hat\varepsilon_{0}^{*}[\hat{q}]}
\label{eq-epsilon-star-0}
\eeqn

Then at $O(\lambda)$ we find,
\begin{widetext}
\beqn
&& -\beta \tilde{F}_{n,1}[\hat{Q},\hat{q}]=-\beta \tilde{G}_{n,1}[\hat\epsilon_{0}^{*}[\hat{Q}],\hat\varepsilon_{0}^{*}[\hat{q}]] 
 +\sum_{a< b,\bs} \left. \frac{\partial(-\beta  G^{\rm bond}_{n,0}[\hat\epsilon])}{\partial \epsilon_{ab,\bs}}
\right|_{\hat\epsilon=\hat\epsilon^{*}_{0}[\hat{Q}]}(\epsilon^{*}_{1})_{ab,\bs} 
+c \sum_{a< b,bs} (\epsilon^{*}_{1})_{ab,\bs}Q_{ab,\bs} \nonumber \\
&+&\sum_{a< b,\bs} \left. \frac{\partial (-\beta  G^{\rm spin}_{n,0}[\hat\varepsilon])}{\partial \varepsilon_{ab,\bs}}
\right|_{\hat\varepsilon=\hat\varepsilon^{*}_{0}[\hat{q}]}(\varepsilon^{*}_{1})_{ab,\bs}
+M \sum_{a< b} (\varepsilon^{*}_{1})_{ab,\bs}q_{ab,\bs} 
 =-\beta \tilde{G}_{n,1}[\hat\epsilon_{0}^{*}[\hat{Q}],\hat\varepsilon_{0}^{*}[\hat{q}]] 
\label{eq-G-1}
\eeqn
\end{widetext}
In the 2nd equation, we used \eq{eq-epsilon-star-0}.

Similarly at $O(\lambda^{2})$ we find,
\begin{widetext}
\beqn
&& -\beta \tilde{F}_{n,2}[\hat{Q},\hat{q}]
=-\beta \tilde{G}_{n,2}[\epsilon_{0}^{*},\varepsilon_{0}^{*}] \nonumber \\
&&+2\sum_{a<b,\bs}\left.\frac{\partial (-\beta \tilde{G}_{n,1}[\epsilon,\varepsilon])}{\partial \epsilon_{ab,\bs}} \right |_{\epsilon=\epsilon_{0}^{*},\varepsilon=\varepsilon^{*}_{0}}(\epsilon^{*}_{1})_{ab,\bs}+2\sum_{a<b,\bs}\left.\frac{\partial\left(-\beta \tilde{G}_{n,1}[\epsilon,\varepsilon]\right)}{\partial \varepsilon_{ab,\bs}} \right |_{\epsilon=\epsilon_{0}^{*},\varepsilon=\varepsilon^{*}}(\varepsilon^{*}_{1})_{ab,\bs} \nonumber \\
&&+\sum_{a<b,\bs}\left. \frac{\partial \left(-\beta G^{\rm bond}_{n,0}[\epsilon]\right)}{\partial \epsilon_{ab,\bs}}\right |_{\epsilon=\epsilon_{0}^{*}}(\epsilon^{*}_{2})_{ab,\bs}+\sum_{a<b,\bs}\left.\frac{\partial\left(-\beta G^{\rm spin}_{n,0}[\varepsilon]\right)}{\partial \varepsilon_{ab,\bs}} \right |_{\varepsilon=\varepsilon^{*}_{0}}(\varepsilon^{*}_{2})_{ab,\bs} \nonumber \\
&&
+\sum_{\bs}\sum_{a<b,\bs}\sum_{c<d}
\left.
\frac{\partial^{2}(-\beta G^{\rm bond}_{n,0})[\epsilon,\varepsilon]}{\partial \epsilon_{ab,\bs}\epsilon_{cd,\bs}}
\right|_{\epsilon=\epsilon_{0}^{*}}
(\epsilon_{1}^{*})_{ab,\bs}(\epsilon_{1}^{*})_{cd,\bs}
+\sum_{\bs}\sum_{a<b}\sum_{c<d}
\left.
\frac{\partial^{2}(-\beta G^{\rm spin}_{n,0})[\epsilon,\varepsilon]}{\partial \varepsilon_{ab,\bs}\varepsilon_{cd,\bs}}
\right|_{\varepsilon=\varepsilon^{*}_{0}}
(\varepsilon_{1}^{*})_{ab,\bs}(\varepsilon_{1}^{*})_{cd,\bs}
\nonumber \\
&&  +c \sum_{a< b,\bs} (\epsilon_{2}^{*})_{ab,\bs} Q_{ab,\bs} +M \sum_{a< b,\bs} (\varepsilon_{2}^{*})_{ab,\bs} q_{ab,\bs} \nonumber \\
%&=&-\beta G_{n,2}[\epsilon_{0}^{*},\varepsilon_{0}^{*}] \nonumber \\
%&+&2\sum_{a<b}\left.\frac{\partial (-\beta G_{n,1}[\epsilon,\varepsilon])}{\partial \epsilon_{ab}} \right |_{\epsilon=\epsilon_{0}^{*},\varepsilon=\varepsilon^{*}_{0}}(\epsilon^{*}_{1})_{ab}+2\sum_{a<b}\left.\frac{\partial\left(-\beta G_{n,1}[\epsilon,\varepsilon]\right)}{\partial \varepsilon_{ab}} \right |_{\epsilon=\epsilon_{0}^{*},\varepsilon=\varepsilon^{*}}(\varepsilon^{*}_{1})_{ab} \nonumber \\
%&&
%+\sum_{a<b}\sum_{c<d}
%\left.
%\frac{\partial^{2}(-\beta G^{\rm bond}_{n,0})[\epsilon,\varepsilon]}{\partial \epsilon_{ab}\epsilon_{cd}}
%\right|_{\epsilon=\epsilon_{0}^{*}}
%(\epsilon_{1}^{*})_{ab}(\epsilon_{1}^{*})_{cd}
%+\sum_{a<b}\sum_{c<d}
%\left.
%\frac{\partial^{2}(-\beta G^{\rm spin}_{n,0})[\epsilon,\varepsilon]}{\partial \varepsilon_{ab}\varepsilon_{cd}}
%\right|_{\varepsilon=\varepsilon_{0}^{*}}
%(\varepsilon_{1}^{*})_{ab}(\varepsilon_{1}^{*})_{cd} \nonumber \\
%&=&-\beta G_{n,2}[\epsilon_{0}^{*},\varepsilon_{0}^{*}] \nonumber \\
%%&+&\sum_{a<b}\left.\frac{\partial (-\beta G_{n,1}[\epsilon,\varepsilon])}{\partial \epsilon_{ab}} \right |_{\epsilon=\epsilon_{0}^{*},\varepsilon=\varepsilon^{*}_{0}}(\epsilon^{*}_{1})_{ab}+\sum_{a<b}\left.\frac{\partial\left(-\beta G_{n,1}[\epsilon,\varepsilon]\right)}{\partial \varepsilon_{ab}} \right |_{\epsilon=\epsilon_{0}^{*},\varepsilon=\varepsilon^{*}}(\varepsilon^{*}_{1})_{ab} \nonumber \\
&=&-\beta \tilde{G}_{n,2}[\epsilon_{0}^{*},\varepsilon_{0}^{*}] \nonumber \\
&-& \sum_{\bs}\sum_{a< b}\sum_{c<d}
\frac{\partial (-\beta \tilde{G}_{n,1})[\hat{\epsilon},\hat{\varepsilon}]}{\partial \epsilon_{ab,\bs}}
\left(
\frac{\partial^{2}(-\beta G^{\rm bond}_{n,0}[\hat{\epsilon}])}{\partial \epsilon_{ab,\bs}\partial \epsilon_{cd,\bs}}
\right)^{-1}
\frac{\partial (-\beta \tilde{G}_{n,1})[\hat{\epsilon},\hat{\varepsilon}]}{\partial \epsilon_{cd,\bs}}
\nonumber \\
 &&
-\sum_{\bs}\sum_{a< b}\sum_{c<d}
\frac{\partial (-\beta \tilde{G}_{n,1})[\hat{\varepsilon},\hat{\varepsilon}]}{\partial \varepsilon_{ab,\bs}}
\left(
\frac{\partial^{2}(-\beta G^{\rm spin}_{n,0}[\hat{\varepsilon}])}{\partial \varepsilon_{ab,\bs}\partial \varepsilon_{cd,\bs}}
\right)^{-1}
\frac{\partial (- \beta \tilde{G}_{n,1})[\hat{\varepsilon},\hat{\varepsilon}]}{\partial \varepsilon_{cd,\bs}} \nonumber \\
\label{eq-G2}
\eeqn
To derive the last line we used \eq{eq-epsilon-star-0} and
\beqn
0=
&&   \left. \frac{\partial (-\beta \tilde{G}_{n,1}[\hat{\epsilon},\hat{\varepsilon}])}{\partial \epsilon_{ab,\bs}} \right|_{\epsilon=\epsilon_{0}^{*},\varepsilon=\varepsilon_{0}^{*}}
+\left. \sum_{c< d}\frac{\partial^{2} (-\beta G^{\rm bond}_{n,0}[\hat{\epsilon}])}{\partial \epsilon_{ab,\bs}\partial \epsilon_{cd,\bs}} \right|_{\epsilon=\epsilon_{0}^{*}}(\epsilon_{1}^{*})_{cd,\bs} \nonumber \\
0 &=& 
  \left. \frac{\partial (-\beta \tilde{G}_{n,1}[\hat{\epsilon},\hat{\varepsilon}])}{\partial \epsilon_{ab,\bs}} \right|_{\epsilon=\epsilon_{0}^{*},\varepsilon=\varepsilon_{0}^{*}}
+\left. \sum_{c< d}\frac{\partial^{2} (-\beta G^{\rm spin}_{n,0}[\hat{\varepsilon}])}{\partial \varepsilon_{ab,\bs}\partial \varepsilon_{cd,\bs}} \right|_{\varepsilon=\varepsilon_{0}^{*}} 
(\varepsilon_{1}^{*})_{cd,\bs}
\eeqn
\end{widetext}
which is obtained by expanding \eq{eq-q-as-saddlepoint} up to $O(\lambda)$ and then using \eq{eq-epsilon-star-0} for the 0-th order terms.
%\beqn
%-c Q_{ab} &=& \left. \frac{\partial \left(-\beta G_{n}[\epsilon,\varepsilon]\right)}{\partial \epsilon_{ab}} \right|_{\epsilon=\epsilon^{*},\varepsilon=\varepsilon^{*}}
%=\left. \frac{\partial (-\beta G^{\rm bond}_{n,0}[\epsilon])}{\partial \epsilon_{ab}}  \right|_{\epsilon=\epsilon^{*}_{0}} \nonumber \\
%&&  +\lambda \left( \left. \frac{\partial (-\beta G_{n,1}[\hat{\epsilon},\hat{\varepsilon}])}{\partial \epsilon_{ab}} \right|_{\epsilon=\epsilon_{0}^{*},\varepsilon=\varepsilon_{0}^{*}}
%+\left. \sum_{c< d}\frac{\partial^{2} (-\beta G^{\rm bond}_{n,0}[\hat{\epsilon}])}{\partial \epsilon_{ab}\partial \epsilon_{cd}} \right|_{\epsilon=\epsilon_{0}^{*}}(\epsilon_{1}^{*})_{cd} 
%\right)+O(\lambda)^{2} \nonumber \\
%-M q_{ab} &=& \left. \frac{\partial \left(-\beta G_{n}[\epsilon,\varepsilon]\right)}{\partial \varepsilon_{ab}} \right|_{\epsilon=\epsilon^{*},\varepsilon=\varepsilon^{*}}
%=\left. \frac{\partial (-\beta G^{\rm spin}_{n,0}[\varepsilon])}{\partial \varepsilon_{ab}}  \right|_{\varepsilon=\varepsilon^{*}_{0}} \nonumber \\
%&&  +\lambda \left( \left. \frac{\partial (-\beta G_{n,1}[\hat{\epsilon},\hat{\varepsilon}])}{\partial \epsilon_{ab}} \right|_{\epsilon=\epsilon_{0}^{*},\varepsilon=\varepsilon_{0}^{*}}
%+\left. \sum_{c< d}\frac{\partial^{2} (-\beta G^{\rm spin}_{n,0}[\hat{\varepsilon}])}{\partial \varepsilon_{ab}\partial \varepsilon_{cd}} \right|_{\varepsilon=\varepsilon_{0}^{*}} 
%(\varepsilon_{1}^{*})_{cd} 
%  \right)+O(\lambda)^{2}
%\eeqn
%Then using \eq{eq-epsilon-star-0} in the last equation we obtain \eq{eq-h-1}.
%

If $O(\lambda)^{2}$ terms and higher order terms vanish (as happens in the dense coupling), we can put $\lambda=1$ and obtain,
\begin{widetext}
\beqn
\tilde{F}_{n}[\hat{Q},\hat{q}]&=&-\beta F_{n,0}[\hat{Q},\hat{q}]-\beta \tilde{F}_{n,1}[\hat{Q},\hat{q}] \nonumber\\
&=&-\beta G_{n,0}[\hat{\epsilon}^{*},\hat{\varepsilon}^{*}]+c \sum_{a<b,\bs}\epsilon^{*}_{ab,\bs}Q_{ab,\bs}
+M \sum_{a<b,\bs}\varepsilon^{*}_{ab,\bs}q_{ab,\bs} -\beta \tilde{G}_{n,1}[\hat{\epsilon}^{*}]
\eeqn
\end{widetext}
where $\hat{\epsilon}^{*}=\hat{\epsilon}^{*}_{0}[\hat{q}]$ and
$\hat{\varepsilon}^{*}=\hat{\varepsilon}^{*}_{0}[\hat{Q}]$
are those determined by \eq{eq-epsilon-star-0}.
%We will refer to $F_{n,0}[\hat{Q},\hat{q}]$
%as the entropic part of the free-energy and $F_{n,1}[\hat{Q},\hat{q}]$ as the interaction part of the free-energy.

\subsection{Summary 1}
\label{sec-replicated-fee-energy-functional}

Here we can wrap up the above results to find the replicated Gardner volume
\eq{eq-replicated-gardner-volume-appendix} expressed as,
\begin{widetext}
        \beqn
        V^{n}\left({\bf S}_{0},{\bf S}_{L}\right)
&=&e^{N M {\cal S}_{n}\left({\bf S}_{0},{\bf S}_{l}\right)} \nonumber \\
&=& \prod_{a <  b,\bs}
 \left\{
   \int_{-\infty}^{\infty}
   dQ_{ab,\bs}\right \}
\prod_{a <  b,\bs\backslash {\rm output}}
 \left\{
   \int_{-\infty}^{\infty}
   dq_{ab,\bs}\right \} 
   e^{-\beta F_{n}[\hat{Q},\hat{q}]}
   \label{eq-wrapup-1}
   \eeqn
   \end{widetext}
The functional $-\beta F_{n}[\hat{Q},\hat{q}]$ may be regarded as replicated the free-energy functional
 \beq
-\beta F_{n}[\hat{Q},\hat{q}]
= -\beta F_{0}[\hat{Q},\hat{q}] -\beta F_{\rm ex}[\hat{Q},\hat{q}]
   \label{eq-wrapup-2}
  \eeq
where $-\beta F_{0}[\hat{Q},\hat{q}]$ given by \eq{eq-F-0}
may be regarded as the entropic part of the free-energy while
$-\beta  F_{\rm ex}$ is the interaction part of the free-energy,
\begin{widetext}
 \beqn
 e^{-\beta  F_{\rm ex}[\hat{Q},\hat{q}]} &=&
   \prod_{\mu,\bs,a}
 \left \{ 
 \int \frac{d\eta_{\mu,\bs,a}}{\sqrt{2\pi}}
 W_{\eta_{\mu,\bs,a}}\right\}
 e^{-\beta\tilde{F}_{\rm ex}[\hat{Q},\hat{q}; \{i\eta_{\mu,\bs,a}\}]
 }\nonumber \\
 &=& e^{-\beta \tilde{F}_{\rm ex}[\hat{Q},\hat{q},\{\partial/\partial_{\mu,\bs,a}\}]}
 \left. \prod_{\mu,\bs, a}e^{-\beta v(h_{\mu,\bs,a})} \right|_{\{h_{\mu,bs,a\}=0}}
 \label{eq-Fex}
 \eeqn
 \end{widetext}
 with 
 \beq
\tilde{F}_{\rm ex}=\tilde{F}_{n,1}+ \tilde{F}_{n,2}+\ldots
\eeq
In the 1st equation of \eq{eq-Fex}
 we recalled that $\tilde{F}_{\rm ex}[\hat{Q},\hat{q}; \{i\eta_{\mu,\bs,a}\}]$ depend on  $\{i\eta_{\mu,\bs,a}\}$.
 In the 2nd equation of \eq{eq-Fex},
 $\tilde{F}_{\rm ex}[\hat{Q},\hat{q}; \{i\eta_{\mu,\bs,a}\}]$ is a differential operator.

\subsection{Cumulant expansion}
\label{sec-cumulant}

Now we turn to the explicit evaluation of the
 $-\beta \tilde{G}_{n}[\hat{\epsilon},\hat{\varepsilon}]$ defined in \eq{eq-def-G}
by a cumulant expansion, introducing the parameter $\lambda$,
\begin{widetext}
\beqn
-\beta \tilde{G}_{n}[\hat{\epsilon},\hat{\varepsilon}] &=& \ln \left \langle
 \exp \left[ i  \sum_{\mu,\bs,a}\eta_{\mu,\bs,a}
(S^{\mu}_{\bs})^{a}
  \sum_{k=1}^{c}\frac{\sqrt{\lambda}}{\sqrt{c}}(J_{\bs}^{k})^{a}
  (S^{\mu}_{\bs(k)})^{a}\right] 
 \right \rangle_{\epsilon,\varepsilon}  \nonumber \\
 && = \ln \left \langle
 1+  \sum_{\mu,\bs,a} i\eta_{\mu,\bs,a}
(S^{\mu}_{\bs})^{a}
  \sum_{k=1}^{c}\frac{\sqrt{\lambda}}{\sqrt{c}}(J_{\bs}^{k})^{a}
  (S^{\mu}_{\bs(k)})^{a} \right. \nonumber \\
&&  \left. +\frac{1}{2!}
  \sum_{\mu,\bs,a} \sum_{\nu,\ws,b} i\eta_{\mu,\bs,a}i\eta_{\nu,\ws,b}
(S^{\mu}_{\bs})^{a}(S^{\nu}_{\ws})^{b}
  \sum_{k=1}^{c}\frac{\sqrt{\lambda}}{\sqrt{c}}(J_{\bs}^{k})^{a} (S^{\mu}_{\bs(k)})^{a}
  \sum_{k'=1}^{c}\frac{\sqrt{\lambda}}{\sqrt{c}}(J_{\ws}^{k'})^{b} (S^{\nu}_{\ws(k')})^{b}
  + \ldots
\right \rangle_{\epsilon,\varepsilon}
\eeqn
\end{widetext}
From \eq{eq-def-av-free} we find averages
$\langle \ldots \rangle_{\epsilon,\varepsilon}$
of terms with odd numbers of spins $(S_{\bs}^{\mu})^{a}$
and bonds $(J_{\bs}^{k})^{a}$ vanish by symmetry.
Consequently we find non-vanishing terms at order $O(\lambda)$, $O(\lambda^{2})$,\ldots
corresponding to the 2nd, 4th order terms of the cumulant expansion
which are represented by connected diagrams.
They define $-\beta \tilde{G}_{n,1}$,$-\beta \tilde{G}_{n,2}$ ... in the Plefka expansion \eq{eq-plefka})
of  $-\beta \tilde{G}_{n}$. 

%\subsection{Entropic part of the free-energy}

%The entropic part of the free-energy can be derived as shown in \cite{yoshino2020complex}.

%\subsection{Interaction part of the freee-energy}

\subsubsection{$O(\lambda)$ term}

We find the 2nd order cumulant yields the $O(\lambda)$, i.~e. $-\beta \hat{G}_{n,1}$.
Then by \eq{eq-G-1} we find this is also $-\beta \tilde{F}_{n,1}$,
\begin{widetext}
\beqn
-\beta \tilde{F}_{n,1}[\hat{Q},\hat{q}] 
&=& -\beta \tilde{G}_{n,1}[\hat{\epsilon}^{*}[\hat{Q}],\hat{\varepsilon}^{*}[\hat{q}]] \nonumber \\
&=&\left \langle \frac{1}{2!}
  \sum_{\mu,\bs,a} \sum_{\nu,\ws,b} i\eta_{\mu,\bs,a}i\eta_{\nu,\ws,b}
(S^{\mu}_{\bs})^{a}(S^{\nu}_{\ws})^{b}
  \sum_{k=1}^{c}\frac{\sqrt{\lambda}}{\sqrt{c}}(J_{\bs}^{k})^{a} (S^{\mu}_{\bs(k)})^{a}
  \sum_{k'=1}^{c}\frac{\sqrt{\lambda}}{\sqrt{c}}(J_{\ws}^{k'})^{b} (S^{\nu}_{\ws(k')})^{b}\right \rangle_{\epsilon^{*}[\hat{Q}],\varepsilon^{*}[\hat{q}]}
  \nonumber \\
  &&  =\frac{\lambda}{2}
  \sum_{\mu,\bs}\sum_{a,b}  i\eta_{\mu,\bs,a}i\eta_{\mu,\bs,b}q_{ab,\bs}Q_{ab,\bs}\frac{1}{c}\sum_{k=1}^{c}
  q_{ab,\bs(k)}
  \label{eq-F1}
  \eeqn
  \end{widetext}
  where we have used \eq{eq-epsilon-and-q}.
Anticipating the homogeneous solution with each layer \eq{eq-order-parameters-at-each-layers}
we find $-\beta \hat{F}_{n,1}/(NM) \sim O(1)$. This term will become the dominant term
that contributes to the interaction part of the free-energy $-\beta F_{\rm ex}$ in the dense limit $N \gg c \gg 1$.

\begin{figure}[h]
    \bc
  \includegraphics[width=0.2\textwidth]{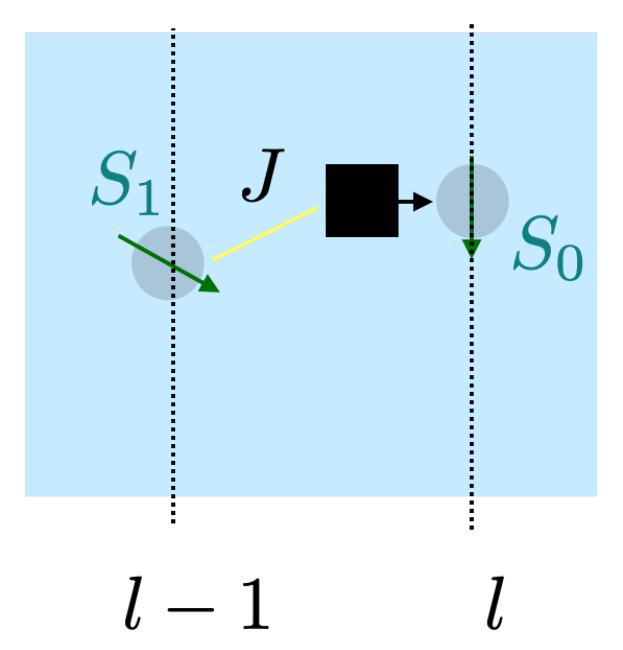}
  \ec
  \caption{Graphical representation of a contribution to $\tilde{G}_{1}$ (and $\tilde{F}_{1}$)
}
   \label{fig_G1}
        \end{figure}

In Fig.~\ref{fig_G1} we show a graphical representation of the term.
$\tilde{G}_{1}$ (and $\tilde{F}_{1}$) is obtained by associating 2 replicas to the diagram.

\subsubsection{$O(\lambda^{2})$ terms}

At the 4th order of the cumulant expansion we easily find a $O(\lambda^{2})$ term
which contributes to $-\beta G_{n,2}$ and thus  $-\beta F_{n,2}$ via \eq{eq-G2}
by associating 4 replicas to the same diagram shown in Fig.~\ref{fig_G1},
\begin{widetext}
\beqn
&&
-\beta \tilde{F}_{n,2}[{\rm Fig.} \ref{fig_G1}]= -\beta \tilde{G}_{n,2}[{\rm Fig.} \ref{fig_G1}]=
\frac{1}{c}\frac{\lambda^{2}}{4!}
\sum_{\mu,\bs}\sum_{a,b,c,d}  i\eta_{\mu,\bs,a}i\eta_{\mu,\bs,b}i\eta_{\mu,\bs,c}i\eta_{\mu,\bs,d} \nonumber \\
&& \frac{1}{c}\sum_{k=1}^{c}
\left[  \langle
  (S_{\bs}^{\mu})^{a}  (S_{\bs}^{\mu})^{b}  (S_{\bs}^{\mu})^{c}  (S_{\bs}^{\mu})^{d}
  (J_{\bs}^{k})^{a} (J_{\bs}^{k})^{b} (J_{\bs}^{k})^{c} (J_{\bs}^{k})^{d}
  (S_{\bs(k)}^{\mu})^{a}  (S_{\bs(k)}^{\mu})^{b}  (S_{\bs(k)}^{\mu})^{c}  (S_{\bs(k)}^{\mu})^{d}\rangle_{\hat{\epsilon},\hat{\varepsilon}} \right. \nonumber \\
&&   -q_{ab,\bs}Q_{ab,\bs}q_{ab,\bs(k)}q_{cd,\bs}Q_{cd,\bs}q_{cd,\bs(k)} \nonumber \\
&&  -q_{ac,\bs}Q_{ac,\bs}q_{ac,\bs(k)}q_{bd,\bs}Q_{bd,\bs}q_{bd,\bs(k)} \nonumber \\
&&  \left.  -q_{ad,\bs}Q_{ad,\bs}q_{ad,\bs(k)}q_{bc,\bs}Q_{bc,\bs}q_{bc,\bs(k)}
 \right]
\eeqn
\end{widetext}
We see that $-\beta \tilde{F}_{n,2}[{\rm Fig.} \ref{fig_G1}]/(NM)  \propto 1/c$
and vanishes in $c \to \infty$ limit.

        \begin{figure}[h]
    \bc
  \includegraphics[width=0.2\textwidth]{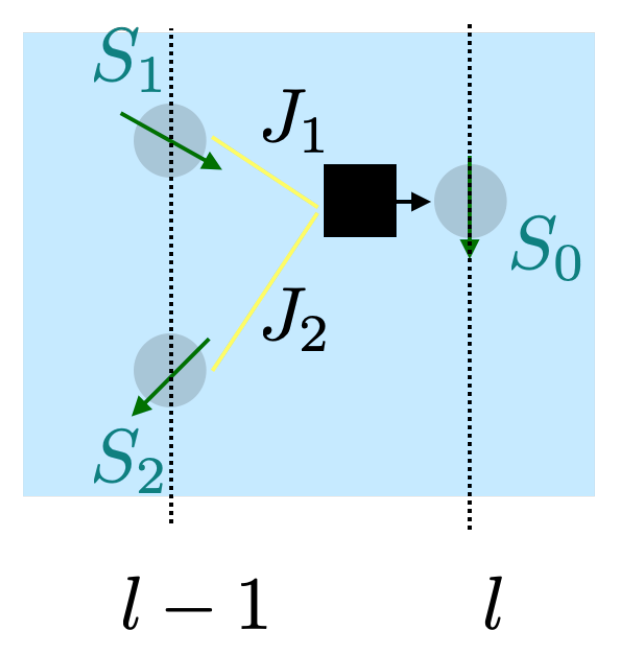}
  \ec
  \caption{
    A contribution to $G_{2}$ which is one-line reducible.
%       We neglect the effects
%    of such loops (and more extended ones) in our theory.
  }
   \label{fig_G2}
        \end{figure}

There is another contribution to $\tilde{G}_{2}$ which is associated with a diagram shown in Fig.~\ref{fig_G2}.
We associate two replicas a,b to branch '1' and replicas c,d to branch '2',
\begin{widetext}
        \beqn
&&    -\beta  \tilde{G}_{n,2}[{\rm Fig.}\ref{fig_G2}] \sim  \sum_{a< b}\sum_{c<d} \left[ \langle (S_{0})_{a}(J_{1})_{a}(S_{1})_{a}(S_{0})_{b}(J_{1})_{b}(S_{1})_{b}(S_{0})_{c}(J_{2})_{c}(S_{2})_{c}(S_{0})_{d}(J_{2})_{d}(S_{2})_{d}
        \rangle_{\epsilon,\varepsilon}- \nonumber \right. \\
&&    \hspace*{5cm} \left. \left\langle (S_{0})_{a}(J_{1})_{a}(S_{1})_{a}(S_{0})_{b}(J_{1})_{b}(S_{1})_{b}\right\rangle_{\epsilon,\varepsilon}
        \left\langle(S_{0})_{c}(J_{2})_{c}(S_{2})_{c}(S_{0})_{d}(J_{2})_{d}(S_{2})_{d}\right\rangle_{\epsilon,\varepsilon} \right]\nonumber \\
        &&
        =  \sum_{a< b}\sum_{c<d}
        %\left[
%          \left\langle (S_{0})_{a}(S_{0})_{b}(S_{0})_{c}(S_{0})_{d}\right\rangle_{\epsilon,\varepsilon}
          %        - \left\langle (S_{0})_{a}(S_{0})_{b}\right\rangle_{\epsilon,\varepsilon}         \left\langle(S_{0})_{c}(S_{0})_{d}\right\rangle_{\epsilon,\varepsilon}
          \left\langle (S_{0})_{a}(S_{0})_{b}(S_{0})_{c}(S_{0})_{d}\right\rangle^{c}_{\epsilon,\varepsilon}
%        \right]
        Q_{ab,\bs}q_{ab,\bs}Q_{cd,\bs}q_{cd,\bs}
        \eeqn
        \end{widetext}
        where $\langle S^{a}S^{b}S^{c}S^{d}\rangle^{c}$'s are connected correlation functions defined as
        \beqn
&&        \langle S^{a}S^{b}S^{c}S^{d}\rangle^{c}=\langle S^{a}S^{b}S^{c}S^{d}\rangle-\langle S^{a}S^{b}\rangle\langle S^{c}S^{d}\rangle \nonumber \\
  &&      =\langle S^{a}S^{b}S^{c}S^{d}\rangle-q_{ab}q_{cd}
        \eeqn
        Note that it involves 4 perceptrons associated with the 4 replicas so that we have a factor
        $(1/\sqrt{c})^{4}$ but there are $c(c-1)$ different ways to choose the endpoints of branch '1' and '2'.
        So that contribution by this type of term survives in $c \to \infty$ limit as $O(1)$ contribution
        to $-\beta G_{n,2}/(NM)$.

        However this does not contribute to $-\beta F_{n,2}$ because it is exactly canceled by 
        the 2nd term in \eq{eq-G2}. To see this let us recall that $G_{n,1}$ is like,
        \beq
        \tilde{G}_{n,1} \sim \sum_{a < b}\langle (S_{0})_{a}(J_{1})_{a}(S_{1})_{a}(S_{0})_{a}(J_{1})_{b}(S_{1})_{b}\rangle
        \eeq
        then we find
        \begin{widetext}
        \beqn
&&        -\sum_{a< b}\sum_{c<d}
\frac{\partial \tilde{G}_{n,1}[\hat{\varepsilon},\hat{\varepsilon}]}{\partial \varepsilon_{ab,\bs}}
\left(
\frac{\partial^{2}(-\beta G^{\rm spin}_{n,0}[\hat{\varepsilon}])}{\partial \varepsilon_{ab,\bs}\partial \varepsilon_{cd,\bs}}
\right)^{-1}
\frac{\partial \tilde{G}_{n,1}[\hat{\varepsilon},\hat{\varepsilon}]}{\partial \varepsilon_{cd,\bs}}\nonumber \\
&&
 \sim -\sum_{a< b}\sum_{c<d}
\frac{\partial }{\partial \varepsilon_{ab,\bs}} (\sum_{e< f} \langle (S_{0})_{e}(J_{1})_{e})(S_{1})_{e}(S_{0})_{f}(J_{1})_{f})(S_{1})_{f} \rangle
\left(
\frac{\partial^{2}(-\beta G^{\rm spin}_{n,0}[\hat{\varepsilon}])}{\partial \varepsilon_{ab,\bs}\partial \varepsilon_{cd,\bs}}
\right)^{-1}
\frac{\partial }{\partial \varepsilon_{cd,\bs}} (\sum_{g< h} \langle (S_{0})_{g}(J_{1})_{g})(S_{1})_{g}(S_{0})_{h}(J_{1})_{h})(S_{1})_{h} \rangle
\nonumber \\
&& =
-\sum_{e< f}\sum_{g<h}
\langle (J_{1})_{e}(S_{1})_{e}(J_{1})_{f})(S_{1})_{f} \rangle  \langle (J_{1})_{g}(S_{1})_{g}(J_{1})_{h})(S_{1})_{h} \rangle
\sum_{a< b}\sum_{c< d} \frac{\partial q_{ef,\bs}}{\partial \varepsilon_{ab,\bs}}
\left(
\frac{\partial^{2}(-\beta G^{\rm spin}_{n,0}[\hat{\varepsilon}])}{\partial \varepsilon_{ab,\bs}\partial \varepsilon_{cd,\bs}}
\right)^{-1}
\sum_{g< h}\frac{\partial q_{gh,\bs}}{\partial \varepsilon_{cd,\bs}} \nonumber \\
&&= -\sum_{e< f}\sum_{g<h}
Q_{ef,\bs}q_{ef,\bs}Q_{gh,\bs}q_{gh,\bs}
\left(
\frac{\partial^{2}(-\beta G^{\rm spin}_{n,0}[\hat{\varepsilon}])}{\partial \varepsilon_{ef,\bs}\partial \varepsilon_{gh,\bs}}
\right)
\nonumber \\
&&
=-\sum_{e< f}\sum_{g<h}Q_{ef,\bs}q_{ef,\bs}Q_{gh,\bs}q_{gh,\bs}
%\left[
  \left\langle (S_{0})_{e}(S_{0})_{f}(S_{0})_{g}(S_{0})_{h}\right\rangle^{c}_{\epsilon,\varepsilon}
%  - \left\langle (S_{0})_{e}(S_{0})_{f}\right\rangle_{\epsilon,\varepsilon}         \left\langle(S_{0})_{g}S_{0})_{h}\right\rangle_{\epsilon,\varepsilon}
  %\right]
  \eeqn
  \end{widetext}
This exactly cancels $-\beta \tilde{G}_{n,2}[{\rm Fig.} \ref{fig_G2}]$. Thus the diagram shown in Fig.~\ref{fig_G2} do not contribute $-\beta \tilde{F}_{n,2}$.

Indeed it is known in diagrammatic expansions that 'one-line (or particle) reducible' diagrams like
the one shown in Fig.~\ref{fig_G2} become canceled after Legendre transform from $-\beta \tilde{G}$ to $-\beta \tilde{F}$
\cite{hansen1990theory,zinn2021quantum} leaving only {\it loop diagrams} which are
     {\it one-line irreducible}, i.~e. diagrams which cannot be separated into two disconnected
diagrams by cutting a line. At $O(\lambda^{2})$  we do not have such a loop diagram.

To sum up we find
\beq
-\beta \tilde{F}_{n,2}/(NM) = -\beta \tilde{F}_{n,2} [{\rm Fig.} \ref{fig_G1}]/(NM)  \propto 1/c 
\eeq
which vanishes in the dense limit $c\to \infty$.
        
\subsubsection{$O(\lambda^{3})$ terms}

At $O(\lambda^{3})$ we will have a term that is obtained by associating 6 replicas to the diagram
Fig.~\ref{fig_G1} whose contribution to $-\beta F_{n,3}/(NM)$
vanishes  as $1/c^{2}$ in $c \to \infty$ limit.

        \begin{figure}[h]
    \bc
  \includegraphics[width=0.3\textwidth]{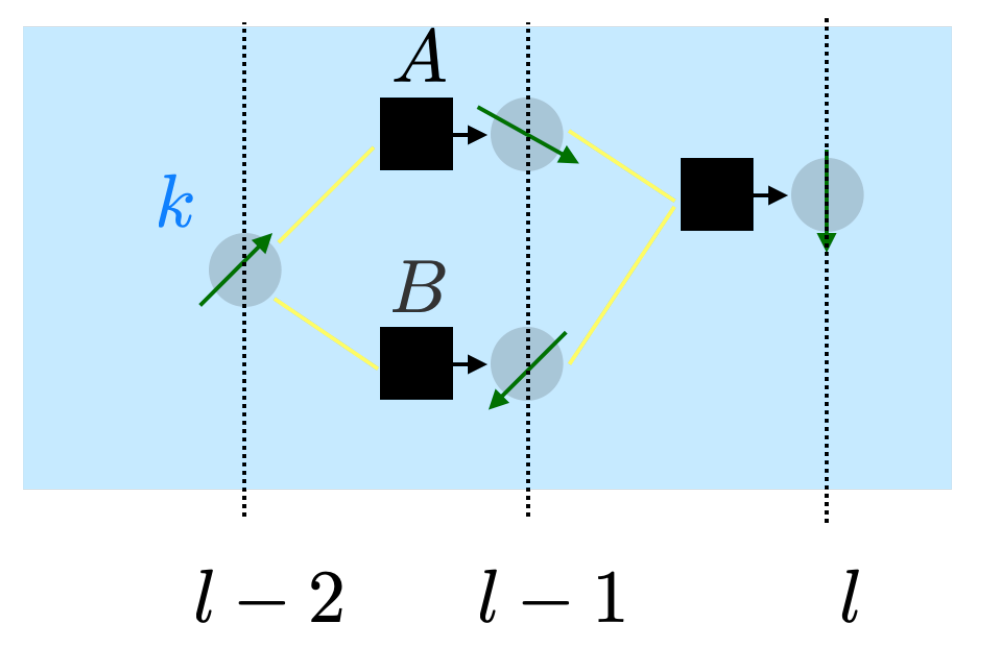}
  \ec
  \caption{
    A loop of interactions in a DNN
      extended over 3 layers, through 3 perceptrons
      and 4 bonds.
%       We neglect the effects
%    of such loops (and more extended ones) in our theory.
  }
   \label{fig_loop_appendix}
        \end{figure}

Apart from that we find contributions of one-loop diagrams. 
As the simplest example, consider the loop shown in Fig.~\ref{fig_loop_appendix}(same one as shown in Fig.~\ref{fig_loop}).
Such a loop contribute to the form,
\begin{widetext}
\beqn
 -\beta \tilde{F}_{n,3}[{\rm Fig.}\ref{fig_loop_appendix}]
=\frac{\lambda^{3}}{6!}
\left(\frac{1}{\sqrt{c}}\right)^{6}  \sum_{\bs,\mu}
\sum_{a,b,c,d,e,f}  
\left(\sum_{\bs_{A},\bs_{B},k} \right)_{\rm loop}
i\eta_{\mu,\bs,a}i\eta_{\mu,\bs,b}
       i\eta_{\mu,\bs_{A},c}i\eta_{\mu,\bs_{A},d}
       i\eta_{\mu,\bs_{B},e}i\eta_{\mu,\bs_{B},f} \nonumber \\
 Q_{ab,\bs}Q_{cd,\bs_{A}}Q_{ef,\bs_{B}}
%\nonumber \\
       %q_{ab,\bs}
       \left[
\langle S^{a}_{\bs}S^{b}_{\bs}S^{c}_{\bs}S^{d}_{\bs}\rangle_{\epsilon,\varepsilon}
\langle S^{a}_{\bs_{A}}S^{b}_{\bs_{A}}S^{c}_{\bs_{A}}S^{d}_{\bs_{A}}\rangle_{\epsilon,\varepsilon}
\langle S^{a}_{\bs_{B}}S^{b}_{\bs_{B}}S^{e}_{\bs_{B}}S^{f}_{\bs_{B}}\rangle_{\epsilon,\varepsilon}
\langle S^{c}_{k}S^{d}_{k}S^{e}_{k}S^{f}_{k}\rangle_{\epsilon,\varepsilon} \right. \nonumber \\
\left. -q_{ab,\bs}q_{cd,\bs}q_{ab,\bs_{A}}q_{cd,\bs_{A}}q_{ab,\bs_{B}}q_{ef,\bs_{B}}q_{cd,k}q_{ef,k} \right]
%\nonumber \\
%&&
% \qquad
       \eeqn
       \end{widetext}
%where $\langle S^{a}S^{b}S^{c}S^{d}\rangle^{c}$'s are connected correlation functions defined as
%$\langle S^{a}S^{b}S^{c}S^{d}\rangle^{c}=\langle S^{a}S^{b}S^{c}S^{d}\rangle-\langle S^{a}S^{b}\rangle\langle S^{c}S^{d}\rangle$.
        Here the factor $(1/\sqrt{c})^{6}$ appears because 6 perceptrons 
        (two replicas for each of the three perceptrons $\bs$,$\bs_{A}$,$\bs_{B}$) are involved.
        The expression $\left(\sum_{\bs_{A},\bs_{B},k} \right)_{\rm loop}$ means to sum over $\bs_{A}$,
        $\bs_{B}$ and $k$ conditioned that the loop $\bs \to \bs_{A} \to \bs_{B} \to \bs$ is closed.
        
        Let us consider how many such loops exist for a given perceptron $\bs$.
        Starting from $0$, there are $c$ choices for $\bs_{A}$ connected to $\bs$
        and $c-1$ choices for $\bs_{B}$ (different from $\bs_{A}$) connected to $\bs$.
        Similarly there are $c$ choices for $k$ connected to $\bs_{A}$.
        Finally the probability (in a given realization of the random network)
        that $k$ happens to be connected to $\bs_{B}$ is $\sim c/N$.
        Thus
        \beq
        \left(\sum_{\bs_{A},\bs_{B},k} \right)_{\rm loop}  \sim c^{2}(c-1)\frac{c}{N}
        \eeq
        Thus the net contribution of the one-loop terms scales as
        \beq
        \frac{-\beta \tilde{F}_{n,3}[{\rm Fig.}\ref{fig_loop_appendix}]}{NM} \propto \frac{c}{N}
        \eeq
        Thus the contribution vanishes in the dense limit because $N \to \infty$ limit is taken before
        $c \to \infty $ limit.  However, in the case of global coupling $c=N$ the contribution cannot be neglected.

        \subsubsection{Higher order terms}

        Similarly to the $O(\lambda^{3})$ terms, higher order terms of  $-\beta \tilde{F}_{\rm ex}/(NM)$
        can be classified into two cases.

        \begin{itemize}
        \item
          At $O(\lambda^{p})$ ($p \geq 3$) we will have a term that is obtained by associating $2p$ replicas to the diagram
Fig.~\ref{fig_G1} whose contribution to $-\beta F_{n,p}/(NM)$ vanishes  as $1/c^{p-1}$ in $c \to \infty$ limit.

\item All other terms are associated with loop diagrams. 
  Similarly to the loop diagram considered at $O(\lambda^{3})$, we can consider more extended one-loops
          as the one shown in Fig.~\ref{fig_loop_appendix_extended} which involves $2p$ perceptrons
          ($2$ replicas for each of $p$ perceptrons) extended over $(p-1)/2+2$ layers.
          It is easy to find a notice that all such one-loops make $O(c/N)$ contributions
          to the higher order terms of $-\beta \tilde{F}_{n,p}/(NM)$ for $p \geq 3$.
          It is interesting to note that the order of the
          correction term is order $O(c/N)$ which is independent of the
          size $p$ of the loop.

          Onto the same one-loop diagram, we can associate 4 replicas: two replicas along one path
          from the right to left and the other two replicas along the other path.
          This yields a contribution to $-\beta \tilde{F}_{n,2p}/(NM)$  of order $O(c^{-p}(c/N))$.

                  \begin{figure}[h]
    \bc
  \includegraphics[width=0.5\textwidth]{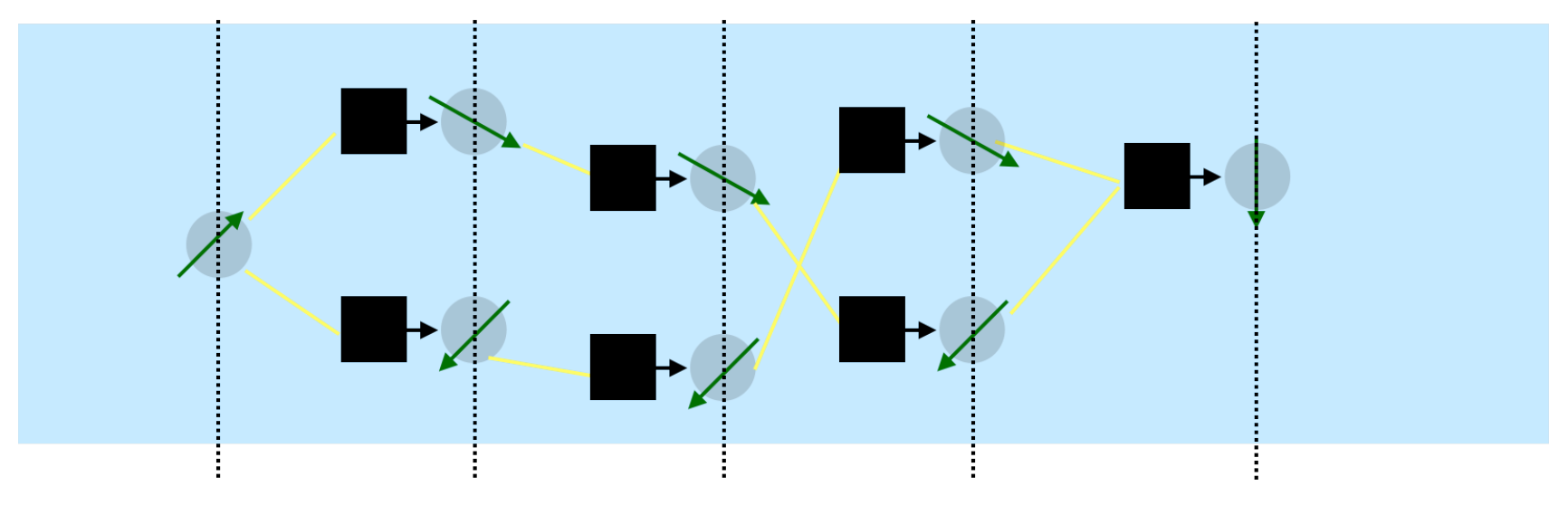}
  \ec
  \caption{
    More extended loop
%       We neglect the effects
%    of such loops (and more extended ones) in our theory.
  }
   \label{fig_loop_appendix_extended}
                  \end{figure}

                \item Contributions of two, three-loops,... can be considered similarly.
                  First one can see that the probability to close two, three loops...
                  scales as $O(c/N)^{2}$,$O(c/N)^{3}$,...
                  By associating two replicas to such diagrams we find contributions to
                  $-\beta \tilde{F}_{n,p}/(NM)$ of order $O(c/N)^{2}$,$O(c/N)^{3}$,...

                \item Note that loop corrections
                  breaks the symmetry with respect to the exchange of input/output sides.   

                \item  In general, by associating more replicas to the same diagram we find contributions
                  which vanish more rapidly increasing $c$.

%                \item The above results tells us that finite size effects due to loop corrections are stronger in denser systems.
%               \item  In th elimit  of global coupling $c=N$,
%                 loop diagrams  are present even in $N \to \infty$ limit
%                 so that the theory becomes ill defined.
%                    Finite width $N$ corrections in the case of global coupling $c=N$ can be
%                    obtained by associating more replicas to the same diagrams.
                  
          \end{itemize}

        \subsection{Summary 2}
        \label{sec-replica-summary}

        Now we can collect the above results to obtain the free-energy functional $-\beta F_{n}[\hat{Q},\hat{q}]$
        defined in sec.~\ref{sec-replicated-fee-energy-functional},
%        We can safely assume that order parameters are homogeneous at the saddle points
        %        \eq{eq-order-parameters-at-each-layers}.
        \begin{widetext}
        \beq
       \frac{ -\beta F_{n}[\{\hat{Q},\hat{q}\}]}{M}= \frac{1}{\alpha}
\sum_{l=1,2,\ldots,L}\sum_{\bs \in l} s_{\rm ent, bond}[\hat{Q}_{\bs}]+
\sum_{l=1,2,\ldots,L-1}\sum_{\bs \in l}
    s_{\rm ent, spin}[\hat{q}_{\bs}]+
    \frac{-\beta F_{\rm ex}[\{\hat{Q}_{\bs},\hat{q}_{\bs}\}]}{M}
    %\ln
%e^{-\beta \tilde{F}_{\rm ex}[\hat{Q},\hat{q},\{\partial/\partial_{\mu,\bs,a}\}]}
%\left. \prod_{\mu,\bs,a}e^{-\beta v(h_{\mu,\bs,a})} \right|_{\{h_{\mu,bs,a\}=0}}
\label{eq-free-energy-functional-summary}
\eeq
\end{widetext}
 with the first two terms being the entropic part of the free-energy due to bonds and spins
(see \eq{eq-def-G0} and \eq{eq-F-0}),
 \beqn
     c s_{\rm ent, bond}[\hat{Q}]&=&-\beta G_{n,0}^{\rm bond}[\hat{\epsilon}_{0}^{*}]
     +c \sum_{a< b} (\epsilon^{*}_{0})_{ab}Q_{ab} \nonumber \\
     M s_{\rm ent, spin}[\hat{Q}]&=&-\beta G_{n,0}^{\rm spin}[\hat{\epsilon}_{0}^{*}]
+M \sum_{a< b} (\varepsilon^{*}_{0})_{ab}q_{ab}
 \eeqn
The last term in \eq{eq-free-energy-functional-summary}
is the interaction part of the free-energy $-\beta F_{\rm ex}[\{\hat{Q}_{\bs},\hat{q}_{\bs}\}]$  (see \eq{eq-Fex}).

In the dense limit $\lim_{c \to \infty}\lim_{N \to \infty}$, we have found that only the 1st order term
$-\beta \tilde{F}_{n,1}$  (see \eq{eq-F1}) in the  Plefka expansion contributes
to  $-\beta \tilde{F}_{\rm ex}[\hat{Q},\hat{q},\{\partial/\partial_{\mu,\bs,a}\}]$. Thus we find,
 \beq
\frac{-\beta F_{\rm ex}[\{\hat{Q}_{\bs},\hat{q}_{\bs}\}]}{M}
= \sum_{l=1,2,\ldots,L-1} \sum_{\bs \in l}
(-{\cal F}_{\rm int})
\left[ \hat{\lambda}_{\bs}\right]
\label{eq-free-energy-functional-summary-ex}
\eeq
with
\beq
-{\cal F}_{\rm int}[ \hat\Lambda]=
\ln
\left. \exp
\left[
\sum_{a,b} \Lambda_{ab}
\frac{\partial^{2}}{\partial h_{a}\partial h_{b}}
\right]
\prod_{a}e^{-\beta v(h_{a})} \right |_{h=0}
\label{eq-free-energy-functional-summary-int}
\eeq
and
\beq
\lambda_{ab,\bs}=Q_{ab,\bs}q_{ab,\bs}\frac{1}{c}\sum_{k=1}^{c}q_{ab,\bs(k)}
\label{eq-def-lambda}
\eeq
On the boundaries we have $q_{ab,\bs}=q_{ab,\bs}=1$ for $\bs \in 0$ and $\bs \in L$.

Finally, assuming that order parameters are homogeneous within the layers \eq{eq-order-parameters-at-each-layers} we find the expression \eq{eq-S-total}.

The saddle point equations are
\begin{widetext}
\beqn
0=\frac{\partial}{\partial Q_{ab,\bs}} (-\beta F_{n}[\{\hat{Q},\hat{q}\}])
&=&\frac{1}{\alpha}\frac{\partial}{\partial Q_{ab,\bs}}{s}_{\rm ent,bond}[\hat{Q}_{\bs}]+\sum_{\ws}\frac{\partial\Lambda_{\ws}}{\partial Q_{ab,\bs}}(-{\cal F}_{\rm int})'\left[\Lambda_{\ws}\right] \nonumber \\
0=\frac{\partial}{\partial q_{ab,\bs}} (-\beta F_{n}[\{\hat{Q},\hat{q}\}])
&=&\frac{1}{\alpha}\frac{\partial}{\partial Q_{ab,\bs}}{s}_{\rm ent,bond}[\hat{Q}_{\bs}]+\sum_{\ws}\frac{\partial\Lambda_{\ws}}{\partial q_{ab,\bs}}(-{\cal F}_{\rm int})'\left[\Lambda_{\ws}\right] 
\label{eq-SP-general}
\eeqn
\end{widetext}

\subsection{Franz-Parisi's potential in the replica symmetric ansatz}
\label{sec-FP-potential-RS}

Here we display the expressions for the Franz-Parisi's potential
within the replica symmetric ansatz needed to evaluate the generalization error.

From (128) of \cite{yoshino2020complex} we find,
\begin{widetext}
\beqn
\hspace*{-1cm} &&
s_{\rm ent,spin}[\hat{\epsilon}^{1+s},\hat{q}^{1+s}]
=s\epsilon_{r}r+\frac{1}{2}\epsilon_{r}+\frac{s}{2}\sum_{i=0}^{k}\epsilon_{i}q_{i}(m_{i}-m_{i+1})
+\frac{s}{2} \epsilon_{k} \nonumber \\
\hspace*{-2cm} && \hspace*{1cm}+
\ln 
\exp \left[\frac{\Lambda^{\rm Ising}_{\rm com}}{2}\sum_{a,b=0}^{s}\frac{\partial^{2}}{\partial h_{a}\partial h_{b}}\right]
\prod_{i=0}^{k}
 \left. \exp \left[
  \frac{\Lambda^{\rm Ising}_{i}}{2}\sum_{a,b=1}^{s}I_{ab}^{m_{i}}\frac{\partial^{2}}{\partial h_{a}\partial h_{b}}\right]
 \prod_{a=0}^{s} (2\cosh(h_{a})) \right |_{\{h_{a}=0\}}
 \nonumber \\
\hspace*{-2cm} &&  \hspace*{1cm}=s\epsilon_{r}r+\frac{1}{2}\epsilon_{r}+\frac{s}{2}\sum_{i=0}^{k}\epsilon_{i}q_{i}(m_{i}-m_{i+1}) 
+\frac{s}{2} \epsilon_{k} \nonumber \\
\hspace*{-2cm} && \hspace*{2cm}+\left. \ln \gamma_{\Lambda_{\rm com}} \otimes (2\cosh(h) \gamma_{\Lambda_{0}^{\rm Ising}} \otimes e^{-s f^{\rm Ising}(m_{1},h)}
 \right |_{h=0} \qquad
 \label{eq-S-ent-q-k-RSB-teacher-student}
 \eeqn
 Then we find
 \beqn
 \partial_{s} \left.  s_{\rm ent,spin}[\hat{\epsilon}^{1+s},\hat{q}^{1+s}] \right|_{s=0}
 &=&\epsilon_{r}r+\frac{1}{2}\sum_{i=0}^{k}\epsilon_{i}q_{i}(m_{i}-m_{i+1})+\frac{1}{2} \epsilon_{k} \nonumber \\
 &+&\frac{
   \int Dz_{\rm com} \left(2\cosh(\sqrt{\Lambda_{\rm com}}z_{\rm com}\right)
      \int Dz_{0} (-f^{\rm Ising}(m_{1},\sqrt{\Lambda_{\rm com}}z_{\rm com}+\sqrt{\Lambda^{\rm Ising}_{0}}z_{0}))
 }{
   \int Dz_{\rm com} \left(2\cosh(\sqrt{\Lambda_{\rm com}}z_{\rm com}\right)
 }
 \eeqn
 For the interaction part of the free-energy we find from (134) of \cite{yoshino2020complex},
   \beqn
&& \left. -\partial_{s}     {\cal F}_{\rm int}[\hat{q}^{1+s}(l-1),\hat{Q}^{1+s}(l),\hat{q}^{1+s}(l)]  \right |_{s=0}
=
-\partial_{s} \ln 
\exp \left[ \frac{\Lambda_{\rm com}(l)}{2} \sum_{a,b=0}^{s}\frac{\partial^{2}}{\partial h_{a}\partial h_{b}}\right]
\exp \left[ \frac{\Lambda_{\rm teacher}(l)}{2} \frac{\partial^{2}}{\partial h^{2}_{0}}\right]
\nonumber \\
&& \left.  \prod_{i=0}^{k+1}
 \left. \exp \left[
  \frac{\Lambda_{i}(l)}{2}\sum_{a,b=1}^{s}I_{ab}^{m_{i}}\frac{\partial^{2}}{\partial h_{a}\partial h_{b}}\right]
 \prod_{a=0}^{s} e^{-\beta v(r(h_{a}))} \right |_{\{h_{a}=0\}}  \right |_{s=0}
\nonumber \\
&&= -\partial_{s} \ln  \int D z_{\rm com} \int D z_{\rm teacher} e^{-\beta v(\sqrt{\Lambda_{\rm com}(l)}z_{\rm com}+\sqrt{\Lambda_{\rm teacher}(l)}z_{\rm teacher})} \nonumber \\
&& \left. \hspace*{2cm}\int D z_{0}e^{-sf(m_{1},\sqrt{\Lambda_{\rm com}(l)}z_{\rm com}+\sqrt{\Lambda_{0}(l)}z_{0})}  \right |_{s=0}  \nonumber \\
&&=\frac{\int Dz_{\rm com}g_{\rm teacher}(\sqrt{\Lambda_{\rm com}(l)}z_{\rm com})
  \int Dz_{0}(-f(m_{1},\sqrt{\Lambda_{\rm com}(l)}z_{\rm com}+\sqrt{\Lambda_{0}(l)}z_{0}))
}{\int Dz_{\rm com}g_{\rm teacher}(\sqrt{\Lambda_{\rm com}(l)}z_{\rm com})}
\label{eq-F-int-k-RSB-teacher-student}
\eeqn
where we introduced (see (147) of \cite{yoshino2020complex}),
\beq
g_{\rm teacher}(h) \equiv \int Dz_{\rm teacher} e^{-\beta v(h-\sqrt{\Lambda_{\rm teacher}}z_{\rm teacher})}
\eeq
 \end{widetext}

\subsection{Quadratic and cubic expansions of the free-energy}
%\subsection{Quadratic expansions of the free-energy}
\label{sec-hessian}

Here expansion of the free-energy functional given by \eq{eq-free-energy-functional-summary} supplemented by \eq{eq-free-energy-functional-summary-ex},
\eq{eq-free-energy-functional-summary-int}
and \eq{eq-def-lambda} around the saddle point given by  \eq{eq-SP-general}.
We can write
\beq
Q_{ab,\bs}=Q^{*}_{ab}(l)+\Delta Q_{ab,\bs} \qquad q_{ab,\bs}=q^{*}_{ab}(l)+\Delta q_{ab,\bs}\eeq
where $Q^{*}_{ab}(l)$ and $q^{*}_{ab}(l)$ are the saddle point values of the order parameters, $l$ is the label of the layer to which $\bs$ belongs to, $\Delta Q_{ab,\bs}$ and $\Delta q_{ab,\bs}$ are fluctuations around the saddle point.

\subsubsection{Quadratic expansion}

The quadratic expansion of the replicated free-energy functional
is specified in the Hessian matrix. It is obtained as,
\begin{widetext}
\beqn
    H^{QQ}_{ab,cd,\bs_{1},\bs_{2}}&=&
    \frac{\partial^{2}}{\partial Q_{ab,\bs_{1}} \partial Q_{cd,\bs_{2}}}
    \frac{(\beta F_{n})[\{\hat{Q},\hat{q}\}]}{M}
    =
    -\delta_{\bs_{1},\bs_{2}} \frac{1}{\alpha}
    \frac{\partial^{2}}{\partial Q^{2}_{ab,\bs_{1}}}{s}_{\rm ent, bond}[\hat{Q}_{\bs_{1}}]
    \nonumber \\
   & -& \sum_{\ws}\frac{\partial^{2}\Lambda_{\ws}}{\partial Q_{ab,\bs_{1}}\partial Q_{ab,\bs_{2}}}(-{\cal F}_{\rm int})'\left[\Lambda_{\ws}\right]
    - \sum_{\ws}
    \frac{\partial\Lambda_{\ws}}{\partial Q_{ab,\bs_{1}}}
    \frac{\partial\Lambda_{\ws}}{\partial Q_{ab,\bs_{2}}}
    (-{\cal F}_{\rm int})''\left[\Lambda_{\ws}\right] \nonumber \\
   & =&
        -\delta_{\bs_{1},\bs_{2}} \left[ \frac{1}{\alpha}
        \frac{\partial^{2}}{\partial Q^{2}_{ab,\bs_{1}}}{s}_{\rm ent, bond}[\hat{Q}_{\bs_{1}}]
        + \frac{\partial^{2}\Lambda_{\bs_{1}}}{\partial Q^{2}_{ab,\bs_{1}}}(-{\cal F}_{\rm int})'\left[\Lambda_{\bs_{1}}\right]
    + 
  \left(  \frac{\partial\Lambda_{\bs_{1}}}{\partial Q_{ab,\bs_{1}}} \right)^{2}
    (-{\cal F}_{\rm int})''\left[\Lambda_{\bs_{1}}\right]\right]
    \nonumber \\
        H^{Qq}_{ab,cd,\bs_{1},\bs_{2}}&=&
        \frac{\partial^{2}}{\partial Q_{ab,\bs_{1}} \partial q_{cd,\bs_{2}}}
        \frac{(\beta F)[\{\hat{Q},\hat{q}\}]}{M}
        \nonumber \\
           & =& -\sum_{\ws}\frac{\partial^{2}\Lambda_{\ws}}{\partial Q_{ab,\bs_{1}}\partial q_{ab,\bs_{2}}}(-{\cal F}_{\rm int})'\left[\Lambda_{\ws}\right]
    - \sum_{\ws}
    \frac{\partial\Lambda_{\ws}}{\partial Q_{ab,\bs_{1}}}
    \frac{\partial\Lambda_{\ws}}{\partial q_{ab,\bs_{2}}}
    (-{\cal F}_{\rm int})''\left[\Lambda_{\ws}\right]
    \nonumber \\
    &=&
    -\frac{\partial^{2}\Lambda_{\bs_{1}}}{\partial Q_{ab,\bs_{1}}\partial q_{ab,\bs_{2}}}(-{\cal F}_{\rm int})'\left[\Lambda_{\bs_{1}}      \right]
    -
    \frac{\partial\Lambda_{\bs_{1}}}{\partial Q_{ab,\bs_{1}}}
    \frac{\partial\Lambda_{\bs_{1}}}{\partial q_{ab,\bs_{2}}}
    (-{\cal F}_{\rm int})''\left[\Lambda_{\bs_{1}}\right]
   \nonumber \\
            H^{qq}_{ab,cd,\bs_{1},\bs_{2}}&=&
            \frac{\partial^{2}}{\partial q_{ab,\bs} \partial q_{cd,\ws}}
            \frac{(\beta F)[\{\hat{Q},\hat{q}\}]}{M}
            =
            -\delta_{\bs_{1},\bs_{2}} 
            \frac{\partial^{2}}{\partial q^{2}_{ab,\bs_{1}}}{s}_{\rm ent, spin}[\hat{q}_{\bs_{1}}]\nonumber \\
               & -& \sum_{\ws} \frac{\partial^{2}\Lambda_{\ws}}{\partial q_{ab,\bs_{1}}\partial q_{ab,\bs_{2}}}(-{\cal F}_{\rm int})'\left[\Lambda_{\ws}\right]
    - \sum_{\ws}
    \frac{\partial\Lambda_{\ws}}{\partial q_{ab,\bs_{1}}}
    \frac{\partial\Lambda_{\ws}}{\partial q_{ab,\bs_{2}}}
    (-{\cal F}_{\rm int})''\left[\Lambda_{\ws}\right]
    \eeqn
where
    \beqn
    \frac{\partial\Lambda_{\bs_{1}}}{\partial Q_{ab,\bs_{1}}}
     = q_{ab,\bs_{1}}\frac{1}{c}\sum_{k=1}^{c} q_{ab,\bs_{1}(k)}
\qquad 
    \frac{\partial^{2}\Lambda_{\bs_{1}}}{\partial Q^{2}_{ab,\bs_{1}}}
     = 0
    \eeqn
    and
        \beqn
  \frac{\partial^{2}\Lambda_{\bs_{1}}}{\partial Q_{ab,\bs_{1}}\partial q_{ab,\bs_{2}}}
   & =&\delta_{\bs_{1},\bs_{2}}\frac{1}{c}\sum_{k=1}^{c}q_{ab,\bs_{1}(k)}
    +\frac{1}{c}q_{ab,\bs_{1}}I_{\partial \bs_{1}}(\bs_{2}) \nonumber \\
    \frac{\partial\Lambda_{\bs_{1}}}{\partial Q_{ab,\bs_{1}}}
    \frac{\partial\Lambda_{\bs_{1}}}{\partial q_{ab,\bs_{2}}}    
    & =&\delta_{\bs_{1},\bs_{2}}q_{ab,\bs_{1}}Q_{ab,\bs_{1}}
    \left(\frac{1}{c}\sum_{k=1}^{c}q_{ab,\bs_{1}(k)}\right)^{2}
    +\frac{1}{c}q^{2}_{ab,\bs_{1}}Q_{ab,\bs_{1}}
    \left(\frac{1}{c}\sum_{k=1}^{c}q_{ab,\bs_{1}(k)}\right)
    I_{\partial \bs_{1}}(\bs_{2}) 
    \eeqn
    and
    \beqn
 \sum_{\ws}   \frac{\partial^{2}\Lambda_{\ws}}{\partial q_{ab,\bs_{1}}\partial q_{ab,\bs_{2}}}
    & =&
 \frac{1}{c}\left (Q_{ab,\bs_{1}}    I_{\partial \bs_{1}}(\bs_{2})+
                   Q_{ab,\bs_{2}}    I_{\partial \bs_{2}}(\bs_{1})
                   \right)
 \nonumber \\
\sum_{\ws}    \frac{\partial\Lambda_{\ws}}{\partial q_{ab,\bs_{1}}}
    \frac{\partial\Lambda_{\ws}}{\partial q_{ab,\bs_{2}}}    
    & =& \delta_{\bs_{1},\bs_{2}} \left(Q_{ab,\bs_{1}}\frac{1}{c}\sum_{k=1}^{c}q_{ab,\bs_{1}(k)} \right)^{2} \nonumber \\
    &&    +\frac{1}{c}Q^{2}_{ab,\bs_{1}}q_{ab,\bs_{1}}\frac{1}{c}\sum_{k=1}^{c}q_{ab,\bs_{1}(k)}I_{\partial \bs_{1}}(\bs_{2})
    %\nonumber \\
    %    &&
    +\frac{1}{c}Q^{2}_{ab,\bs_{2}}q_{ab,\bs_{2}}\frac{1}{c}\sum_{k=1}^{c}q_{ab,\bs_{2}(k)}I_{\partial \bs_{2}}(\bs_{1}) \nonumber \\
    &+& \frac{1}{c^{2}}\sum_{\ws}\left(q_{ab,\ws}Q_{ab,\ws}\right)^{2}
    I_{\partial \ws}(\bs_{1})    I_{\partial \ws}(\bs_{2})
    \eeqn
    where $I_{A}(x)$ is the indicator function, i.~e. $I_{a}(x)=1$ if $x \in a$ and $0$ otherwise.

    Let us note that in the liquid phase where $Q_{ab}=q_{ab}=0$ for $a \neq b$, the Hessian matrix  become simplified as,
    \beqn
    H^{QQ}_{ab,cd,\bs_{1},\bs_{2}}&=&
    -\delta_{\bs_{1},\bs_{2}} \left. \frac{1}{\alpha}
    \frac{\partial^{2}}{\partial Q_{ab,\bs_{1}}\partial Q_{cd,\bs_{1}}}s_{\rm ent, bond}[\hat{Q}_{\bs_{1}}]\right |_{\hat{Q}_{\bs_{1}}=0}  \nonumber \\
    H^{Qq}_{ab,cd,\bs_{1},\bs_{2}}&=& 0 \nonumber \\
    H^{qq}_{ab,cd,\bs_{1},\bs_{2}}&=&
                -\delta_{\bs_{1},\bs_{2}}  \left.
            \frac{\partial^{2}}{\partial q_{ab,\bs_{1}}\partial q_{cd,\bs_{1}}}s_{\rm ent, spin}[\hat{q}_{\bs_{1}}] \right |_{\hat{q}_{\bs_{1}}=0}
\label{eq-hessian-liquid-phase}
\eeqn

    \end{widetext}

            \subsubsection{Cubic expansion}
            \label{sec-cubic-expansion}

Here let us analyze the cubic expansion. For simplicity let us only consider the liquid phase where $Q_{ab}=q_{ab}=0$ for $a\neq b$. We find the only non-vanishing contribution
is due to,
\begin{widetext}
\beqn
   W^{qQq}_{ab,cd,ef,\bs_{1},\bs_{2},\bs_{3}} &=&
   \left.     \frac{\partial^{3}}{\partial q_{ab,\bs_{1}} \partial Q_{cd,\bs_{2}}\partial _{ef,\bs_{3}}} \frac{(\beta F_{n})[\{\hat{Q},\hat{q}\}]}{M}  \right |_{\hat{Q}=\hat{q}=0}
\nonumber \\
&   = &  \left.
   -\frac{\partial^{3}\Lambda_{\bs_{2}}}{\partial q_{ab,\bs_{1}}\partial Q_{ab,\bs_{2}}\partial q_{ab,\bs_{3}}}(-{\cal F}_{\rm int})'\left[\Lambda_{\bs_{2}}      \right]
   \right |_{\hat{Q}=\hat{q}=0} \delta_{(ab),(cd)}\delta_{(cd),(ef)} \nonumber \\
   &  = & -\frac{1}{c} \left[ I_{\partial \bs_{2}}(\bs_{1}) \delta_{\bs_{2},\bs_{3}}
     +\delta_{\bs_{2},\bs_{1}}I_{\partial \bs_{2}}(\bs_{3}) \right]
   \left. (-{\cal F}_{\rm int})'\left[\Lambda_{\bs_{2}}      \right]
   \right |_{\hat{Q}=\hat{q}=0}
   \delta_{(ab),(cd)}\delta_{(cd),(ef)}
   \label{eq-cubic-liquid-phase}
   \eeqn
   \end{widetext}
It is interesting to note that this cubic term breaks the symmetry concerning the exchange of input/output sides.   

   \subsubsection{Correction to the saddle point}
   \label{section-correction-to-SP}
   Now let us turn to corrections due to fluctuations around the saddle point.
These give finite connectivity $c$ or $M=c\alpha$ corrections ($\alpha$ is fixed).

We can write
    \beq
    Q_{ab,\bs}=Q^{*}_{ab}(l)+\Delta Q_{ab,\bs} \qquad q_{ab,\bs}=q^{*}_{ab}(l)+\Delta Q_{ab,\bs} 
    \eeq
    where $Q^{*}_{ab}(l)$ and $q^{*}_{ab}(l)$ are the saddle point values of the order parameters, $l$ is the label of the
    layer to which $\bs$ belongs to, $\Delta Q_{ab,\bs}$ and $\Delta q_{ab,\bs}$ are fluctuations around the saddle point.
    Including the correction due to the fluctuations around the saddle point,
    the replicated Gardner volume \eq{eq-replicated-gardner-volume} can be written as,
    \begin{widetext}
    \beq
    \overline{V^{1+s}({\bf S}_{0},{\bf S}_{L}({\bf S}_{0},{\cal J}_{\rm teacher})))}^{{\bf S}_{0},{\cal J}_{\rm teacher}}
    =e^{NM s_{1+s}[\{{\hat Q}^{*},{\hat q}^{*}\}]}Z_{\rm fluctuation}
    \eeq
where
    \beqn
Z_{\rm fluctuation}&=&\int \prod_{\bs}\prod_{a < b}d\Delta Q_{ab,\bs}d\Delta q_{ab,\bs}
\exp \left[      - \frac{M}{2}\sum_{a < b}\sum_{c < d}\sum_{\bs,\ws} [
    H^{QQ}_{ab,cd,\bs,\ws} \Delta Q_{ab,\bs}\Delta Q_{cd,\ws} \right. 
  \nonumber \\
  &&
%  \left.
  \left. \left.
  +
  H^{Qq}_{ab,cd,\bs,\ws} \Delta Q_{ab,\bs}\Delta q_{cd,\ws}
  +
  H^{qq}_{ab,cd,\bs,\ws} \Delta q_{ab,\bs}\Delta q_{cd,\ws}
  \right] \right. \nonumber \\
  &&
  \left.  \left.
  -\frac{M}{3!}\sum_{a < b}\sum_{c < d}\sum_{e < f}\sum_{\bs_{1},\bs_{2},\bs_{3}}
  \left[W_{ab,cd,ef,\bs_{1},\bs_{2},\bs_{3}}^{qQq}\Delta q_{ab,\bs_{1}}\Delta Q_{cd,\bs_{2}}\Delta q_{ef,\bs_{3}} + \ldots\right]
    \right]
  \right]
    \eeqn
    \end{widetext}
where $H^{QQ}_{ab,cd,\bs,\ws}...$ are the Hessian matrices given in sec.~\ref{sec-hessian}.

For the following discussion, we do not need to perform a complete analysis of the correction.
We restrict ourselves in the liquid phase $Q=q=0$.
Then as shown in sec.~\ref{sec-hessian}, the Hessian matrices become completely local, i.~e.
$H^{QQ}_{\bs,\ws}=\delta_{\bs,\ws}H^{QQ}_{\bs,\bs}$ and
$H^{qq}_{\bs,\ws}=\delta_{\bs,\ws}H^{qq}_{\bs,\bs}$ while $H^{Qq}_{\bs,\ws}=0$
(see \eq{eq-hessian-liquid-phase}). Thus at the quadratic level of fluctuations,
there is no correlation between different layers in the liquid phase.
In the cubic order, we find
$W^{qQq}_{\bs_{1},\bs_{2},\bs_{3}} \propto \frac{1}{c} \left[ I_{\partial \bs_{2}}(\bs_{1}) \delta_{\bs_{2},\bs_{3}}+\delta_{\bs_{2},\bs_{1}}I_{\partial \bs_{2}}(\bs_{3}) \right]$ (see   \eq{eq-cubic-liquid-phase}).
This will induce correlations between different layers even in the liquid phase.
And this will be enhanced next to the frozen wall and enhanced by correlation in the frozen wall (the hidden manifold model).

To understand the key point it is sufficient to consider a simplified model.
\begin{widetext}
    \beq
   Z=\int \prod_{\bs} dx_{\bs} dy_{\bs}
   \exp \left[-\frac{M}{2}\sum_{\bs} h_{xx}x_{\bs}^{2}
-\frac{M}{2}\sum_{\bs} h_{yy}y_{\bs}^{2}
     -\alpha w\sum_{\bs}\sum_{\ws \in \partial \bs} y_{\bs}x_{\bs}y_{\ws}  \right]
   \eeq
   Then by introducing 
    \beq
    Z_{0}=
    \int \prod_{\bs \in (1,2,\ldots,L)} dx_{\bs}
    \exp \left[-\frac{M}{2}\sum_{\bs} h_{xx}x_{\bs}^{2}\right]
        \int \prod_{\bs \in (1,2,\ldots,L-1)} dy_{\bs}
    \exp \left[-\frac{M}{2}\sum_{\bs} h_{yy}y_{\bs}^{2}\right]
    = \left(\sqrt{\frac{2\pi}{Mh_{xx}}}\right)^{NL}
    \left(\sqrt{\frac{2\pi}{Mh_{yy}}}\right)^{N(L-1)}
    \eeq
    and
    \beq
    \langle \cdots \rangle_{x,y}
    =\frac{
    \int \prod_{\bs} dx_{\bs}\prod_{\bs} dy_{\bs}
    \exp \left[-M\frac{h_{xx}}{2}\sum_{\bs} x_{\bs}^{2}
      -M\frac{h_{yy}}{2}\sum_{\bs} y_{\bs}^{2}
      \right] \ldots
    }{    \int \prod_{\bs} dx_{\bs} \prod_{\bs} dy_{\bs}
      \exp \left[-M\frac{h_{xx}}{2}\sum_{\bs} x_{\bs}^{2}
        -M\frac{h_{yy}}{2}\sum_{\bs} y_{\bs}^{2}\right]}
    \eeq
we can write
\beqn
\ln Z - \ln Z_{0} &= &\ln \left \langle 
\exp \left[
  -\alpha w \sum_{\bs}\sum_{\ws \in \partial \bs} y_{\bs}x_{\bs}y_{\ws}   
  \right]
\right \rangle_{x} \nonumber \\
&=&
-\alpha  w\sum_{\bs}\sum_{\ws \in \partial \bs} \langle y_{\bs}x_{\bs}y_{\ws}
\rangle_{xy}
+\frac{1}{2}\left(\alpha w\right)^{2}
\sum_{\bs_{1}}\sum_{\ws_{1} \in \partial \bs_{1}}
\sum_{\bs_{2}}\sum_{\ws_{2} \in \partial \bs_{2}}
\langle
y_{\bs_{1}}x_{\bs_{1}}y_{\ws_{1}}
y_{\bs_{2}}x_{\bs_{2}}y_{\ws_{2}}
]\rangle_{xy} + \ldots \nonumber \\
&=&
\frac{1}{2}\left(\alpha w \right)^{2}
\sum_{\bs}\sum_{\ws \in \partial \bs}
\langle
\underbrace{y^{2}_{\bs}x^{2}_{\bs}y^{2}_{\ws}}_{M^{3}(h_{yy}h_{xx}h_{yy})^{-1}}
\rangle_{xy}+\ldots
\eeqn
    \end{widetext}

 \end{appendix}

%\bibliography{ref_yoshino}
\bibliography{teacher_student.v7}

\end{document}